\documentclass{article}

\usepackage[utf8]{inputenc}

\usepackage{hyperref}
\hypersetup{colorlinks=false}
\usepackage{lscape}

\usepackage{amsmath}
\usepackage{graphicx}
\usepackage{caption}
\usepackage{subcaption}

\usepackage[colorinlistoftodos]{todonotes}
\usepackage{float}
\usepackage{multirow} 
\usepackage{slashed}
\usepackage{booktabs}
\usepackage{latexsym}

\usepackage{fancyhdr}
\usepackage{graphics,psfrag}
\usepackage{mathrsfs}
\usepackage{braket}
\usepackage{amssymb}
\usepackage{authblk}
\usepackage{csquotes}

\graphicspath{{media/}} 
\usepackage[left=2.5cm,right=2.5cm,top=2cm,bottom=2cm]{geometry}
\numberwithin{equation}{section}

\usepackage[
backend=biber,
style=numeric,
sorting=none
]{biblatex}

\title{Lepton flavor violation within the Simplest Little Higgs model}

\author[1]{Enrique Ramírez}
\author[1]{Pablo Roig} 
\affil[1]{Departamento de Física, Centro de Investigación y de Estudios Avanzados del Instituto Politécnico Nacional.

Ciudad de México, México.

\texttt{\{eramirez, proig\}@fis.cinvestav.mx}
}

\date{\today}

\hypersetup{
    breaklinks=true,
    linktocpage=true,
    colorlinks=true,
    linkcolor=blue,
    filecolor=magenta,      
    urlcolor=cyan,
    citecolor=blue,
    }
\DeclareUnicodeCharacter{2212}{-}

\begin{document}

\maketitle

\begin{abstract}
    We carry out an exhaustive analysis of lepton flavor violating processes within the Simplest Little Higgs model. Its discovery could be expected from either $\mu\to e$ conversion in nuclei,  $\mu\to e\gamma$ or $\mu\to3e$ decays. Then, the tau sector could help discriminating this model not only via $\tau\to\ell\gamma$ ($\ell=\mu,e$) and $\tau\to3\ell$  decays, but also by means of $\ell\to\tau$ conversion in nuclei, which is promising in this respect. Although the model violates slightly custodial symmetry, acommodating the recent CDF $M_W$ measurement is in tension with electroweak precision data.
\end{abstract}

\clearpage

\tableofcontents

\clearpage

\section{Introduction}
With the discovery of the Higgs boson \cite{ATLAS:2012yve, CMS:2012qbp} the Standard Model of particle physics (SM) \cite{Glashow:1961tr, Weinberg:1967tq, Salam:1968rm} was completed. Powerful and praiseworthy as it is, a Higgs mass value in the electroweak scale, $v$, calls for a deeper understanding  of the hierarchy concept. 

Assuming the SM is a low-energy effective theory of a more general high-energy theory, generalizing it at a high-energy scale $\Lambda>>v$,  raises the question: Is there new physics between these two energy scales? \textit{Hierarchy problem} or \textit{the fine tuning problem} \cite{Georgi:1974yf} is the reason motivating the existence of new physics, that lies between $v$ and $\Lambda$. In the context of the SM, it means that the Higgs boson mass receives quadratically divergent loop contributions which are much larger than its measured value, and require a correspondingly large bare mass value so that the fine-tuned cancellation between both yields the observed  $m_H\sim125$ GeV. These leading quantum corrections are only cancelled when the parameters are fine-tuned. Nevertheless after the LEP experiment and their Electroweak Precision Data (EWPD) \cite{LEPWorkingGroupforHiggsbosonsearches:2003ing} a Little hierarchy problem emerged \cite{Barbieri:2000gf}, given that LEP measurements prevented new physics near $v$. There must be then a little hierarchy between $v$ and the lightest new physics scale, which should lie  above the TeV. 

Many beyond SM theories have been used to alleviate the hierarchy problem, like supersymmetry, technicolor, extra dimensions and Little Higgs. Our approach belongs to the last one. Little Higgs models \cite{Arkani-Hamed:2001kyx, Arkani-Hamed:2001nha,Arkani-Hamed:2002sdy,Arkani-Hamed:2002ikv,Schmaltz:2005ky,Perelstein:2005ka} postpone the hierarchy problem in the SM, introducing adequate new particles under an enlarged symmetry at an energy scale of some TeVs. All these models are based on the idea that the Higgs is a Pseudo Nambu-Goldstone boson (pNGB), which arises from some approximate spontaneously broken symmetry at a scale $f \gtrsim$ TeV. This new symmetry is introduced  to protect the Higgs mass from large quantum corrections and the Higgs fields are taken to be NGBs corresponding to a spontaneously broken global symmetry of a new strongly interacting sector. As a result of this novel non-perturbative dynamics, additional new physics is expected at a scale $\sim 4\pi f$.

Through the years many Little Higgs models have already been constructed, in which the new particles
depend on the particular symmetry of the model. We can divide Little Higgs models into two categories~\cite{Han:2005ru}: \textit{product group models}, where the SM gauge group arises from the diagonal breaking of two or more gauge groups, i.e, $\left( SU(2) \times U(1)\right)^{N}$ and \textit{Simple group models} where the SM gauge group stems from the breaking of a single larger group, i.e, $SU(N) \times U(1)$. One the most important product group model realizations of the Little Higgs model is the T-parity extension proposed by Cheng and Low \cite{Cheng:2004yc}. LFV has been extensively studied within this model \cite{delAguila:2008zu, delAguila:2010nv, delAguila:2017ugt, delAguila:2019htj, DelAguila:2019xec, Illana:2021uwu, Pacheco:2021djh, Yang:2016hrh}. In these models, there is no need to enlarge the SM matter sector and the collective breaking can be realized with just one sigma model, although there is more freedom related to the extra gauge couplings, and a discrete symmetry needs to be imposed to comply with EWPD. 
On the other hand, a simple group model that is popular by its minimality ($N=3$ above)  is the Simplest Little Higgs (SLH) that was proposed by Kaplan and Schmaltz \cite{Schmaltz:2004de, Kaplan:2003uc}, which we will use. In this case an additional fermion field is needed for the SM doublets to become triplets. In the lepton sector this is a heavy quasi-Dirac neutrino, which drives LFV. The situation is more involved in the quark counterpart, where there are two possible embeddings, as we will explain.  LFV has already been studied within the SLH in slightly  different approaches~\cite{delAguila:2011wk, Lami:2016vrs, Lami:2016mjf, Han:2011aq, Wang:2012zza}~\footnote{Only ref. \cite{Lami:2016mjf} considered  three heavy neutrinos with general mixing, as we do here.}. However, leptonic tau decays and $\ell - \tau$ conversion in nuclei did not receive much attention because they are less restrictive than the analogous muon processes, according to  experimental limits, several orders of magnitude weaker for taus. We will include them here for the first time, mainly to increase the model-discriminating power adding these observables to our toolkit. We will not discuss Z \cite{Han:2011aq} or Higgs \cite{Lami:2016mjf} LFV decays  as their branching fractions turn out to be $\leq10^{-11}$ and  $\leq10^{-12}$, respectively~\footnote{These upper bounds correspond to the range of the SLH model parameters that we study, see section \ref{sec:Num}.}, far away from current or near-future bounds. Similarly, we will not address semileptonic LFV $\tau$ decays as purely lepton LFV $\tau$ decays have always a few orders of magnitude larger branching fractions \cite{delAguila:2011wk, Lami:2016vrs}.

LFV in the charged lepton sector is long sought as it will surely be due to new physics, given the GIM-like \cite{Glashow:1970gm} suppression of SM contributions in presence of massive neutrinos \cite{Petcov:1976ff, Bilenky:1977du, Cheng:1977nv}. There are very stringent bounds \cite{ParticleDataGroup:2020ssz} from MEG \cite{MEG:2013oxv}, SINDRUM \cite{SINDRUM:1987nra}, SINDRUM-II \cite{SINDRUMII:2006dvw}, BaBar \cite{BaBar:2009hkt} and Belle \cite{Belle:2021ysv}. There also is and will be a plethora of experiments contributing to this quest: MEG-II \cite{MEGII:2018kmf}, PRISM/PRIME \cite{PRISM:2013fvg}, Mu2e \cite{Mu2e:2014fns}, Mu3e \cite{Hesketh:2022wgw}, COMET \cite{Lee:2022moh}, DeeMe \cite{Teshima:2019orf}, Belle II \cite{Belle-II:2018jsg}, ...  enhancing the case for studying the related phenomenology. In the case of $\ell - \tau$ nuclei conversion there are still no experimental limits for this phenomenon  (recently been studied \cite{Abada:2016vzu, Husek:2020fru, Husek:2020kxu}). Ref. \cite{Gninenko:2018num} pointed out that the NA64  experiment could be able to search for it,   as well as the proposed muon collider \cite{Delahaye:2013jla} or the electron-ion collider \cite{Deshpande:2013paa, Cirigliano:2021img}, among others. Indeed, as $\mu\to e$ conversion in nuclei is synergic with the LFV $\mu$ decays in $\mu\to e$ transitions; we will find that in the $\tau\leftrightarrow\ell$ ($\ell=e,\mu$) processes, conversion in nuclei will put significant constraints together with the purely leptonic $\tau$ LFV decays.

This work is divided into the following parts, in  section \ref{sec:feynman} we quote all necessary Feynman rules using the 't Hooft-Feynman gauge. After that, in section \ref{sec:LFV} we develop the full structure of the LFV processes. Then, in  section \ref{sec:Num} we show our numerical results for all LFV lepton decays and conversions in nuclei. Next, section \ref{sec:MW} discusses the implications of the recent CDF $M_W$ measurement in light of the SLH model. Finally,  in section \ref{sec:Con} we present our conclusions.

\section{ Particle Content and Feynman Rules in the SLH}
\label{sec:feynman}

We are going to develop the main characteristics of this model first introduced in refs. \cite{Schmaltz:2004de, Kaplan:2003uc}. The Higgs fields are Goldstone bosons which are associated with a new global symmetry breaking at a high scale $f\sim\mathcal{O}(10$ TeV). The Higgs fields will acquire a mass and become pseudo-Goldstone bosons via collective symmetry breaking at the electroweak scale, $v$. This mass will be light compared to $f$, since it is protected by the approximate global symmetry and is free from quadratic divergences at one-loop. Through this section we develop the fields expansion of the theory. We remind this was done in ref. \cite{Han:2005ru} using the unitary gauge, though we follow the notation of ref. \cite{delAguila:2011wk} and use the 't Hooft-Feynman gauge
. 

The SLH model is constructed by expanding the SM $SU(3)_{c} \times SU(2)_{L} \times U(1)_{Y}$ gauge group to $SU(3)_{c} \times SU(3)_{L} \times U(1)_{x}$. In this case  the $SU(2)$ doublets of the SM have to be enlarged to $SU(3)$ triplets and  additional $SU(3)_L$ gauge bosons appear. The subscript $x$ indicates a new \textit{x-hypercharge}. Following the usual convention, the quantum numbers of the fundamental fermions in the model will be indicated using the notation:
 
 \begin{equation}
     \left( \textrm{color representation}, \textrm{weak multiplet representation} \right)_{\textrm{x-Hypercharge}}. 
     \end{equation}
     
The $SU(3)_{L} \times U(1)_{x}$ gauge symmetry is broken down to the SM electroweak gauge group by two scalar non-linear sigma fields $\Phi_{1}$ and $\Phi_{2}$, which transform as complex triplets. The model contains a global $[SU(3) \times U(1)]^2$ symmetry. The diagonal subgroup is gauged so the gauge symmetry is $[SU(3) \times U(1)]$. The symmetry is spontaneously broken, $[SU(3) \times U(1)]^2 \longrightarrow [SU(2) \times U(1)]^2$, by the vacuum in which the scalar fields acquire vacuum non-vanishing expectation values. These are chosen to be aligned but not necessarily equal in magnitude:

\begin{equation}
     \begin{aligned}
 \langle \Phi_{1}  \rangle = \;
   \begin{pmatrix}
      0 \\
      0  \\
      f c_{\beta}
   \end{pmatrix}_{(\textbf{3},\textbf{1})} \hspace{5mm}  \langle \Phi_{2}  \rangle = \;
   \begin{pmatrix}
      0 \\
      0  \\
      f s_{\beta}
   \end{pmatrix}_{(\textbf{1},\textbf{3})}\,.
     \end{aligned}
 \end{equation}
 
The gauge symmetry above is also broken: $[SU(3) \times U(1)] \longrightarrow [SU(2) \times U(1)]$, where the latter is the SM gauge symmetry. Originally, $f \sim 1$ TeV was considered but larger values are assumed nowadays, according to LHC searches \cite{ParticleDataGroup:2020ssz}. Subscripts in the column vectors above indicate the $[SU(3) \times U(1)]_{1} \times [SU(3) \times U(1)]_{2}$ transformation properties of each condensate. We note that under the full new gauge group, the scalar fields have quantum numbers $\left( \textbf{1}, \textbf{3}  \right)_{- \frac{1}{3}}$.

The SM fermions are embedded into $SU(3)_{L}$ triplets. For the lepton sector case the enlarging is straightforward,  but for the quark sector this is not obvious. There are two choices of representations for the quarks. In the \textit{Universal Embedding}, the representation is the same for each generation, but not all gauge anomalies are cancelled within the model~\footnote{In this case there must be new physics, beyond the SLH, obviously. In any case, the  sensitivity of the Higgs mass to the cutoff at two loops, within the SLH,  requires this additional new physics at scales not much larger than $f$ (typically $\Lambda\sim4\pi f$).}. The other representation is \textit{the Anomaly-free embedding} where all gauge  anomalies are cancelled. The cost, however, is placing the first and second generation quarks in a different representation than the third generation quarks. In both embeddings the lepton sector remains equal,  but right-handed neutrinos are omitted, so  neutrinos are treated as massless~\footnote{Ref. \cite{delAguila:2005yi} extended the SLH accounting for the measured neutrino masses.}.

\subsection{Expansion of the Scalar Fields}
The two scalar triplets are introduced as non-linear sigma fields and they can be parameterized in the following manner, to realize the spontaneous global symmetry breaking pattern:

\begin{equation}\label{qq}
    \begin{aligned}
        & \Phi_{1} = \exp \left( \frac{i \Theta^{'}}{f}\right) \exp \left( \frac{i t_{\beta} \Theta}{f} \right) \begin{pmatrix}
      0 \\
      0 \\
      f c_{\beta}
   \end{pmatrix},
    \end{aligned}
\end{equation}
\begin{equation}\label{q1}
    \begin{aligned}
        & \Phi_{2} = \exp \left( \frac{i \Theta^{'}}{f}\right) \exp \left(- \frac{i  \Theta}{f t_{\beta}} \right) \begin{pmatrix}
      0 \\
      0 \\
      f s_{\beta}
   \end{pmatrix},
    \end{aligned}
\end{equation}

where we have introduced the short notation: $s_{\beta} = \sin \beta$, $c_{\beta} = \cos \beta$, $t_{\beta} = \tan \beta$. This parametrization has the form of an $SU(3)$ (broken) transformation. $\Theta$ and $\Theta^{'}$ are $3\times3$ matrix fields, parametrized as:

\begin{equation}
    \begin{aligned}
       & \Theta= \frac{\eta}{\sqrt{2}} \mathbf{1_{3 \times 3}} + \begin{pmatrix}
      \mathbf{0_{2\times 2}} & h \\
      h^{\dagger} & 0 \\
   \end{pmatrix}, \\
   & \Theta^{'}= \frac{\xi}{\sqrt{2}} \mathbf{1_{3 \times 3}} + \begin{pmatrix}
      \mathbf{0_{2\times 2}} & k \\
      k^{\dagger} & 0 \\
   \end{pmatrix},
    \end{aligned}
\end{equation}

where

\begin{equation}
    \begin{aligned}
       & h = \begin{pmatrix}
      h^0 \\
      h^{-}
       \end{pmatrix}, & h^0 &= \frac{1}{\sqrt{2}}\left( v+H -i \chi \right), & h^{\pm} &=-\phi^{\pm}\,, \\
       & k = \begin{pmatrix}
      k^0 \\
      x^{-}
       \end{pmatrix}, &  k^0 &= \frac{\sigma-i \omega}{\sqrt{2}}.
    \end{aligned}
\end{equation}

Here $h$ is an $SU(2)$ doublet,  becoming the SM Higgs doublet, and $\eta$ is a real $SU(2)$ singlet, that will play no role in the next development (see \cite{Han:2005ru,Cheung:2006nk,Cheung:2008zu, He:2017jjx, He:2018iyd} for details). We will assume that only the real part of $h^{0}$ may acquire a non-zero vacuum expectation value. In the Unitary gauge the nonphysical eaten fields $(\Theta^{'})$ must be rotated away through a $SU(3)_{L} \times U(1)_{x}$ transformation, giving only the physical particle spectrum. Within the ’t Hooft-Feynman gauge nonphysical fields are preserved and the fields expansion are the same that in equations \eqref{qq} and  \eqref{q1}. In this gauge, loop calculations are easier than in the unitary gauge but there exist more Feynman diagrams. In the following we work with ’t Hooft-Feynman gauge.

The fields $\Phi_{i}$ can be expanded in powers of $\frac{v}{f}$, in this work only precision of $\mathcal{O}\left( \frac{v^2}{f^2} \right)$ is desired, for which it is necessary to expand the scalars triplets up to the fourth order:

\begin{equation}
\Phi_{1}= \exp \left( \frac{i \Theta^{'}}{f}\right) \begin{pmatrix}
      \mathbf{1}_{2 \times 2}-\frac{t^2_\beta}{2f^2} \mathbf{hh^{\dagger}}_{2 \times 2} + \frac{t^4_\beta}{24f^4} (\mathbf{hh^{\dagger}})^{2}_{2 \times 2} & \frac{it_\beta}{f} \mathbf{h}_{2 \times 1}- \frac{it^3_\beta}{6f^3} \mathbf{hh^{\dagger}h}_{2 \times 1}\\
      \frac{it_\beta}{f} \mathbf{h}^{\dagger}_{1 \times 2}- \frac{it^3_\beta}{6f^3} \mathbf{h^{\dagger}hh^{\dagger}}_{1 \times 2} & 1-\frac{t^2_\beta}{2f^2} h^{\dagger}h +\frac{t^4_\beta}{24f^4} (h^{\dagger}h)^2  \\
   \end{pmatrix} \begin{pmatrix}
      0 \\
      0 \\
      f c_{\beta}
   \end{pmatrix}\,,
\end{equation}
\begin{equation}
\Phi_{2}= \exp \left( \frac{i \Theta^{'}}{f}\right) \begin{pmatrix}
      \mathbf{1}_{2 \times 2}-\frac{1}{2f^2 t^2_\beta} \mathbf{hh^{\dagger}}_{2 \times 2} + \frac{1}{24f^4 t^4_\beta} (\mathbf{hh^{\dagger}})^{2}_{2 \times 2} &- \frac{i}{ft_\beta} \mathbf{h}_{2 \times 1}+ \frac{i}{6f^3t^3_\beta} \mathbf{hh^{\dagger}h}_{2 \times 1}\\
      -\frac{i}{f t_\beta} \mathbf{h}^{\dagger}_{1 \times 2}+ \frac{i}{6f^3 t^3_\beta} \mathbf{h^{\dagger}hh^{\dagger}}_{1 \times 2} & 1-\frac{1}{2f^2 t^2_\beta} h^{\dagger}h +\frac{1}{24f^4 t^4_\beta} (h^{\dagger}h)^2  \\
   \end{pmatrix} \begin{pmatrix}
      0 \\
      0 \\
      f s_{\beta}
   \end{pmatrix}\,.
\end{equation}

\subsection{Gauge Sector}

The $SU(3)_{L} \times U(1)_{x}$ is promoted to a local symmetry by the introduction of the gauge-covariant derivative:

 \begin{equation}\label{20}
     D_{\mu} = \partial_{\mu}- igA^{a}_{\mu}T_{a}+ig_{x}Q_{x}B^{x}_{\mu}, \hspace{5mm} g_{x} = \frac{g t_{W}}{\sqrt{1-t^{2}_{W}/3}},
 \end{equation}
 
where $g$ is the Standard Model weak coupling constant and $g_{x}$ is a new $U(1)_{x}$ coupling constant. $A^{a}_{\mu}$ and $B^{x}_{\mu}$ denote $SU(3)_{L}$ and $U(1)_{x}$ gauge fields, respectively. 

The kinetic terms for the $\phi$ field can be written as:

\begin{equation}\label{21}
    \mathcal{L}_{\Phi}  = \left( D^{\mu} \Phi_{1} \right)^{\dagger} \left( D_{\mu} \Phi_{1} \right) + \left( D^{\mu} \Phi_{2} \right)^{\dagger} \left( D_{\mu} \Phi_{2} \right) .
\end{equation}

The $SU(3)_{L}$ gauge fields, in the fundamental representation, read:

\begin{equation}\label{22}
    \begin{aligned}
        A^{a}_{\mu}T_{a} &=  \frac{A^{3}_{\mu}}{2} \begin{pmatrix}
      1 & 0 & 0 \\
      0 & -1 & 0 \\
      0 & 0 & 0
       \end{pmatrix} + \frac{A^{8}_{\mu}}{2 \sqrt{3}} \begin{pmatrix}
      1 & 0 & 0 \\
      0 & 1 & 0 \\
      0 & 0 & -2
       \end{pmatrix} +\frac{1}{\sqrt{2}} \begin{pmatrix}
      0 & W^{+}_{\mu}  & Y^{0}_{\mu} \\
      W^{-}_{\mu} & 0 & X^{-}_{\mu}\\
     Y^{0 \dagger}_{\mu} & X^{+}_{\mu} & 0
       \end{pmatrix}.
    \end{aligned}
\end{equation}

The diagonal terms will join with the $U(1)_{x}$ generator to form the neutral gauge bosons $A_{\mu}$, $Z_{\mu}$ and $Z^{'}_{\mu}$. In the third term three pairs of conjugate particles can be recognized. Since the upper left $2 \times 2$ sub-matrix contains the unbroken $SU(2)$, we can identify the SM gauge bosons $W^{\pm}$. From the gauge invariant Lagrangian in \eqref{21} we obtain the masses of the unmixed gauge bosons up to order $\mathcal{O}(v^4/f^4)$ that can now be read directly from this Lagrangian term \cite{delAguila:2011wk}:

 \begin{equation}\label{33}
     \begin{aligned}
       \mathcal{L}_{mass} \supset &  \frac{g^{2} v^{2}}{4} \left[1- \frac{v^{2}}{6 f^{2}} \left( \frac{s^{4}_{\beta}}{c^{2}_{\beta}}+ \frac{c^{4}_{\beta}}{s^{2}_{\beta}}  \right) \right] W^{+}_{\mu} W^{- \mu} + \frac{g^{2} f^{2}}{2} \left[1- \frac{v^{2}}{2f^{2}} + \frac{v^{4}}{12 f^{4}} \left( \frac{s^{4}_{\beta}}{c^{2}_{\beta}}+ \frac{c^{4}_{\beta}}{s^{2}_{\beta}}  \right) \right] X^{+}_{\mu} X^{- \mu} \\ 
       & + \left[ \frac{i v^3}{6 \sqrt{2} f^3} \left( \frac{c^3_{\beta}}{s_{\beta}}- \frac{s^3_{\beta}}{c_{\beta}} \right) W^{-}_{\mu} X^{+ \mu} +h.c   \right]\,.
     \end{aligned}
 \end{equation}
 
We need to rotate the original fields to eliminate the mixed terms as follows:

\begin{equation}\label{34}
    \begin{aligned}
      &  W^{\pm} \longrightarrow  W^{\pm} \pm \frac{i v^3}{3 \sqrt{2} f^3} \left( \frac{c^3_{\beta}}{s_{\beta}}- \frac{s^3_{\beta}}{c_{\beta}} \right) X^{\pm}, \\
      & X^{\pm} \longrightarrow  X^{\pm} \pm \frac{i v^3}{3 \sqrt{2} f^3} \left( \frac{c^3_{\beta}}{s_{\beta}}- \frac{s^3_{\beta}}{c_{\beta}} \right) W^{\pm}.
    \end{aligned}
\end{equation}

The physical states $W$ and $X$ differ from the interaction states only by a term of order $v^3/f^3$~\cite{delAguila:2011wk}. This does not matter for the following calculations, but is important in determining the Goldstone bosons states. The masses of the physical fields are: 

\begin{equation}\label{35}
    \begin{aligned}
       & M_{W} = \frac{gv}{2}\left[1- \frac{v^2}{12 f^2} \left(  \frac{c^4_{\beta}}{s^2_{\beta}}+ \frac{s^4_{\beta}}{c^2_{\beta}} \right) \right], \\
       & M_{X} = \frac{gf}{\sqrt{2}} \left[1- \frac{v^2}{4 f^2} + \frac{v^4}{24 f^4} \left(  \frac{c^4_{\beta}}{s^2_{\beta}}+ \frac{s^4_{\beta}}{c^2_{\beta}} \right) \right] \sim \frac{gf}{\sqrt{2}} \left[1- \frac{v^2}{4 f^2} \right].
    \end{aligned}
\end{equation}

The neutral gauge bosons sector is more complicated because it is non-diagonal at order $\mathcal{O}(v^2/f^2)$:

\begin{equation}\label{36}
\begin{aligned}
    \mathcal{L}_{mass} \supset & M^2_{Y} Y^{0 \mu} Y^{0 \dagger}_{\mu} + \left( A_{3} \hspace{2mm}, A_{8},\hspace{2mm} B_{x} \right) \mathcal{M} \left( \begin{array}{c}
A_{3}  \\
A_{8} \\
B_{x}
\end{array}
\right),
\end{aligned}
\end{equation}

with

\begin{equation}\label{37}
    \begin{aligned}
        \mathcal{M} = \left( \begin{array}{ccc}
\frac{g^2 \Delta}{4} & \frac{g^2 \Delta}{4 \sqrt{3}} & \frac{g g_{x} \Delta }{6} \\
\frac{g^2 \Delta}{4 \sqrt{3}} & \frac{g^2 f^2}{3}-\frac{g^2 \Delta}{4} & \frac{g g_{x} \Delta}{2 \sqrt{3}}-\frac{g g_{x} f^2}{3 \sqrt{3}} \\
\frac{g g_{x} \Delta}{6} & \frac{g g_{x} \Delta}{2 \sqrt{3}}-\frac{g g_{x} f^2}{3 \sqrt{3}} & \frac{g^2_{x} f^2}{9}
\end{array}
\right), \hspace{4mm} \Delta = \frac{v^2}{2} -  \frac{v^4}{12f^2}\left(  \frac{c^4_{\beta}}{s^2_{\beta}}+ \frac{s^4_{\beta}}{c^2_{\beta}} \right). 
    \end{aligned}
\end{equation}

The matrix $\mathcal{M} $ needs to be diagonalized to get the physical fields. Masses at order $\mathcal{O}(v^2/f^2)$ are \cite{delAguila:2011wk}:

\begin{equation}\label{38}
    \begin{aligned}
        \mathcal{L}_{mass} \supset &   M^2_{Y} Y^{0 \mu} Y^{0 \dagger}_{\mu} + \frac{1}{2} M^2_{Z} Z^{ \mu} Z_{\mu} + \frac{1}{2} M^2_{Z^{'}} Z^{' \mu} Z^{'}_{\mu} + \frac{1}{2} M^2_{A} A^{ \mu} A_{\mu}\,,
    \end{aligned}
\end{equation}

\begin{equation}\label{39}
    \begin{aligned}
        & M_{A} = 0, \\
        & M_{Y} = \frac{g f}{\sqrt{2}}, \\
        & M_{Z^{'}} = \frac{\sqrt{2} f g }{\sqrt{3-t^{2}_{W}}}\left(  1- \frac{\left(3-t^{2}_{W} \right) v^2}{16 c^{2}_{W} f^2} \right), \\
        & M_{Z} = \frac{g v}{2 c_{W}} \left(1- \frac{v^2}{16f^2}\left(1-t^{2}_{W} \right)^2 -\frac{v^2}{12f^2} \left( \frac{s^{4}_{\beta}}{c^{2}_{\beta}}+\frac{c^{4}_{\beta}}{s^{2}_{\beta}}  \right) \right), \\ 
    \end{aligned}
\end{equation}

where the first order mixing matrix for gauge bosons is: 

\begin{equation}\label{40}
    \begin{aligned}
        \left( \begin{array}{c}
A_{3}  \\
A_{8} \\
B_{x}
\end{array}
\right) = \left( \begin{array}{ccc}
0 & c_{W} & -s_{W} \\
- \sqrt{\frac{3-t^{2}_{W}}{3}} & \frac{s^{2}_{W} }{c_{W}\sqrt{3}} &  \frac{s_{W}}{ \sqrt{3}} \\
\frac{t_{W}}{\sqrt{3}} & s_{W} \sqrt{\frac{3-t^{2}_{W}}{3}} & c_{W} \sqrt{\frac{3-t^{2}_{W}}{3}}
\end{array}
\right)  \left( \begin{array}{c}
Z^{'}  \\
Z \\
A
\end{array}
\right).
    \end{aligned}
\end{equation}

It is important to recall that the SLH model has no custodial symmetry \cite{Chang:2003un, Chang:2003zn}, i.e. there cannot be a $SU(2)_{L} \times SU(2)_{R}$ embedded into the $SU(2)_{L} \times U(1)_{Y}$ to which the $SU(3)_{L} \times U(1)_{x}$ breaks spontaneously. However the parameter $\rho \equiv \frac{M^{2}_{W}}{c^{2}_{W} M^{2}_{Z}} \simeq 1$ only gets corrections at $\mathcal{O}(v^2/f^2)$ and the breaking of this symmetry is very small (a model with custodial symmetry has been proposed in ref. \cite{Schmaltz:2010ac}), see section \ref{sec:MW}. In the SLH there is a mixing between the $Z$ and $Z'$ particles due to the quadratic coupling of the Higgs boson with them, so getting to the physical $Z$ and $Z'$ requires the replacements:

\begin{equation}\label{43}
    \begin{aligned}
        Z^{'}_{\mu} \longrightarrow Z^{'}_{\mu} + \delta_{Z} Z_{\mu}, \hspace{5mm}  Z_{\mu} \longrightarrow Z_{\mu} - \delta_{Z} Z^{'}_{\mu},
    \end{aligned}
\end{equation}

where

\begin{equation}\label{44}
    \begin{aligned}
        \delta_{Z} = \frac{\left(  1-t^{2}_{W} \right) \sqrt{3-t^{2}_{W}} v^2 }{8 c_{W} f^2}. 
    \end{aligned}
\end{equation}

Now we only need the charged Goldstone eigenstates since neutral pNGBs do not contribute to LFV processes. The mixing of pNGBs and gauge bosons are of the form $V^{\mu}\partial_{\mu}\phi$. The kinetic terms for the pNGBs and the Goldstones-gauge mixing terms read~\cite{delAguila:2011wk}:

\begin{equation}
\begin{aligned}
\mathcal{L}_{\Phi} \supset & \left[1-\frac{v^2}{6f^2} \left( \frac{c^4_\beta}{s^2_\beta}+\frac{s^4_\beta}{c^2_\beta}  \right) \right] \partial_{\mu} \phi^{+}\partial^{\mu} \phi^{-} + \left[1-\frac{v^2}{2f^2}  \right] \partial_{\mu} x^{+}\partial^{\mu} x^{-} \\
& -\frac{v^2}{3f^2}\left( \frac{c^3_\beta}{s_\beta}-\frac{s^3_\beta}{c_\beta}  \right) \left( \partial_{\mu} x^{+}\partial^{\mu} \phi^{-} + \partial_{\mu} x^{-}\partial^{\mu} \phi^{+}\right),
\end{aligned}
\end{equation}

\begin{equation}
\begin{aligned}
\mathcal{L}_{\Phi} \supset &     i W^{-}_{\mu}\frac{gv}{2} \left[ \left(1-\frac{v^2}{6f^2} \left( \frac{c^4_\beta}{s^2_\beta}+\frac{s^4_\beta}{c^2_\beta} \right)   \right) \partial^{\mu} \phi^{+} - \frac{v^2}{3f^2} \left( \frac{c^3_\beta}{s_\beta}-\frac{s^3_\beta}{c_\beta} \right) \partial^{\mu}x^{+}   \right] \\
& + X^{-}_{\mu}\frac{gf}{\sqrt{2}} \left[ \frac{v^2}{3f^2} \left( \frac{c^3_\beta}{s_\beta}-\frac{s^3_\beta}{c_\beta} \right) \partial^{\mu}\phi^{+} - \left( 1-\frac{v^2}{2f^2} \right) \partial^{\mu}x^{+}   \right] +  \textrm{h.c.}
\end{aligned}
\end{equation}  

As in the case of  $W$ and $X$ gauge bosons, it is necessary to  rotate their  would-be longitudinal degrees of freedom, to express the interaction eigenstates in terms of the final pNGB eigenstates up to order $\mathcal{O} (v^2/f^2)$: 

\begin{equation}\label{ppp}
\begin{aligned}
& x^{\pm} \rightarrow - \left(1+ \frac{v^2}{4f^2} \right)x^{\pm} \mp i \frac{v^2}{3f^2} \left( \frac{c^3_\beta}{s_\beta}-\frac{s^3_\beta}{c_\beta} \right) \phi^{\pm}, \\
& \phi^{\pm} \rightarrow \mp i \left[1+\frac{v^2}{12 f^2}\left( \frac{c^4_\beta}{s^2_\beta}+\frac{s^4_\beta}{c^2_\beta} \right)   \right] \phi^{\pm}.
\end{aligned}
\end{equation}

For the calculation of these states we use the relations \eqref{34} to obtain the $v^2/f^2$ corrections. Taking eqs. \eqref{40}, \eqref{43} and \eqref{ppp} into account, the relevant Feynman rules can now be obtained.

\subsubsection{Vector-Boson Lagrangian}

The kinetic Lagrangian of the gauge bosons gives rise to the trilinear\footnote{Quartic gauge bosons couplings also arise, but are irrelevant for our work.} gauge bosons couplings necessary for our calculation. It can be written as: 

\begin{equation}\label{46}
     \begin{aligned}
         \mathcal{L}_{V} = - \frac{1}{2} \textrm{Tr} \left[ \tilde{G}_{\mu \nu} \tilde{G}^{\mu \nu} \right] - \frac{1}{4} B^{\mu \nu}_{x}B_{x \mu \nu}, \hspace{5mm} \tilde{G}_{\mu \nu} = \frac{i}{g} \left[D_{\mu}, D_{\nu} \right], 
     \end{aligned}
\end{equation}

the results\footnote{Through this work we have found a few typos in the Feynman rules and form factors given previously.} to order $\mathcal{O}(v^2/f^2)$ are given in tables \ref{tabla1a} and  \ref{tabla1} \cite{delAguila:2011wk}:

\begin{table}[H]
\begin{center}
\begingroup
\renewcommand{\arraystretch}{1.4}
\begin{tabular}{c  c || c  c}
\toprule
\toprule
$V_{i}V_{j}V_{j}$ & $g^{V_{i} V_{j} V_{j}}$ & $V_{i}V_{j}V_{j}$  & $g^{V_{i} V_{j} V_{j}}$  \\
\midrule
\midrule
$AX^{+}X^{-}$ & $-1$ & $AW^{+}W^{-}$ & $-1$ \\

$ZX^{+}X^{-}$ & $\frac{1}{2c_{W}s_{W}} \left[c^{2}_{W}-s^{2}_{W} + c_{W} \delta_{Z} \sqrt{3-t^{2}_{W}} \right]$ & $ZW^{+}W^{-}$ & $\frac{c_{W}}{s_{W}}$ \\

$Z'X^{+}X^{-}$ & $\frac{1}{2c_{W}s_{W}} \left[- \delta_{Z} \left( c^{2}_{W}-s^{2}_{W} \right) + c_{W} \sqrt{3-t^{2}_{W}} \right]$ & $Z'W^{+}W^{-}$ & $-\frac{\delta_{Z} c_{W}}{s_{W}}$\\
\bottomrule
\bottomrule
\end{tabular}
\endgroup
\caption{Feynman rules for the trilinear gauge boson couplings $V_{\mu} \left(p_{1} \right)$ $V^{+}_{\nu} \left(p_{2} \right)$ $V^{-}_{\rho} \left(p_{3} \right)$. All these couplings have the generic form: $ie g^{V_{i} V_{j} V_{j}} \left[ g_{\mu \nu} \left( p_{2}-p_{1} \right)_{\rho} + g_{\nu \rho} \left( p_{3}-p_{2} \right)_{\mu} + g_{\mu \rho} \left( p_{1}-p_{3} \right)_{\nu}  \right]$ ($j$ labels the particle-antiparticle gauge boson pair in the vertex). All  four-momenta are taken incoming.}
\label{tabla1a}
\end{center}
\end{table}

\begin{table}[H]
\begin{center}
\begingroup
\renewcommand{\arraystretch}{1.4}
\begin{tabular}{c  c || c  c}
\toprule
\toprule 
$SVV$ & $K$ & $VSS$  & $G$  \\
\midrule
\midrule
$x^{\pm}X^{\mp} \gamma$ & $\pm i M_{X}$ & $\gamma x^{\pm}x^{\mp}$ & $\pm 1$ \\

$\phi^{\pm}W^{\mp}\gamma$ &$\pm i M_{W}$ & $\gamma \phi^{\pm} \phi^{\mp}$ & $\pm 1$ \\

$x^{\pm}X^{\mp}Z$ & $\mp i M_X \frac{c^2_W-s^2_W}{2c_W s_W} \pm i \delta_Z \frac{M_X \left(1+t^2_W\right)}{2s_W \sqrt{3-t^2_W}}$ & $Z x^{\pm}x^{\mp}$ & $\mp \frac{c^2_W-s^2_W}{2s_W c_W} \mp \delta_Z\frac{1-t^2_W}{2s_W \sqrt{3-t^2_W}}$\\

$\phi^{\pm}W^{\mp}Z$ & $\pm i M_W t_W \mp i \delta_Z  \frac{M_W(1-t^2_W)}{s_W \sqrt{3-t^2_W}}$ & $Z \phi^{\pm} \phi^{\mp}$ & $\mp \frac{c^2_W-s^2_W}{2s_W c_W} \mp \delta_Z \frac{1-t^2_W}{2s_W \sqrt{3-t^2_W}}$ \\

$x^{\pm}X^{\mp}Z'$& $\pm i \frac{M_X \left(1+t^2_W \right)}{2s_W \sqrt{3-t^2_W}} \pm i \delta_Z M_X \frac{c^2_W-s^2_W}{2s_W c_W}$ & $Z' x^{\pm}x^{\mp}$ & $\mp \frac{1-t^2_W}{2s_W \sqrt{3-t^2_W}} \pm \delta_Z \frac{c^2_W-s^2_W}{2s_W c_W}$\\

$\phi^{\pm}W^{\mp}Z'$ & $\mp i M_W \frac{1-t^2_W}{s_W \sqrt{3-t^2_W}} \mp i \delta_Z M_W t_W$ & $Z' \phi^{\pm} \phi^{\mp}$ & $\mp \frac{1-t^2_W}{2s_W \sqrt{3-t^2_W}} \pm \delta_Z \frac{c^2_W -s^2_W}{2s_W c_W}$\\
\bottomrule
\bottomrule
\end{tabular}
\endgroup
\caption{Vertices $[SV^{\mu}V^{\nu}]=ieK g^{\mu \nu}$ and   $[V^{\mu}S(p_1)S(p_2)]=ieG \left(p_1-p_2 \right)^{\mu}$. }
\label{tabla1}
\end{center}
\end{table}

\subsection{Fermion sector}

As anticipated, the SM fermion $SU(2)$ doublets  must be enlarged to $SU(3)$ triplets. In addition, in order to give mass to the new third components of the $SU(3)$-triplet fermions, new $SU(3)$-singlet fermions must be introduced. 

Then each lepton family consists of an $SU(3)$ left-handed triplet $\mathbf{3}$ and two right-handed singlets $\mathbf{1}$. A right-handed neutrino is not included, leaving them massless as in the SM:

\begin{equation}\label{47}
    \begin{aligned}
        L^{T}_{m} = \left( \nu_{L} \hspace{3mm} \ell_{L} \hspace{3mm} iN_{L} \right)_{m}, \hspace{5mm} \ell_{Rm}, \hspace{5mm} N_{Rm},
    \end{aligned}
\end{equation}

where $m$ is the generation index. There are three new heavy neutral states $N_{m}$, defined with a phase $i$,  necessary to get real masses and lepton mixing angles. In the case that we want to give mass to the SM neutrinos, one would need extra singlets to define Dirac neutrinos or new terms that break lepton number to introduce Majorana masses for the SM neutrinos, as shown in refs. \cite{Lee:2005mba}, \cite{delAguila:2005yi}. The structure of the quarks fields depends on the embedding we select: 

\begin{itemize}
    \item \textbf{Universal embedding}
\end{itemize}

All generations carry identical gauge quantum numbers and the $SU(3)_{L} \times U(1)_{x}$ gauge group is anomalous (the SM $SU(2) \times U(1)_{Y}$ gauge group remains anomaly-free). Each quark family consists of an $SU(3)$ left-handed triplet $\mathbf{3}$  and three right-handed singlets $\mathbf{1}$: 
    
\begin{equation}\label{47a}
  \mathcal{Q}^{T}_{m} = \left( u_{L} \hspace{3mm} d_{L} \hspace{3mm} iU_{L} \right)_{m}, \hspace{5mm} u_{Rm}, \hspace{5mm} d_{Rm}, \hspace{5mm} U_{Rm}.
\end{equation}

The new massive quarks $U$, $C$ and $T$ have charge $+ \frac{2}{3}$. 

\begin{itemize}    \item \textbf{Anomaly free embedding}
\end{itemize}

In this configuration we take different charge assignments for the different generations of quarks triplets, the third generation of quarks is the same as in the Universal representation, but the first two generations are in the anti-fundamental representation of $SU(3)$ \cite{Kong:2004cv}: 

\begin{equation}\label{48}
 \begin{aligned}
 &  \mathcal{Q}^{T}_{1} = \left( d_{L} \hspace{3mm} -u_{L} \hspace{3mm} i D_{L} \right), & d_{R}, & &  u_{R}, & & D_{R}, \\
 & \mathcal{Q}^{T}_{2} = \left( s_{L} \hspace{3mm} -c_{L} \hspace{3mm} i S_{L} \right), & s_{R}, & & c_{R}, & & S_{R}, \\
  & \mathcal{Q}^{T}_{3} = \left( t_{L} \hspace{3mm} b_{L} \hspace{3mm} i T_{L} \right), & t_{R}, & & b_{R}, & & T_{R},
\end{aligned}
\end{equation}

such that with this new charge assignment all anomalies cancel \cite{Kong:2003tf, Kong:2003vm}. The new massive quarks are now labeled $D$ and $S$ because of their charge of $-\frac{1}{3}$, and we have again a massive quark $T$. In both embeddings, the phase $i$ is needed to produce real masses and mixing angles. Table \ref{tabla2} collects the gauge representations and hypercharges for the fermion sector in both embeddings.

\begin{table}[H]
\begin{center}
\begingroup
\renewcommand{\arraystretch}{1.2}
\begin{tabular}{c c c c c c c c}
\toprule
\toprule
\multicolumn{8}{c}{Universal Embedding} \\
\midrule
\midrule
Fermion & $\mathcal{Q}_{1,2}$ & $\mathcal{Q}_{3}$  & $u_{Rm}$, $U_{Rm}$ & $d_{Rm}$ & $L_{m}$ & $N_{Rm}$ & $e_{Rm}$   \\
$Q_{x}$ charge & $1/3$ & $1/3$ & $2/3$ & $-1/3$ & $-1/3$  & $0$ & $-1$ \\
$SU(3)$ rep. & $\mathbf{3}$ & $\mathbf{3}$ & $\mathbf{1}$ & $\mathbf{1}$ & $\mathbf{3}$ & $\mathbf{1}$ & $\mathbf{1}$ \\
\toprule
\toprule
\multicolumn{8}{c}{Anomaly free Embedding} \\
\midrule
\midrule
Fermion & $\mathcal{Q}_{1,2}$ & $\mathcal{Q}_{3}$  & $u_{Rm}$, $T_{Rm}$ & $d_{Rm}$, $D_{Rm}$, $S_{Rm}$ & $L_{m}$ & $N_{Rm}$ & $e_{Rm}$   \\
$Q_{x}$ charge & $0$ & $1/3$ & $2/3$ & $-1/3$ & $-1/3$  & $0$ & $-1$ \\
$SU(3)$ rep. & $\mathbf{\bar{3}}$ & $\mathbf{3}$ & $\mathbf{1}$ & $\mathbf{1}$ & $\mathbf{3}$ & $\mathbf{1}$ &  $\mathbf{1}$ \\
\bottomrule
\bottomrule
\end{tabular}
\endgroup
\caption{Quark quantum numbers in different embeddings.}
\label{tabla2}
\end{center}
\end{table}

\subsubsection{Lepton-Yukawa and Lepton-Gauge Lagrangian}

The Yukawa sector of the SLH model collects the flavour structure  of the theory. Lepton masses follow from the Yukawa Lagrangian, and are generated by two types of  terms: linear and bilinear in the $\Phi$ fields. This Lagrangian can be written as: 
\begin{equation}\label{49}
    \begin{aligned}
        \mathcal{L}_{Y} \supset i \lambda^{m}_{N} \bar{N}_{Rm} \Phi^{\dagger}_{2}L_{m} + \frac{ \lambda^{mn}_{\ell}}{\Lambda} \bar{\ell}_{Rm} \epsilon_{ijk} \Phi^{i}_{1} \Phi^{j}_{2} L^{k}_{n} + \textrm{h.c.},
    \end{aligned}
\end{equation}

where $\Lambda $ is the ultraviolet cut-off of the theory. Here $m$ and $n$ are generation indices, whereas $i,j,k$ are $SU(3)$ indices. Notice that $\lambda_{N}$ has been taken diagonal. However $\lambda_{\ell}$ does not need to be aligned in flavor space. After spontaneous electroweak symmetry breaking, this Lagrangian yields the lepton masses and the heavy masses up to $\mathcal{O} (v^2/f^2)$ \cite{delAguila:2011wk}:

\begin{equation}\label{50}
    \begin{aligned}
         \mathcal{L}_{Y} \supset - f s_{\beta} \lambda^{m}_{N} \left[ \left(1- \frac{\delta^{2}_{\nu}}{2} \right) \bar{N}_{Rm} N_{Lm} - \delta_{\nu} \bar{N}_{Rm} \nu_{Lm} \right] + \xi_{\beta} \frac{fv}{\sqrt{2} \Lambda} \lambda^{mn}_{\ell} \bar{\ell}_{Rm} \ell_{Ln} + \textrm{h.c.}, 
    \end{aligned}
\end{equation}
where 
\begin{equation}\label{51}
    \begin{aligned}
        \delta_{\nu} = - \frac{v}{\sqrt{2} f t_{\beta}}, \hspace{5mm} \xi_{\beta} = \left[1- \frac{v^2}{4f^2}- \frac{v^2}{12f^{2}}\left( \frac{s^{4}_{\beta}}{c^{2}_{\beta}} + \frac{c^{4}_{\beta}}{s^{2}_{\beta}}  \right) \right],
    \end{aligned}
\end{equation}

here $\delta_{\nu}$ represents the mixing angle between a heavy neutrino and a SM neutrino of the same generation. Notice that the rotation that diagonalizes $\lambda_{N}$ does not necessarily diagonalize $\lambda_{\ell}$, meaning that there is a mixing between the charged leptons and heavy neutrinos mediated by the charged gauge bosons. Charged leptons mass eigenstates and flavour eigenstates are related by the rotation:

\begin{equation}\label{52}
    \begin{aligned}
        \ell_{Lm} \longrightarrow \left( V_{\ell} \ell_{L} \right)_{m} = V^{mi}_{\ell} \ell_{Li}, 
    \end{aligned}
\end{equation}

where $V^{mi}$ is a CKM-like matrix. Furthermore, according to \eqref{50} each heavy neutrino is mixed just with the light neutrino of the same family. To separate them, we rotate only the left-handed sector. To order $\mathcal{O}(v^2/f^2)$, the physical states for the neutrinos are given by:

\begin{equation}\label{53}
    \begin{aligned}
        \left( \begin{array}{c}
\nu_{L}  \\
N_{L} 
\end{array}
\right)_{m} = \left[  \left( \begin{array}{cc}
1-\frac{\delta^{2}_{\nu}}{2} & - \delta_{\nu}  \\
\delta_{\nu} & 1-\frac{\delta^{2}_{\nu}}{2}  \\
\end{array}
\right)  \left( \begin{array}{c}
V_{\ell} \nu_{L}  \\
N_{\ell}
\end{array}
\right) \right]_{m}.
    \end{aligned}
\end{equation}
After the Spontaneous Symmetry Breaking, in the mass eigenstates basis, the matrix $\lambda^{mn}_{\ell}$ is diagonal. The lepton masses up to $\mathcal{O}(v^2/f^2)$ are \cite{delAguila:2011wk}:

\begin{equation}\label{54}
    \begin{aligned}
        m_{\ell_{i}} = - \xi_{\beta} \frac{fv}{\sqrt{2} \Lambda} y_{\ell_{i}},
    \end{aligned}
\end{equation}

where $y_{\ell}$ is the eigenvalue of the $\lambda_{\ell}$ matrix, and we rotate in the same way the SM charged leptons and neutrinos, because in this work we consider massless SM neutrinos. We note that, in the physical basis, Higgs LFV interactions arise at one loop \cite{Lami:2016mjf}, which makes Higgs-mediated contributions negligible in the processes under study. Finally, the heavy neutrino masses are:

\begin{equation}\label{55}
    m_{N_{i}}= f s_{\beta} \lambda^{i}_{N}\,.
\end{equation}

For a complete description of the leptons sector it is necessary to calculate the vertices of a  Goldstone boson with a lepton pair. These vertices are obtained from the lepton kinetic Lagrangian, which can be written as: 

\begin{equation}\label{56}
    \begin{aligned}
        \mathcal{L}_{F} = \bar{\psi}_{m} i \slashed{D} \psi_{m} , \hspace{7mm} \psi= \left( L_{m}, \ell_{Rm}, N_{Rm} \right)\,.
    \end{aligned}
\end{equation}

The covariant derivative was given in eq.~\eqref{20} with the $Q_{x}$ charges in Table \ref{tabla2}. The vertices of Goldstone bosons and leptons are collected in the following \cite{delAguila:2011wk}:

\begin{table}[H]
\begin{center}
\begingroup
\renewcommand{\arraystretch}{1.4}
\begin{tabular}{c c c}
\toprule
\toprule
$SFF$&  $g_{L}$ & $g_R$ \\
\midrule
\midrule
$x^{+} \bar{N}_{m} \ell_{i}$ & $-\frac{1}{\sqrt{2}s_W} \frac{M_{N_m}}{M_X} \left(1-\delta^2_{\nu}/2 \right)V^{mi}_{\ell}$ & $\frac{1}{\sqrt{2}s_W} \frac{m_{\ell_i}}{M_X} \left(1-\delta^2_{\nu}/2 \right)V^{mi}_{\ell}$ \\

$x^{-} \bar{\ell}_{i} N_{m}$ & $\frac{1}{\sqrt{2}s_W} \frac{m_{\ell_i}}{M_X} \left(1-\delta^2_{\nu}/2 \right)V^{im*}_{\ell}$ & $-\frac{1}{\sqrt{2}s_W} \frac{M_{N_m}}{M_X} \left(1-\delta^2_{\nu}/2 \right)V^{im*}_{\ell}$ \\

$\phi^{+} \bar{N}_{m} \ell_{i}$ & $ \delta_{\nu} \frac{i}{\sqrt{2}s_{W}} \frac{M_{N_m}}{M_W} V^{mi}_{\ell}$  &  $\delta_{\nu} \frac{i}{\sqrt{2}s_W}\frac{m_{\ell_i}}{M_W} V^{mi}_{\ell}$  \\

$\phi^{-}\bar{\ell}_{i}N_{m}$& $-\delta_{\nu}\frac{i}{\sqrt{2}s_W} \frac{m_{\ell_i}}{M_W} V^{im*}_{\ell}$ & $-\delta_{\nu}\frac{i}{\sqrt{2}s_W} \frac{M_{N_m}}{M_W} V^{im*}_{\ell}$\\

$x^{+} \bar{\nu}_{i} \ell_i$ & $0$ & $\delta_{\nu}\frac{1}{\sqrt{2}s_W}\frac{m_{\ell_i}}{M_X}$\\

$x^{-} \bar{\ell}_{i}\nu_{i}$& $\delta_{\nu}\frac{1}{\sqrt{2}s_W}\frac{m_{\ell_i}}{M_X}$ & $0$\\

$\phi^{+}\bar{\nu}_{i}\ell_{i}$ & $0$ & $\frac{i}{\sqrt{2}s_W}\frac{m_{\ell_i}}{M_W}\left(1-\delta^{2}_{\nu}/2 \right)$\\

$\phi^{-}\bar{\ell}_{i}\nu_{i}$ & $-\frac{i}{\sqrt{2}s_W}\frac{m_{\ell_i}}{M_W}\left(1-\delta^{2}_{\nu}/2 \right)$ & $0$\\
\bottomrule
\bottomrule
\end{tabular}
\endgroup
\caption{Vertices $\left[ SFF \right]= ie \left(g_{L}P_{L}+g_{R}P_{R} \right)$ for the lepton sector.}
\label{tabla4}
\end{center}
\end{table}

We highlight that the non-chirally suppressed couplings of the heavy neutrinos  showcase their non-decoupling behaviour, which was stressed before (see e.g.~\cite{delAguila:2008zu, delAguila:2010nv}).

To get those couplings it is necessary to use  eqs. \eqref{ppp}, \eqref{52} and \eqref{53}. Some couplings vanish because they would be proportional to SM neutrino masses, that we neglect.  

\begin{table}[H]
\begin{center}
\begingroup
\renewcommand{\arraystretch}{1.4}
\begin{tabular}{c c c}
\toprule
\toprule
$V_{\mu} \bar{f}_{i} f_{m} $ Vertex &  $g^{V \bar{f}_{i} f_{m}}_{L}$ & $g^{V \bar{f}_{i} f_{m}}_{R}$ \\
\midrule
\midrule
$A \bar{\ell}_{i} \ell_{i}$ & $1$ & $1$ \\
$W^{+} \bar{\nu}_{i} \ell_{i}$ & $\frac{1}{\sqrt{2} s_{W}} \left(1-\frac{\delta^{2}_{\nu}}{2} \right)$ & $0$ \\
$W^{+} \bar{N}_{m} \ell_{i}$ & $- \delta_{\nu} \frac{1}{\sqrt{2}s_{W}} V^{mi}_{\ell}$  & $0$  \\
$Z \bar{\ell}_{i} \ell_{i}$ & $\frac{2 s^{2}_{W}-1}{2c_{W}s_{W}} +  \frac{ \delta_{Z} \left(2s^{2}_{W}-1\right)}{2s_{W}c^{2}_{W} \sqrt{3-t^{2}_{W}}} $ & $t_{W}+ \frac{\delta_{Z} s_{W}}{c^{2}_{W}\sqrt{3-t^{2}_{W}}}$ \\ 
$Z \bar{\nu}_{i} \nu_{i}$ & $\frac{1-\delta^{2}_{\nu}}{2c_{W}s_{W}}- \frac{\delta_{Z} \left(1-2s^{2}_{W} \right)}{2s_{W}c^{2}_{W}\sqrt{3-t^{2}_{W}}}$ & $0$ \\
$Z \bar{N}_{i} N_{i}$ & $\frac{\delta_{Z}}{s_{W} \sqrt{3-t^{2}_{W}}} + \frac{\delta^{2}_{\nu}}{2c_{W}s_{W}}$ & $0$ \\
$Z \bar{N}_{m} \nu_{i}$ & $-\delta_{\nu} \frac{1}{2c_{W}s_{W}} V^{mi}_{\ell}$ & $0$ \\
$X^{+}\bar{\nu}_{i} \ell_{i}$ & $-\delta_{\nu} \frac{i}{\sqrt{2}s_{W}}$ & $0$ \\
$X^{+} \bar{N}_{m}\ell_{i}$ & $- \frac{i}{\sqrt{2}s_{W}}\left(1-\frac{\delta^{2}_{\nu}}{2} \right) V^{mi}_{\ell}$ & $0$ \\
$Y^{0} \bar{\nu}_{i} \nu_{i}$ & $\delta_{\nu} \frac{i}{\sqrt{2}s_{W}}$ & $0$ \\
$Y^{0} \bar{N}_{i} N_{i}$ & $- \delta_{\nu} \frac{i}{\sqrt{2}s_{W}}$ & $0$ \\
$Y^{0} \bar{\nu}_{i} N_{m}$ & $\frac{i}{\sqrt{2}s_{W}}\left(1-\delta^{2}_{\nu} \right)V^{mi \dagger}_{\ell}$ & $0$ \\
$Y^{0} \bar{N}_{m}\nu_{i}$ & $- \delta^{2}_{\nu} \frac{i}{\sqrt{2}s_{W}}V^{mi}_{\ell}$ & $0$\\
$Z^{'} \bar{\ell}_{i} \ell_{i}$ & $\frac{2s^{2}_{W}-1}{2s_{W}c^{2}_{W}\sqrt{3-t^{2}_{W}}} + \frac{\delta_{Z}\left(1-2s^{2}_{W} \right)}{2s_{W}c_{W}} $ & $\frac{s_{W}}{c^{2}_{W}\sqrt{3-t^{2}_{W}}}- \delta_{Z} t_{W}$ \\
$Z^{'}\bar{\nu}_{i}\nu_{i}$ & $\frac{2s^{2}_{W}-1}{2s_{W}c^{2}_{W}\sqrt{3-t^{2}_{3}}} \left( 1- \frac{\left(3-t^{2}_{W}\right) \delta^{2}_{\nu}c^{2}_{W}}{1-2s^{2}_{W}} \right)- \frac{\delta_{Z}}{2c_{W}s_{W}} $ & $0$ \\
$Z^{'} \bar{N}_{i}N_{i}$ & $\frac{1}{2s_{W}\sqrt{3-t^{2}_{W}}} \left[2-\delta^{2}_{\nu} \left(3-t^{2}_{W} \right) \right]$ & $0$ \\
$Z^{'}\bar{N}_{m}\nu_{i}$ & $\frac{\delta_{\nu} \sqrt{3-t^{2}_{W}}}{2s_{W}}V^{mi}_{\ell}$ & $0$ \\
\bottomrule
\bottomrule
\end{tabular}
\endgroup
\caption{Vertices $\left[ V^\mu ff \right]= ie \gamma^{\mu} \left(g_{L}P_{L}+g_{R}P_{R} \right)$ for the lepton sector~\cite{delAguila:2011wk}.}
\label{tabla3}
\end{center}
\end{table}

It is possible to find the SM couplings with $v^2/f^2$ corrections (in $\delta_{\nu}$ and $\delta_{Z}$), new couplings of the heavy gauge bosons to leptons, couplings of the SM gauge bosons to the new heavy neutral leptons and couplings of the new gauge bosons with the new heavy neutral leptons. Entries of the table \ref{tabla3} were obtained using equations \eqref{52} and \eqref{53}.

\subsubsection{Quarks in the Anomaly free Embedding}
The Yukawa couplings are found by contracting the fermion fields with the scalar sets into singlets in all possible ways. For the \textit{anomaly free embedding}, the basic Yukawa Lagrangian reads \cite{delAguila:2011wk}: 

\begin{equation}\label{57}
    \begin{aligned}
        \mathcal{L}_{Y} \supset & \hspace{2mm} i \lambda^{t}_{1} \bar{u}^{1}_{R3} \Phi^{\dagger}_{1}Q_{3} +i \lambda^{t}_{2} \bar{u}^{2}_{R3} \Phi^{\dagger}_{2}Q_{3} + i \frac{\lambda^{m}_{b}}{\Lambda} \bar{d}_{Rm} \epsilon_{ijk}\Phi^{i}_{1}\Phi^{j}_{2}Q^{k}_{3} \\
        & + i \lambda^{dn}_{1} \bar{d}^{1}_{Rn} Q^{T}_{n}\Phi_{1} + i \lambda^{dn}_{2} \bar{d}^{2}_{Rn} Q^{T}_{n}\Phi_{2} + i \frac{\lambda^{mn}_{u}}{\Lambda} \bar{u}_{Rm} \epsilon_{ijk}\Phi^{*i}_{1}\Phi^{*j}_{2}Q^{k}_{n},
    \end{aligned}
\end{equation}

where $n=1,2$; $i,j,k=1,2,3$ are $SU(3)$ indices; $d_{Rm}= \{d_{R},s_{R},b_{R},D_{R},S_{R} \}$; $u_{Rm}= \{u_{R},c_{R},t_{R},T_{R} \}$; $u^{1}_{R3}$ and $u^{2}_{R3}$ are linear combinations of $t_{R}$ and $T_{R}$; $d^{n}_{R1}$ and $d^{n}_{R2}$ are linear combinations of $d_{R}$ and $D_{R}$ for $n=1$ and of $s_{R}$ and $S_{R}$ for $n=2$:

\begin{equation}\label{58}
    \begin{aligned}
        & T_{R} = \frac{\lambda^{t}_{1}c_{\beta}u^{1}_{R3}+ \lambda^{t}_{2}s_{\beta}u^{2}_{R3} }{\sqrt{\left( \lambda^{t}_{1} \right)^{2}c^{2}_{\beta}+\left( \lambda^{t}_{2} \right)^{2}s^{2}_{\beta}}}, \hspace{9mm} t_{R}= \frac{- \lambda^{t}_{2}s_{\beta}u^{1}_{R3}+ \lambda^{t}_{1}c_{\beta}u^{2}_{R3} }{\sqrt{\left( \lambda^{t}_{1} \right)^{2}c^{2}_{\beta}+\left( \lambda^{t}_{2} \right)^{2}s^{2}_{\beta}}}, \\
        & D_{R} =  \frac{\lambda^{d1}_{1}c_{\beta}d^{1}_{R1}+ \lambda^{d1}_{2}s_{\beta}d^{2}_{R1} }{\sqrt{\left( \lambda^{d1}_{1} \right)^{2}c^{2}_{\beta}+\left( \lambda^{d1}_{2} \right)^{2}s^{2}_{\beta}}}, \hspace{5mm} d_{R}= \frac{- \lambda^{d1}_{2}s_{\beta}d^{1}_{R1}+ \lambda^{d1}_{1}c_{\beta}d^{2}_{R1} }{\sqrt{\left( \lambda^{d1}_{1} \right)^{2}c^{2}_{\beta}+\left( \lambda^{d1}_{2} \right)^{2}s^{2}_{\beta}}}, \\
        & S_{R} =  \frac{\lambda^{d2}_{1}c_{\beta}d^{1}_{R2}+ \lambda^{d2}_{2}s_{\beta}d^{2}_{R2} }{\sqrt{\left( \lambda^{d2}_{1} \right)^{2}c^{2}_{\beta}+\left( \lambda^{d2}_{2} \right)^{2}s^{2}_{\beta}}}, \hspace{5mm} s_{R}= \frac{- \lambda^{d2}_{2}s_{\beta}d^{1}_{R2}+ \lambda^{d2}_{1}c_{\beta}d^{2}_{R2} }{\sqrt{\left( \lambda^{d2}_{1} \right)^{2}c^{2}_{\beta}+\left( \lambda^{d2}_{2} \right)^{2}s^{2}_{\beta}}}\,.
    \end{aligned}
\end{equation}

We have obtained heavy states with  corresponding large mass and light orthogonal states which remain massless at this point. In general, $\lambda^{d}_{1}$ can be taken diagonal and, to avoid large quark flavour changing effects, we also assume $\lambda^{d}_{2}$ to be diagonal \cite{Han:2005ru}. Corrections of the order $v^2/f^2$ to vertices are only needed for particles involved in triangle diagrams and, since quarks only appear in box diagrams, $\mathcal{O}(v/f)$ precision is sufficient. Then, before the SEWSB we obtain the following masses for the heavy quarks:

\begin{equation}\label{59}
    \begin{aligned}
        & m_{T} = f \sqrt{\left( \lambda^{t}_{1} \right)^{2} c^{2}_{\beta}+ \left( \lambda^{t}_{2} \right)^{2} s^{2}_{\beta}}, \\
        & m_{D} = f \sqrt{\left( \lambda^{d1}_{1} \right)^{2} c^{2}_{\beta}+ \left( \lambda^{d1}_{2} \right)^{2} s^{2}_{\beta}}, \\
       & m_{S} = f \sqrt{\left( \lambda^{d2}_{1} \right)^{2} c^{2}_{\beta}+ \left( \lambda^{d2}_{2} \right)^{2} s^{2}_{\beta}}. \\
    \end{aligned}
\end{equation}
After the SEWSB, the quark mass terms work out as follows to leading order \cite{delAguila:2011wk}: 

\begin{equation}\label{60}
    \begin{aligned}
        \mathcal{L}^{mass}_{Y} \supset & -m_{T}\bar{T}_{R}T_{L} + \frac{v}{\sqrt{2}} \frac{s_{\beta}c_{\beta} \left[ \left( \lambda^{t}_{1} \right)^{2} -\left( \lambda^{t}_{2} \right)^{2}  \right] }{\sqrt{\left( \lambda^{t}_{1} \right)^{2} c^{2}_{\beta} + \left( \lambda^{t}_{2} \right)^{2} s^{2}_{\beta}}} \bar{T}_{R}t_{L} - \frac{v}{\sqrt{2}} \frac{\lambda^{t}_{1} \lambda^{t}_{2}}{\sqrt{\left( \lambda^{t}_{1} \right)^{2} c^{2}_{\beta} + \left( \lambda^{t}_{2} \right)^{2} s^{2}_{\beta}}} \bar{t}_{R}t_{L} \\
        & -m_{D}\bar{D}_{R}D_{L} - \frac{v}{\sqrt{2}} \frac{s_{\beta}c_{\beta} \left[ \left( \lambda^{d1}_{1} \right)^{2} -\left( \lambda^{d1}_{2} \right)^{2}  \right] }{\sqrt{\left( \lambda^{d1}_{1} \right)^{2} c^{2}_{\beta} + \left( \lambda^{d1}_{2} \right)^{2} s^{2}_{\beta}}} \bar{D}_{R}d_{L} + \frac{v}{\sqrt{2}} \frac{\lambda^{d1}_{1} \lambda^{d1}_{2}}{\sqrt{\left( \lambda^{d1}_{1} \right)^{2} c^{2}_{\beta} + \left( \lambda^{d1}_{2} \right)^{2} s^{2}_{\beta}}} \bar{d}_{R}d_{L} \\
         & -m_{S}\bar{S}_{R}S_{L} - \frac{v}{\sqrt{2}} \frac{s_{\beta}c_{\beta} \left[ \left( \lambda^{d2}_{1} \right)^{2} -\left( \lambda^{d2}_{2} \right)^{2}  \right] }{\sqrt{\left( \lambda^{d2}_{1} \right)^{2} c^{2}_{\beta} + \left( \lambda^{d2}_{2} \right)^{2} s^{2}_{\beta}}} \bar{S}_{R}s_{L} + \frac{v}{\sqrt{2}} \frac{\lambda^{d2}_{1} \lambda^{d2}_{2}}{\sqrt{\left( \lambda^{d2}_{1} \right)^{2} c^{2}_{\beta} + \left( \lambda^{d2}_{2} \right)^{2} s^{2}_{\beta}}} \bar{s}_{R}s_{L}\\
         & + \frac{vf}{\sqrt{2}\Lambda} \lambda^{mn}_{u} \bar{u}_{Rm}u_{Ln} + \frac{vf}{\sqrt{2}\Lambda} \lambda^{m}_{b} \bar{d}_{Rm}b_{L} + \textrm{h.c.}
    \end{aligned}
\end{equation}

In general, the couplings $\lambda^{m}_{b}$ and $\lambda^{mn}_{u}$ generate a misalignment between the up and down sectors in the mass basis, causing the CKM matrix to appear, these couplings also provoke a misalignement between heavy and SM quarks, but since in this work we are interested in LFV, we will assume no flavour mixing in the quark sector for simplicity, so we demand $\lambda^{m}_{b} = \lambda^{mn}_{u} \equiv 0$ for all the couplings that mix different families or heavy and light quarks. SEWSB also induces mixing between heavy left-handed
quarks and the SM quarks, mixing that we keep. We rotate the left-handed fields to obtain the physical quarks states~\cite{delAguila:2011wk}:

\begin{equation}\label{61}
    \begin{aligned}
        & T_{L} \rightarrow T_{L} + \delta_{t}t_{L}, \\
        & t_{L} \rightarrow t_{L} - \delta_{t}T_{L}, \\
        & D_{L} \rightarrow D_{L} + \delta_{d}d_{L}, \\
        & d_{L} \rightarrow d_{L} - \delta_{d}D_{L}, \\
        & S_{L} \rightarrow S_{L} + \delta_{s}s_{L}, \\
        & s_{L} \rightarrow s_{L} - \delta_{s}S_{L}, \\
    \end{aligned}
\end{equation}

where

\begin{equation}\label{62}
    \begin{aligned}
       & \delta_{t}= \frac{v}{\sqrt{2}f} \frac{s_{\beta}c_{\beta}\left[ \left( \lambda^{t}_{1} \right)^{2}-\left(\lambda^{t}_{2} \right)^{2} \right]}{\left( \lambda^{t}_{1} \right)^{2}c^{2}_{\beta} + \left( \lambda^{t}_{2} \right)^{2}s^{2}_{\beta}},  \\
       & \delta_{d}= - \frac{v}{\sqrt{2}f} \frac{s_{\beta}c_{\beta}\left[ \left( \lambda^{d1}_{1} \right)^{2}-\left(\lambda^{d1}_{2} \right)^{2} \right]}{\left( \lambda^{d1}_{1} \right)^{2}c^{2}_{\beta} + \left( \lambda^{d1}_{2} \right)^{2}s^{2}_{\beta}},  \\
        & \delta_{s}= - \frac{v}{\sqrt{2}f} \frac{s_{\beta}c_{\beta}\left[ \left( \lambda^{d2}_{1} \right)^{2}-\left(\lambda^{d2}_{2} \right)^{2} \right]}{\left( \lambda^{d2}_{1} \right)^{2}c^{2}_{\beta} + \left( \lambda^{d2}_{2} \right)^{2}s^{2}_{\beta}},
    \end{aligned}
\end{equation}

are complex in general. Taking all this into account we get the SM quark masses:

\begin{equation}\label{63}
    \begin{aligned}
       & m_{u} = -\frac{vf}{\sqrt{2}\Lambda} \lambda^{11}_{u}, \\
         & m_{c} = -\frac{vf}{\sqrt{2}\Lambda} \lambda^{22}_{u}, \\
         & m_{b} = -\frac{vf}{\sqrt{2}\Lambda} \lambda^{3}_{b},\\
        & m_{t} = \frac{v}{\sqrt{2}} \frac{\lambda^{t}_{1} \lambda^{t}_{2}}{\sqrt{\left( \lambda^{t}_{1} \right)^{2}c^{2}_{\beta}+\left( \lambda^{t}_{2} \right)^{2}s^{2}_{\beta}}}, \\
         & m_{d} = - \frac{v}{\sqrt{2}} \frac{\lambda^{d1}_{1} \lambda^{d1}_{2}}{\sqrt{\left( \lambda^{d1}_{1} \right)^{2}c^{2}_{\beta}+\left( \lambda^{d1}_{2} \right)^{2}s^{2}_{\beta}}}, \\
         & m_{s} = - \frac{v}{\sqrt{2}} \frac{\lambda^{d2}_{1} \lambda^{d2}_{2}}{\sqrt{\left( \lambda^{d2}_{1} \right)^{2}c^{2}_{\beta}+\left( \lambda^{d2}_{2} \right)^{2}s^{2}_{\beta}}}. \\
    \end{aligned}
\end{equation}

Like for the lepton sector we need the quark-gauge Lagrangian to complete the review of the quark couplings. In the anomaly free embedding we have:

\begin{equation}\label{65}
    \begin{aligned}
        \mathcal{L}_{F} = \bar{Q}_{m}i \slashed{D}^{L}_{m}Q_{m} + \bar{u}_{Rm}i \slashed{D}^{u}u_{Rm} + \bar{d}_{Rm}i \slashed{D}^{d}d_{Rm} + \bar{T}_{R}i \slashed{D}^{u}T_{R} + \bar{D}_{R}i \slashed{D}^{d}D_{R} + \bar{S}_{R}i \slashed{D}^{d}S_{R}.
    \end{aligned}
\end{equation}
Remembering that the first two families are in the anti-fundamental representation:

\begin{equation}\label{66}
    \begin{aligned}
       & D^{L}_{(1,2) \mu} = \partial_{\mu} + igA^{a}_{\mu}T^{*}_{a}, \\
       & D^{L}_{3\mu} = \partial_{\mu} - igA^{a}_{\mu}T_{a}+ \frac{ig_{x}}{3}B^{x}_{\mu}, \\
       & D^{u}_{ \mu} = \partial_{\mu} + \frac{2ig_{x}}{3}B^{x}_{\mu}, \\ 
       & D^{d}_{ \mu} = \partial_{\mu} - \frac{ig_{x}}{3}B^{x}_{\mu}.
    \end{aligned}
\end{equation}

With this information and redefining the Goldstone fields as given in \eqref{ppp}, we can obtain the relevant quark-Goldstone boson couplings for our processes, which are given in table \ref{tabla5}~\cite{delAguila:2011wk}:

\begin{table}[H]
\begin{subtable}[H]{0.5\textwidth}
\centering
\begingroup
\renewcommand{\arraystretch}{1.4}
\begin{tabular}{c  c  c}
\toprule
\toprule
$SFF$ & $g_L$ & $g_R$ \\
\midrule
\midrule
$x^{-} \bar{D}_{m}u_{m}$ & $-\frac{M_{D_{m}}}{M_X} \frac{1}{\sqrt{2} s_W}$ & $\frac{m_{u_{m}}}{M_X} \frac{1}{\sqrt{2}s_W}$\\

$x^{-} \bar{d}_{m} u_{m}$ & $0$ & $\delta^{*}_{d_{m}}\frac{m_{u_{m}}}{M_X} \frac{1}{\sqrt{2}s_W}$\\

$\phi^{-} \bar{D}_{m}u_{m}$ & $\delta_{d_{m}}\frac{i}{\sqrt{2}s_W} \frac{M_{D_{m}}}{M_W}$ & $- \delta^{*}_{d_{m}}\frac{i m_{u_{m}}}{M_W} \frac{1}{\sqrt{2}s_W}$\\

$\phi^{-} \bar{d}_{m} u_{m}$ & $-\frac{i m_{d_{m}}}{M_W} \frac{1}{\sqrt{2}s_W}$ & $\frac{i m_{u_{m}}}{M_W} \frac{1}{\sqrt{2}s_W}$\\
\bottomrule
\bottomrule
\end{tabular}
\endgroup
\caption{First and second family,  where $u_{m} = u, c$ and $d_{m} (D_{m}) = d, s (D,S).$}
\label{tabla5.1}
\end{subtable}
\hfill
\begin{subtable}[H]{0.5\textwidth}
\centering
\begingroup
\renewcommand{\arraystretch}{1.4}
\begin{tabular}{c  c  c}
\toprule
\toprule
$SFF$ & $g_L$ & $g_R$ \\
\midrule
\midrule
$x^{+} \bar{T}b$ & $-\frac{M_{T}}{M_X} \frac{1}{\sqrt{2} s_W}$ & $\frac{m_{b}}{M_X} \frac{1}{\sqrt{2}s_W}$\\

$x^{+} \bar{t} b$ & $0$ & $\delta^{*}_{t}\frac{m_{b}}{M_X} \frac{1}{\sqrt{2}s_W}$\\

$\phi^{+} \bar{T}b$ & $\delta_{t}\frac{i}{\sqrt{2}s_W} \frac{M_{T}}{M_W}$ & $ \delta^{*}_{t}\frac{i m_{b}}{M_W} \frac{1}{\sqrt{2}s_W}$\\

$\phi^{+} \bar{t} b$ & $-\frac{i m_{t}}{M_W} \frac{1}{\sqrt{2}s_W}$ & $-\frac{i m_{b}}{M_W} \frac{1}{\sqrt{2}s_W}$\\
\bottomrule
\bottomrule
\end{tabular}
\endgroup
\caption{Third family.}
\label{tabla5.2}
\end{subtable}
\caption{Vertices $[S F F]=ie \left(g_L P_L + g_R P_R \right)$ for the quark sector in the anomaly-free embedding entering in our calculation.}
\label{tabla5}
\end{table}

We remind that  all quark flavor changing vertices were removed, so there is no CKM-like matrix. For the anomaly-free embedding the vector-quark interactions are given in table \ref{tabla6}~\cite{delAguila:2011wk}:

\begin{table}[H]
\begin{subtable}[H]{0.5\textwidth}
\centering
\begingroup
\renewcommand{\arraystretch}{1.4}
\begin{tabular}{c  c  c}
\toprule
\toprule
$VFF$ & $g_L$ & $g_R$ \\
\midrule
\midrule
$\gamma \bar{u}_{m} u_{m}$ & $-\frac{2}{3}$ & $-\frac{2}{3}$ \\

$\gamma \bar{d}_{m} d_{m}$ & $\frac{1}{3}$ & $\frac{1}{3}$\\

$W^{-} \bar{D}_{m}u_{m}$ & $-\delta^{*}_{d_{m}} \frac{1}{\sqrt{2} s_W}$ & $0$\\

$W^{-} \bar{d}_{m}u_{m}$ & $\frac{1}{\sqrt{2} s_W}$ & $0$\\

$Z \bar{u}_{m}u_{m}$ & $\frac{4 c^{2}_{W}-1}{6 c_W s_W}$ & $-\frac{2 s_W}{3c_W}$\\

$Z \bar{d}_{m} d_{m}$ & $\frac{-1-2c^2_W}{6 c_W s_W}$ & $\frac{s_W}{3 c_W}$\\

$X^{-} \bar{D}_{m} u_{m}$ & $-\frac{i}{\sqrt{2} s_W}$ & $0$\\

$X^{-} \bar{d}_{m} u_{m}$ & $-\delta^{*}_{d_{m}}\frac{i}{\sqrt{2} s_W}$ & $0$\\

$Z' \bar{u}_{m}u_{m}$ & $\frac{\sqrt{3-t^{2}_W}}{6 s_W}$ & $\frac{-2 t^2_W}{3 s_W \sqrt{3-t^2_W}}$\\

$Z' \bar{d}_{m}d_{m}$ & $\frac{\sqrt{3-t^2_W}}{6s_W}$ & $\frac{t^2_W}{3 s_W \sqrt{3-t^2_W}}$\\
\bottomrule
\bottomrule
\end{tabular}
\endgroup
\caption{First and second family, where $u_{m} = u, c$ and $d_{m} (D_{m}) = d, s (D,S).$}
\label{tab:week1}
\end{subtable}
\hfill
\begin{subtable}[H]{0.5\textwidth}
\centering
\begingroup
\renewcommand{\arraystretch}{1.4}
\begin{tabular}{c  c  c}
\toprule
\toprule
$VFF$ & $g_L$ & $g_R$ \\
\midrule
\midrule
$\gamma \bar{t} t$ & $-\frac{2}{3}$ & $-\frac{2}{3}$ \\

$\gamma \bar{b} b$ & $\frac{1}{3}$ & $\frac{1}{3}$\\

$W^{+} \bar{T}b$ & $-\delta^{*}_{t} \frac{1}{\sqrt{2} s_W}$ & $0$\\

$W^{+} \bar{t}b$ & $\frac{1}{\sqrt{2} s_W}$ & $0$\\

$Z \bar{t}t$ & $\frac{4 c^{2}_{W}-1}{6 c_W s_W}$ & $-\frac{2 s_W}{3c_W}$\\

$Z \bar{b} b$ & $\frac{-1-2c^2_W}{6 c_W s_W}$ & $\frac{s_W}{3 c_W}$\\

$X^{+} \bar{T} b$ & $-\frac{i}{\sqrt{2} s_W}$ & $0$\\

$X^{+} \bar{t} b$ & $-\delta^{*}_{t}\frac{i}{\sqrt{2} s_W}$ & $0$\\

$Z' \bar{t}t$ & $\frac{-3-t^{2}_W}{6 s_W\sqrt{3-t^2_W}}$ & $\frac{-2 t^2_W}{3 s_W \sqrt{3-t^2_W}}$\\

$Z' \bar{b}b$ & $\frac{-3-t^{2}_W}{6 s_W\sqrt{3-t^2_W}}$ & $\frac{t^2_W}{3 s_W \sqrt{3-t^2_W}}$\\
\bottomrule
\bottomrule
\end{tabular}
\endgroup
\caption{Third family.}
\label{tab:week2}
\end{subtable}
\caption{Vertices $[V^{\mu} F F]=ie \gamma^{\mu} \left(g_L P_L + g_R P_R \right)$ for the quark sector in the Anomaly-free embedding.}
\label{tabla6}
\end{table}

\subsubsection{Quarks in the Universal Embedding}

The situation is similar in the \textit{universal embedding} although the Yukawa Lagrangian is different:

\begin{equation}
\begin{aligned}
\mathcal{L}^{mass}_{Y} \supset i \lambda^{un}_1 \bar{u}^1_{Rn} \Phi^{\dagger}_1 Q_n + i \lambda^{un}_2 \bar{u}^2_{Rn} \Phi^{\dagger}_2 Q_n + i \frac{\lambda^{mn}_d}{\Lambda} \bar{d}_{Rm} \epsilon_{ijk} \Phi^{i}_1 \Phi^{j}_2 Q^k_n + \textrm{h.c.}
\end{aligned}
\end{equation}

Here $m,n = 1,2,3$ are generation indices and $i,j,k = 1,2,3$ are $SU(3)$ indices; $d_m$ runs over the down quarks $(d, s, b)$ and $u^{1,2}_n$ are linear combinations of the orthogonal light and heavy up quarks states:

\begin{equation}
\begin{aligned}
& U_{Rn} = \frac{\lambda^{un}_1 c_{\beta} u^{1}_{Rn}+ \lambda^{un}_{2} s_{\beta} u^2_{Rn}}{\sqrt{\left( \lambda^{un}_1  \right)^2 c^2_{\beta} + \left( \lambda^{un}_2 \right)^2 s^2_{\beta} }},\\
& u_{Rn} = \frac{-\lambda^{un}_2 s_{\beta} u^{1}_{Rn}+ \lambda^{un}_{1} c_{\beta} u^2_{Rn}}{\sqrt{\left( \lambda^{un}_1  \right)^2 c^2_{\beta} + \left( \lambda^{un}_2 \right)^2 s^2_{\beta} }}.
\end{aligned}
\end{equation}

Analogously to the anomaly free case, $\lambda^{u}_1$ can be made diagonal by a field redefinition and $\lambda^{u}_2$ is also taken diagonal to avoid large quark flavor effects. The mass terms are \cite{delAguila:2011wk}:

\begin{equation}
\begin{aligned}
 \mathcal{L}^{mass}_{Y} \supset & -f \sqrt{\left( \lambda^{un}_1  \right)^2 c^2_{\beta} + \left( \lambda^{un}_2 \right)^2 s^2_{\beta} } \bar{U}_{Rn}U_{Ln} + \frac{v}{\sqrt{2}} \frac{s_{\beta}c_{\beta} \left[ \left( \lambda^{un}_1 \right)^2- \left( \lambda^{un}_2 \right)^2 \right]}{\sqrt{\left( \lambda^{un}_1  \right)^2 c^2_{\beta} + \left( \lambda^{un}_2 \right)^2 s^2_{\beta} }} \bar{U}_{Rn} u_{Ln} \\
 & - \frac{v}{\sqrt{2}} \frac{\lambda^{un}_1 \lambda^{un}_2}{\sqrt{\left( \lambda^{un}_1  \right)^2 c^2_{\beta} + \left( \lambda^{un}_2 \right)^2 s^2_{\beta} }} \bar{u}_{Rn}u_{Ln} + \frac{vf}{\sqrt{2} \Lambda} \lambda^{ij}_d \bar{d}_{Ri}d_{Lj} + \textrm{h.c.}
\end{aligned}
\end{equation}

We have neglected terms proportional to $v^2/f^2$. We will again ignore all generation mixing terms. This means setting $\lambda^{ij}_d = \lambda^{i}_d \delta_{ij}$. The only mixing effect in which we are interested corresponds to terms involving the light and heavy up quarks of each generation. The following rotation of the left-handed fields is required to obtain diagonal mass terms:

\begin{equation}
\begin{aligned}
& U_{Ln} \rightarrow U_{Ln}+\delta_{u_{n}} u_{Ln}, \\
& u_{Ln} \rightarrow U_{Ln}-\delta_{u_{n}} U_{Ln},
\end{aligned}
\end{equation}

where

\begin{equation}
\begin{aligned}
\delta_{u_{n}} = \frac{v}{\sqrt{2}f} \frac{s_{\beta}c_{\beta} \left[ \left( \lambda^{un}_1 \right)^2- \left( \lambda^{un}_2 \right)^2 \right]}{\left( \lambda^{un}_1  \right)^2 c^2_{\beta} + \left( \lambda^{un}_2 \right)^2 s^2_{\beta} }.
\end{aligned}
\end{equation}

The quark masses to order $v/f$ are:

\begin{equation}
\begin{aligned}
& M_{U_{n}}= f \sqrt{\left( \lambda^{un}_1  \right)^2 c^2_{\beta} + \left( \lambda^{un}_2 \right)^2 s^2_{\beta} } , \\
& m_{u_{n}}= \frac{v}{\sqrt{2}} \frac{\lambda^{un}_1 \lambda^{un}_2}{\sqrt{\left( \lambda^{un}_1  \right)^2 c^2_{\beta} + \left( \lambda^{un}_2 \right)^2 s^2_{\beta} }}, \\
& m_{d_{n}}= \frac{vf}{\sqrt{2} \Lambda} \lambda^{n}_{d}.
\end{aligned}
\end{equation}

The quark-gauge Lagrangian is more symmetric for the universal embedding:

\begin{equation}
\mathcal{L}= \bar{Q}_{m}i \slashed{D}^{L} Q_m + \bar{u}_{Rm}i \slashed{D}^{u} u_{Rm} + \bar{d}_{Rm}i \slashed{D}^{d} d_{Rm} + \bar{U}_{Rm}i \slashed{D}^{u} U_{Rm}
\end{equation}

where

\begin{equation}
\begin{aligned}
& D^{L}_{\mu} = \partial_{\mu} - ig A^{a}_{\mu}T^{a}+ \frac{ig_x}{3}B^{x}_{\mu}, \\
& D^{u}_{\mu} = \partial_{\mu}+ \frac{2ig_x}{3}B^{x}_{\mu}, \\
& D^{d}_{\mu} = \partial_{\mu}- \frac{ig_x}{3}B^{x}_{\mu}.
\end{aligned}
\end{equation}

The Feynman rules for quark-Goldstone couplings to order $\mathcal{O}(v/f)$ are given in table \ref{tabla7}~\cite{delAguila:2011wk}:

\begin{table}[H]
\begin{center}
\begingroup
\renewcommand{\arraystretch}{1.4}
\begin{tabular}{c  c  c}
\toprule
\toprule
$SFF$ & $g_L$ & $g_R$ \\
\midrule
\midrule
$x^{+} \overline{U}_{m}d_{m}$ & $-\frac{M_{U_{m}}}{M_X} \frac{1}{\sqrt{2} s_W}$ & $\frac{m_{d_{m}}}{M_X} \frac{1}{\sqrt{2}s_W}$\\

$x^{+} \bar{u}_{m} d_{m}$ & $0$ & $\delta^{*}_{u_{m}} \frac{m_{d_{m}}}{M_X} \frac{1}{\sqrt{2}s_W}$\\

$\phi^{+} \overline{U}_{m}d_{m}$ & $\delta_{u_{m}}\frac{i}{\sqrt{2}s_W} \frac{M_{U_{m}}}{M_W}$ & $ \delta^{*}_{u_{m}}\frac{i m_{d_{m}}}{M_W} \frac{1}{\sqrt{2}s_W}$\\

$\phi^{+} \bar{u}_{m} d_{m}$ & $-\frac{i m_{u_{m}}}{M_W} \frac{1}{\sqrt{2}s_W}$ & $-\frac{i m_{d_{m}}}{M_W} \frac{1}{\sqrt{2}s_W}$\\

\bottomrule
\bottomrule
\end{tabular}
\endgroup
\caption{Vertices $[S F F]=ie \left(g_L P_L + g_R P_R \right)$ for the quark sector in the universal embedding entering  our calculations,  where $u_{m}(U_{m})= u,c,t (U,C,T)$ and $d_{m} = d,s,b$.}
\label{tabla7}
\end{center}
\end{table}

For the universal embedding the vector-quark interactions are given in table \ref{tabla8}~\cite{delAguila:2011wk}:

\begin{table}[H]
\begin{center}
\begingroup
\renewcommand{\arraystretch}{1.4}
\begin{tabular}{c  c  c}
\toprule
\toprule
$VFF$ & $g_L$ & $g_R$ \\
\midrule
\midrule
$\gamma \bar{u}_{m} u_{m}$ & $-\frac{2}{3}$ & $-\frac{2}{3}$ \\

$\gamma \bar{d}_{m} d_{m}$ & $\frac{1}{3}$ & $\frac{1}{3}$\\

$W^{+} \overline{U}_{m}d_{m}$ & $-\delta^{*}_{u_{m}} \frac{1}{\sqrt{2} s_W}$ & $0$\\

$W^{+} \bar{u}_{m}d_{m}$ & $\frac{1}{\sqrt{2} s_W}$ & $0$\\

$Z \bar{u}_{m}u_{m}$ & $\frac{4 c^{2}_{W}-1}{6 c_W s_W}$ & $-\frac{2 s_W}{3c_W}$\\

$Z \bar{d}_{m} d_{m}$ & $\frac{-1-2c^2_W}{6 c_W s_W}$ & $\frac{s_W}{3 c_W}$\\

$X^{+} \overline{U}_{m} d_{m}$ & $-\frac{i}{\sqrt{2} s_W}$ & $0$\\

$X^{+} \bar{u}_{m} d_{m}$ & $-\delta^{*}_{u_{m}}\frac{i}{\sqrt{2} s_W}$ & $0$\\

$Z' \bar{u}_{m}u_{m}$ & $-\frac{3+t^{2}_W}{6 s_W \sqrt{3-t^2_W}}$ & $\frac{-2 t^2_W}{3 s_W \sqrt{3-t^2_W}}$\\

$Z' \bar{d}_{m}d_{m}$ & $-\frac{3+t^2_W}{6s_W \sqrt{3-t^2_W}}$ & $\frac{t^2_W}{3 s_W \sqrt{3-t^2_W}}$\\

\bottomrule
\bottomrule
\end{tabular}
\endgroup
\caption{Vertices $[V^{\mu} F F]=ie \gamma^{\mu} \left(g_L P_L + g_R P_R \right)$ for the quark sector in the universal embedding, where $u_{m}(U_{m})= u,c,t (U,C,T)$ and $d_{m} = d,s,b$.}
\label{tabla8}
\end{center}
\end{table}

\section{Lepton Flavour Violating processes in the SLH}
\label{sec:LFV}

LFV decays in the SLH model arise at one loop level and  are driven by the  heavy neutrinos $N_{i}$ in presence of the induced rotation of light lepton fields  $V^{ij}_{\ell}$. There are two generic topologies participating in this amplitude:

\begin{itemize}
    \item \textbf{Penguin diagrams}, namely $\ell \rightarrow \ell_k \{\gamma, Z, Z'\}$, followed by $\{ \gamma, Z, Z' \} \rightarrow \ell_a \bar{\ell_b}$,
    \item \textbf{Box diagrams}.
\end{itemize}

Penguin diagrams of the form $\ell \rightarrow \ell_k H $ should be added, however the couplings $H \rightarrow \ell_a \bar{\ell_a}$ are suppressed by the light mass of the fermions and, therefore, we do not take those diagrams into account.

\subsection{General Structure for the LFV Processes}

\begin{figure}[htp]
    \centering
    \includegraphics[width=4.5cm]{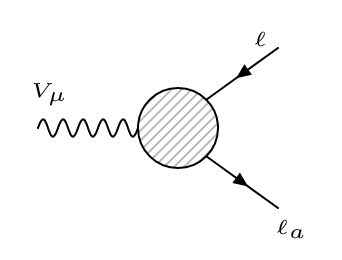}
    \caption{Effective LFV vertex, where $V_{\mu} = \gamma, Z, Z'$.}
    \label{fig:general}
\end{figure}

The contributions of the SM to the \textit{LFV} processes $\ell \rightarrow \ell_{a} \gamma$ and $\ell \rightarrow \ell_{k} \ell_{a} \bar{\ell}_{b}$ are negligible for they are proportional to the observed neutrino masses \cite{Petcov:1976ff, Bilenky:1977du, Cheng:1977nv, Hernandez-Tome:2018fbq},  nevertheless the new Little Higgs contributions can be a priori large. The effective LFV $V_{\mu} \ell \ell_{a}$ vertex with $V_{\mu} = \gamma, Z, Z'$ is sketched in  figure \ref{fig:general}.

\begin{figure}[htp]
    \centering
    \includegraphics[width=9cm]{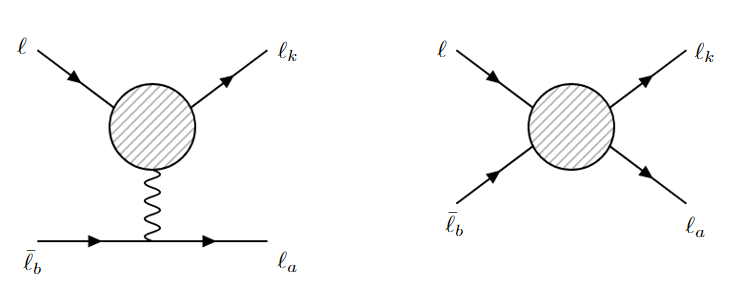}
    \caption{Generic penguin and box diagrams for  $\ell \rightarrow \ell_{k} \ell_{a} \bar{\ell}_{b}$. }
    \label{fig:general1}
\end{figure}

The most general structure for on-shell fermions can be written as :

\begin{equation}\label{71}
    \begin{aligned}
        i \Gamma^{\mu}(p,p_{1}) = ie \left[ \gamma^{\mu}\left( F^{V}_{L}P_{L}+F^{V}_{R}P_{R}\right) - \left(iF^{V}_{M}+F^{V}_{E}\gamma_{5}\right) \sigma^{\mu \nu} Q_{\nu} +  \left(iF^{V}_{S}+F^{V}_{P}\gamma_{5}\right) Q^{\mu}  \right],
    \end{aligned}
\end{equation}
where $Q=p-p_{1}$ is the vector boson momentum. Three body lepton decays $\ell \rightarrow \ell_{k} \ell_{a} \bar{\ell}_{b}$ receive contributions from penguin and box diagrams as we show in figure \ref{fig:general1}. 

There are seven possible decays:

\begin{equation*}\label{deca}
\begin{aligned}
& \textrm{(a)} \hspace{4mm} \tau^{-} \rightarrow \mu^{-} \mu^{-} \mu^{+}, \\
& \textrm{(b)} \hspace{4mm} \tau^{-} \rightarrow \mu^{-} \mu^{-} e^{+}, \\
& \textrm{(c)} \hspace{4mm} \tau^{-} \rightarrow e^{-} \mu^{-} \mu^{+}, \\
& \textrm{(d)} \hspace{4mm} \tau^{-} \rightarrow e^{-} e^{-} \mu^{+}, \\
& \textrm{(e)} \hspace{4mm} \tau^{-} \rightarrow \mu^{-} e^{-} e^{+}, \\
& \textrm{(f)} \hspace{4mm} \tau^{-} \rightarrow e^{-} e^{-} e^{+}, \\
& \textrm{(g)} \hspace{4mm} \mu^{-} \rightarrow e^{-} e^{-} e^{+}.
\end{aligned}
\end{equation*}

We divide them into three categories according to the leptonic flavours in the final state: Category (\textit{i}) comprises all the decays where $\ell_k = \ell_a = \ell_b$ (i.e. the decays (a), (f) and (g)), this kind of decays receive the name of \textit{same-flavors} decays. Category (\textit{ii}) contains all the decays where either $\ell_{k} \neq \ell_{b}$ and $\ell_a = \ell_b$, or $\ell_k = \ell_b$ and $\ell_a \neq \ell_b$ (i.e. the decays (c) and (e)), this category is known as \textit{same-sign} decays. And lastly, all the decays with final leptons having $\ell_k \neq \ell_b$, $\ell_a \neq \ell_b$ belong to the category (\textit{iii}) (i.e. the decays (b) and (d)),  the so-called \textit{wrong-sign} decays. Finally, we studied the $\ell N - \ell_{a} N$ conversion processes ($\ell = \mu, e$, $\ell_{a} = \tau, e$) whose form factors look very similar to the same flavors category decays.

\subsubsection{ \texorpdfstring{$\ell \rightarrow \gamma \ell_{a}$}{TEXT} decays}

The amplitude $\ell \rightarrow \ell_a \gamma$ is proportional to the vertex in figure \ref{fig:general}, however as shown in refs.~\cite{delAguila:2008zu, Hollik:1998vz} the form factors $F^{V}_{L, R}=0$ when $V$ is an on-shell photon. The scalar and pseudoscalar form factors $F^{V}_{S,P}$ do not contribute for real $V$ and are negligible for virtual $V$ in the processes under study. Neglecting $m_{\ell_{a}} \ll m_{\ell}  $ the total width for $\ell \rightarrow \ell_{a} \gamma$ is given by:

\begin{equation}\label{73}
    \Gamma \left( \ell \rightarrow \ell_{a} \gamma \right) = \frac{\alpha m^{3}_{\ell}}{2} \left( \big | F^{\gamma}_{M} \big |^2 + \big | F^{\gamma}_{E} \big |^2 \right).
\end{equation}

The branching ratio is obtained dividing by the SM decay width which, at leading order, is: 

\begin{equation}
    \Gamma \left( \ell_{j} \rightarrow \ell_{i} \nu_{j} \bar{\nu}_{i} \right) = \frac{G^{2}_{F} m^{5}_{\ell_{j}}}{192 \pi^3}, \hspace{4mm} G_{F} = \frac{\pi \alpha_{W}}{\sqrt{2}M^{2}_{W}}, \hspace{4mm} \alpha_{W} = \frac{\alpha}{s^{2}_{W}}.
\end{equation}

In the case of $\tau$ decays the SM branching ratio must be multiplyied by $\sim0.17$ to take into account both lepton Michel and  hadron decay channels. For these calculations we have approximated all the integrations until $\mathcal{O}\left(v^2/f^2 \right)$ and then neglected the ratios\footnote{In principle the neutral Goldstone boson $y^{0(\dagger)}$ could contribute to the $\ell \rightarrow \gamma \ell_{a}$ decay, but it would be a two-loop process, which goes beyond the desired order in this work.}:

\begin{equation}\label{81}
\begin{aligned}
\frac{m^{2}_{\ell}}{M^{2}_{N_{i}}}= \frac{m^{2}_{\ell}}{M^{2}_{W}} = \frac{m^{2}_{\ell}}{M^{2}_{X}} = 0\,.
\end{aligned}
\end{equation}

\begin{figure}[htp]
    \centering
    \includegraphics[width=14cm]{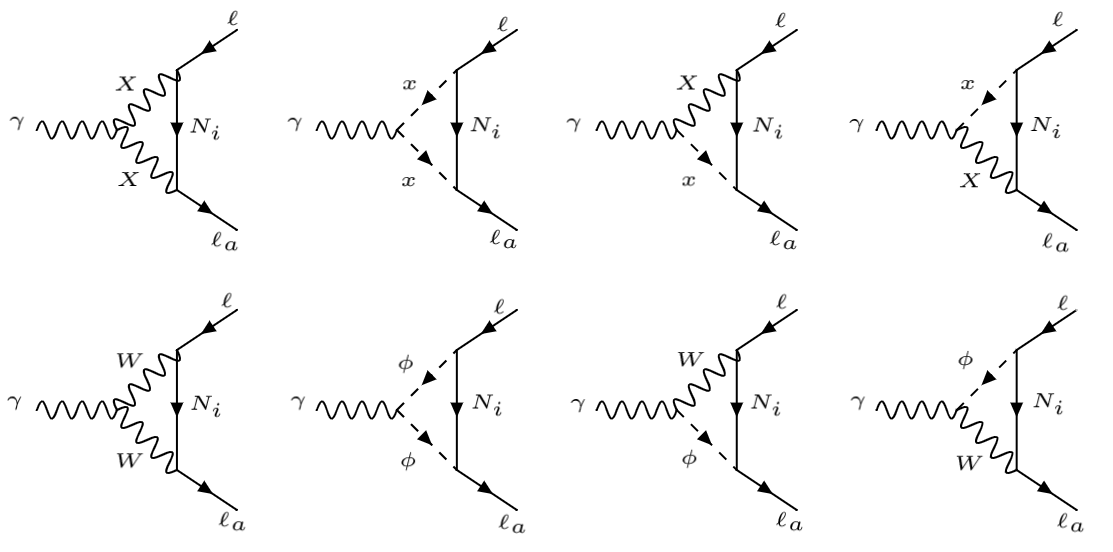}
    \caption{Feynman diagrams for $\ell \rightarrow \gamma \ell_{a}$ }
    \label{fig:mueg}
\end{figure}

We can classify the contributions to $\ell \rightarrow \ell_a \gamma$ into two types of topologies (see figure \ref{fig:mueg}): loop diagrams with heavy $X$ gauge bosons and with $W$ bosons (with corresponding Goldstone bosons $x$ and $\phi$, equivalent to their longitudinal polarizations). Since only dipole form factors contribute to this process, we have:

\begin{equation}\label{82}
\begin{aligned}
& F^{\gamma}_{M}= F^{\gamma}_{M}\big|_{X} + F^{\gamma}_{M}\big|_{W}, \\
& F^{\gamma}_{E}= F^{\gamma}_{E}\big|_{X} + F^{\gamma}_{E}\big|_{W}.
\end{aligned}
\end{equation}

Defining the mass ratios:

\begin{equation}\label{83}
x_{i}= \frac{M^2_{N_{i}}}{M^{2}_{X}}\simeq \mathcal{O}\left(1\right), \hspace{3mm} \omega = \frac{M^2_{W}}{M^{2}_{X}} \simeq \mathcal{O}\left(v^2/f^2 \right),
\end{equation}

we find the following contribution to the dipole form factors for the $X$-mediated diagrams~\cite{delAguila:2011wk}:

\begin{equation}\label{84}
 F^{\gamma}_{M}\big|_{X} =-i  F^{\gamma}_{E}\big|_{X} = \frac{\alpha_{W}}{16 \pi} \frac{m_{\ell}}{M^{2}_{X}} \left(1-\delta^{2}_{\nu} \right) \sum_{i}V^{\ell_{a}i*}V^{i\ell} F_{X}\left( x_{i} \right),
 \end{equation}

where
 
\begin{equation}\label{85}
 \begin{aligned}
  F_{X}(x) &= M^{2}_{X} \left[2 \bar{C}_{1}-3\bar{C}_{11}- x\left(\bar{C}_{0}+3\bar{C}_{1}+ \frac{3}{2}\bar{C}_{11} \right) \right] \\
  & = \frac{5}{6}-\frac{3x-15x^2-6x^3}{12(1-x)^3}+ \frac{3x^2}{2(1-x)^4} \log [x].
 \end{aligned}
\end{equation}

These contributions are equal to those of the SM with massive neutrinos, replacing $X \rightarrow W$,  $N_{i} \rightarrow \nu_{i}$ and  $V^{im} \rightarrow V_{PMNS}$. For tiny neutrino masses, $x_{i}= m^{2}_{\nu_{i}}/M_{W} \ll 1$, 

\begin{equation}\label{85a}
\begin{aligned}
F_{W}(x) \rightarrow \frac{5}{6}-\frac{x}{4}+\mathcal{O}(x^2)
\end{aligned}
\end{equation}

and we recover a well known result \cite{Bilenky:1977du, Cheng:1977nv, Marciano:1977wx} bounded by neutrino oscillation experiments:

\begin{equation}\label{85b}
\mathcal{B}\left( \mu \rightarrow e \gamma \right)_{SM}=\frac{3 \alpha}{32 \pi} \Bigg |  \sum_{i=2,3} V^{ei*}_{PMNS} V^{i\mu}_{PMNS} \frac{\Delta m^{2}_{i1}}{M^2_{W}} \Bigg |^{2} \leq 10^{-54}.
\end{equation}

Expressions for the loop functions are collected in Appendix \ref{sec:jiji} and take the value with $Q^2=0$ for an on-shell photon. For the $W$-based diagrams, we obtain:

\begin{equation}\label{86}
 F^{\gamma}_{M}\big|_{W} =-i  F^{\gamma}_{E}\big|_{W} = \frac{\alpha_{W}}{16 \pi} \frac{m_{\ell}}{M^{2}_{W}} \delta^{2}_{\nu} \sum_{i}V^{\ell_{a}i*}V^{i \ell} F_{W}\left( x_{i}/ \omega \right),
 \end{equation}

with 

\begin{equation}\label{87}
\begin{aligned}
F_{W}(x) = M^2_{W} \left( -2 \bar{C}_{1}+3\bar{C}_{11}\right)+ M^{2}_{N_{i}} \left(\bar{C}_{0}+\bar{C}_{1}- \frac{3}{2}\bar{C}_{11} \right),
\end{aligned}
\end{equation}

that we rewrite as:

\begin{equation}\label{88}
    \begin{aligned}
    F_{W}(x) = \frac{M^{2}_{W}}{M^{2}_{X}} M^{2}_{X} \left(-2 \bar{C}_{1}+3\bar{C}_{11}\right) + \frac{M^{2}_{N_{i}}}{M^{2}_{X}} M^{2}_{X} \left( \bar{C}_{0}+\bar{C}_{1}-\frac{3}{2}\bar{C}_{11}\right),
    \end{aligned}
\end{equation}

to keep leading order terms at  $\mathcal{O}(v^2/f^2)$. The first term of the form factor in \eqref{88} is already of order $\mathcal{O}(v^2/f^2)$, so when multiplied by $\delta^{2}_{\nu}$ the result becomes order $\mathcal{O}(v^4/f^4)$, which we neglect.
 Then, this  contribution to the form factor is:

\begin{equation}\label{89}
      \begin{aligned}
    F_{W}(x) & = x M^{2}_{X} \left( \bar{C}_{0}+\bar{C}_{1}-\frac{3}{2}\bar{C}_{11}\right) \\
    &= \frac{x \left( 8x^2+5x -7\right)}{12(1-x)^3}+\frac{x^2 \left(3x-2 \right)}{2 (1-x)^4} \log [x],
    \end{aligned}
 \end{equation}
in agreement with ref. \cite{delAguila:2011wk}.

The whole dipole form factors are thus:

\begin{equation}\label{90}
\begin{aligned}
F^{\gamma}_{M} = -iF^{\gamma}_{E} = \frac{\alpha_{W}}{4 \pi} \frac{m_{\ell}}{M^{2}_{W}} \sum_{i} V^{\ell_{a}i*}_{\ell}V^{i\ell}_{\ell} \left[ \frac{v^2}{2f^2}F_{X}\left( x_{i}\right)+ \delta^{2}_{\nu} F_{W}\left( x_{i}/\omega \right) \right].
\end{aligned}
\end{equation}

\subsubsection{\texorpdfstring{$\ell \rightarrow \ell_{a} \ell_{a} \bar{\ell}_{a} $}{TEXT} decays}

The contributions to the transition amplitude of the \textit{LFV} three body decays can be summarized as \cite{delAguila:2011wk}:

\begin{equation}\label{74}
    \mathcal{M} = \mathcal{M}_{\gamma\, penguin}+\mathcal{M}_{Z\, penguin}+\mathcal{M}_{Z'\, penguin}+\mathcal{M}_{boxes}\,.
\end{equation}
 We define the amplitudes and form factors as:

\begin{equation}\label{75}
     \begin{aligned}
          \mathcal{M}_{\gamma penguin}= & \frac{e^2}{Q^2}\bar{u}(p_{1}) \left[Q^2 \gamma^{\mu} \left(A^{L}_{1}P_{L}+A^{R}_{1}P_{R} \right) +m_{\ell}i\sigma^{\mu \nu}Q_{\nu}\left(A^{L}_{2}P_{L}+A^{R}_{2}P_{R} \right) \right]u(p) \\
         & \times \bar{u}(p_{2}) \gamma_{\mu}v(p_{3})-\left( p_{1} \leftrightarrow p_{2}\right),\\ 
         \mathcal{M}_{Z penguin}= & \frac{e^2}{M^2_{Z}}\bar{u}(p_{1}) \left[\gamma^{\mu} \left(F_{L}P_{L}+F_{R}P_{R} \right) \right]u(p)\bar{u}(p_{2}) \left[\gamma_{\mu} \left(Z^{a}_{L}P_{L}+Z^{a}_{R}P_{R} \right) \right]v(p_{3}) \\
         &- \left( p_{1} \leftrightarrow p_{2}\right), \\
         \mathcal{M}_{Z' penguin}= & \frac{e^2}{M^2_{Z'}}\bar{u}(p_{1}) \left[\gamma^{\mu} \left(F'_{L}P_{L}+F'_{R}P_{R} \right) \right]u(p)\bar{u}(p_{2}) \left[\gamma_{\mu} \left(Z^{'a}_{L}P_{L}+Z^{'a}_{R}P_{R} \right) \right]v(p_{3}) \\
         &- \left( p_{1} \leftrightarrow p_{2}\right), \\
         \mathcal{M}_{boxes}= &   e^2 B^{L}_{1}\left[\bar{u}(p_{1})\gamma^{\mu}P_{L}u(p) \right]\left[\bar{u}(p_{2})\gamma_{\mu}P_{L}v(p_{3}) \right] \\
        +& e^2B^{R}_{1}\left[\bar{u}(p_{1})\gamma^{\mu}P_{R}u(p) \right]\left[\bar{u}(p_{2})\gamma_{\mu}P_{R}v(p_{3}) \right] \\
        +& e^2 B^{L}_{2} \{ \left[\bar{u}(p_{1})\gamma^{\mu}P_{L}u(p) \right]\left[\bar{u}(p_{2})\gamma_{\mu}P_{R}v(p_{3}) \right] - (p_{1} \leftrightarrow p_{2}) \} \\
         +& e^2 B^{R}_{2} \{ \left[\bar{u}(p_{1})\gamma^{\mu}P_{R}u(p) \right]\left[\bar{u}(p_{2})\gamma_{\mu}P_{L}v(p_{3}) \right] - (p_{1} \leftrightarrow p_{2}) \} \\
         +& e^2 B^{L}_{3} \{ \left[\bar{u}(p_{1})P_{L}u(p) \right]\left[\bar{u}(p_{2})P_{L}v(p_{3}) \right] - (p_{1} \leftrightarrow p_{2}) \} \\
          +& e^2 B^{R}_{3} \{ \left[\bar{u}(p_{1})P_{R}u(p) \right]\left[\bar{u}(p_{2})P_{R}v(p_{3}) \right] - (p_{1} \leftrightarrow p_{2}) \} \\
           +& e^2 B^{L}_{4} \{ \left[\bar{u}(p_{1})\sigma^{\mu \nu}P_{L}u(p) \right]\left[\bar{u}(p_{2})\sigma_{\mu \nu}P_{L}v(p_{3}) \right] - (p_{1} \leftrightarrow p_{2}) \} \\
           +& e^2 B^{R}_{4} \{ \left[\bar{u}(p_{1})\sigma^{\mu \nu}P_{R}u(p) \right]\left[\bar{u}(p_{2})\sigma_{\mu \nu}P_{R}v(p_{3}) \right] - (p_{1} \leftrightarrow p_{2}) \},
     \end{aligned}
\end{equation}

where

\begin{equation}\label{76}
     \begin{aligned}
         & A^{L}_{1}= F^{\gamma}_{L}/Q^2, \hspace{3mm} A^{R}_{1}= F^{\gamma}_{R}/Q^2, \hspace{3mm} A^{L}_{2}= -\left(F^{\gamma}_{M}+i F^{\gamma}_{E} \right)/m_{\ell}, \hspace{3mm} A^{R}_{2}= - \left(F^{\gamma}_{M}-i F^{\gamma}_{E} \right)/m_{\ell} \\
         & F_{L}=-F^{Z}_{L}, \hspace{3mm} F_{R}=-F^{Z}_{R}, \hspace{3mm} F'_{L}=-F^{Z'}_{L}, \hspace{3mm} F'_{R}=-F^{Z'}_{R}. 
     \end{aligned}
\end{equation}

We can use eqs. \eqref{75} to obtain the partial decay width for the \textit{same flavors} decays~\cite{delAguila:2011wk}:

\begin{equation}\label{77}
     \begin{aligned}
        & \Gamma \left( \ell \rightarrow \ell_{a} \ell_{a} \bar{\ell}_{a} \right) = \frac{\alpha^2 m^{5}_{\ell}}{32 \pi}   \Bigg[  \big| A^{L}_{1} \big|^2 + \big| A^{R}_{1} \big|^2 - 2 \left( A^{L}_{1}A^{*R}_{2}+ A^{L}_{2}A^{*R}_{1}+ \textrm{h.c.} \right) \\
         & + \left( \big| A^{L}_{2} \big|^2 + \big| A^{R}_{2} \big|^2 \right) \left( \frac{16}{3} \log \left[ \frac{m_{\ell}}{m_{\ell_{a}}} \right] - \frac{22}{3}\right) + \frac{1}{6} \left( \big| \hat{B}^{L}_{1} \big|^2 + \big| \hat{B}^{R}_{1} \big|^2 \right)  + \frac{1}{3} \left( \big| \hat{B}^{L}_{2} \big|^2 + \big| \hat{B}^{R}_{2} \big|^2 \right) \\
         &  + \frac{1}{24} \left( \big| B^{L}_{3} \big|^2 + \big| B^{R}_{3} \big|^2 \right) +6 \left( \big| B^{L}_{4} \big|^2 + \big| B^{R}_{4} \big|^2 \right) - \frac{1}{2}\left( B^{L}_{3}B^{L*}_{4}+ B^{R}_{3}B^{R*}_{4}+ \textrm{h.c.} \right) \\
         & + \frac{1}{3} \left( A^{L}_{1}\hat{B}^{L*}_{1}+ A^{R}_{1}\hat{B}^{R*}_{1}+ A^{L}_{1}\hat{B}^{L*}_{2}+ A^{R}_{1}\hat{B}^{R*}_{2}+ \textrm{h.c.} \right) \\
         & -\frac{2}{3} \left( A^{R}_{2}\hat{B}^{L*}_{1}+ A^{L}_{2}\hat{B}^{R*}_{1}+ A^{L}_{2}\hat{B}^{R*}_{2}+ A^{R}_{2}\hat{B}^{L*}_{2}+ \textrm{h.c.} \right)\\
         &+ \frac{1}{3} \Big\{ 2\left( \big| F_{LL} \big|^2 + \big| F_{RR} \big|^2 \right) + \big| F_{LR} \big|^2 + \big| F_{RL} \big|^2 \\
         & + \left( \hat{B}^{L}_{1} F^{*}_{LL}+ \hat{B}^{R}_{1}F^{*}_{RR}+ \hat{B}^{L}_{2}F^{*}_{LR}+ \hat{B}^{R}_{2}F^{*}_{RL}+ \textrm{h.c.} \right)+2 \left( A^{L}_{1}F^{*}_{LL} + A^{R}_{1}F^{*}_{RR} + \textrm{h.c.} \right) \\
         & + \left( A^{L}_{1}F^{*}_{LR} + A^{R}_{1}F^{*}_{RL} + \textrm{h.c.} \right)-4 \left( A^{R}_{2}F^{*}_{LL} + A^{L}_{2}F^{*}_{RR} + \textrm{h.c.} \right) \\
         &-2 \left( A^{L}_{2}F^{*}_{RL} + A^{R}_{2}F^{*}_{LR} + \textrm{h.c.} \right)\Big\}
        \Bigg],
     \end{aligned}
 \end{equation} 

where 
 
\begin{equation}\label{78}
     \begin{aligned}
         F_{LL}= \frac{F_{L}Z^{a}_{L}}{M^{2}_{Z}},\hspace{3mm}  F_{RR}= \frac{F_{R}Z^{a}_{R}}{M^{2}_{Z}}, \hspace{3mm}  F_{LR}= \frac{F_{L}Z^{a}_{R}}{M^{2}_{Z}}, \hspace{3mm}  F_{RL}= \frac{F_{R}Z^{a}_{L}}{M^{2}_{Z}}.
     \end{aligned}
\end{equation}

Some box form factors have been redefined to include the contributions from the $Z'$ penguins: 

 \begin{equation}\label{79}
 \begin{aligned}
 & B^{L}_{1} \rightarrow \hat{B}^{L}_{1} = B^{L}_{1}+2F'_{LL}, \\
 & B^{R}_{1} \rightarrow \hat{B}^{R}_{1} = B^{R}_{1}+2F'_{RR}, \\
 & B^{L}_{2} \rightarrow \hat{B}^{L}_{2} = B^{L}_{2}+ F'_{LR}, \\
 & B^{R}_{2} \rightarrow \hat{B}^{R}_{2} = B^{R}_{2}+ F'_{RL},
 \end{aligned}
 \end{equation}

with

\begin{equation}\label{80}
     \begin{aligned}
         F'_{LL}= \frac{F'_{L}Z^{'a}_{L}}{M^{2}_{Z'}},\hspace{3mm}  F'_{RR}= \frac{F'_{R}Z^{'a}_{R}}{M^{2}_{Z'}}, \hspace{3mm}  F'_{LR}= \frac{F'_{L}Z^{'a}_{R}}{M^{2}_{Z'}}, \hspace{3mm}  F'_{RL}= \frac{F'_{R}Z^{'a}_{L}}{M^{2}_{Z'}}.
     \end{aligned}
 \end{equation}

\begin{figure}[H]
    \centering
    \includegraphics[width=\textwidth]{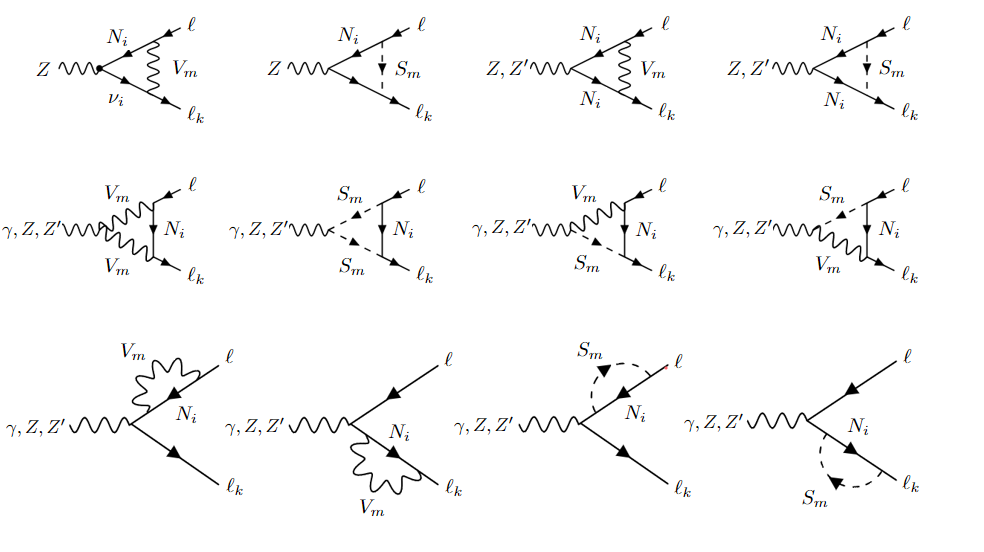}
    \caption{Relevant triangle and self-energy diagrams for $\ell \rightarrow \ell_{k} \ell_{a} \bar{\ell}_{b} $ decays, where $V_{m} = X, W$ and $S_{m} = x, \phi$.}
    \label{fig:mu3e}
\end{figure}

In our case many of the form factors vanish. The relevant penguin diagrams are listed in  figure \ref{fig:mu3e}.

\begin{description}
   \item[Photon penguins] 
\end{description}

The dipole form factors are the same as in the $\ell \rightarrow \ell_{a} \gamma$ case and for those terms we can set $Q^2 = 0$, since $Q^2$ is small in these processes. The form factors $F_{L,R}$ are linear in $Q^2$ and we neglect terms of order $m^2_{\ell}/M^2$, which means that $F_R \simeq 0$. The contribution of diagrams with $X$ bosons is:

\begin{equation}\label{91}
\begin{aligned}
F^{\gamma}_{L}|_X=\frac{\alpha_W}{4 \pi} (1-\delta^2_\nu)\sum_{i} V^{\ell_{a}i*} V^{i \ell}G_{X}(x_i),
\end{aligned}
\end{equation}

where

\begin{equation}\label{92}
\begin{aligned}
G_{X}(x)&= -\frac{1}{2} + \overline{B}_1+6\overline{C}_{00}+x \left(\frac{1}{2} \overline{B}_1 + \overline{C}_{00}- M^2_X \overline{C}_{0} \right)-Q^2 \left(2\overline{C}_{1}+\frac{1}{2}\overline{C}_{11} \right)\\
&= \Delta^{X} + \frac{Q^2}{M^2_{X}}G^{(1)}_{X}(x) + \mathcal{O}\left( \frac{Q^2}{M^4_X}  \right),
\end{aligned}
\end{equation}

where ($\Delta^W$ is defined analogously below)
\begin{equation}
\Delta^{X}=\frac{2}{\epsilon}-\gamma_E+\log(4\pi)-\log\left(\frac{M_X^2}{\mu^2}\right)=\Delta_\epsilon-\log\left(\frac{M_X^2}{\mu^2}\right)\,
\end{equation}
diverges in four dimensions.

The terms that are not proportional to $Q^2$ are cancelled by GIM mechanism, therefore the contribution to the form factor is~\cite{delAguila:2011wk}:

\begin{equation}\label{93}
\begin{aligned}
G_{X}(x)= \frac{Q^2}{M^{2}_{X}} G^{(1)}_{X}(x)
+\mathcal{O} \left(\frac{Q^4}{M^4_{X}} \right)\,,
\end{aligned}
\end{equation}
\begin{equation}\label{94}
G^{(1)}_{X}(x) =-\frac{5}{18}+ \frac{x(12+x-7 x^2)}{24(1-x)^{3}}+ \frac{x^2 (12-10x+x^2)}{12(1-x)^4} \log[x]\,.
\end{equation}

Then, the contributions of the $X$-diagrams is:

\begin{equation}\label{95}
F^{\gamma}_{L}\big|_{X} = \frac{\alpha_{W}}{4\pi} \frac{Q^{2}}{M^{2}_{X}}\left( 1-\delta^{2}_{\nu}\right)\sum_{i} V^{\ell_{a}i*}_{\ell}V^{i\ell}_{\ell} G^{(1)}_{X}(x_{i})\,.
\end{equation}

The $W$-based diagrams contribute with~\cite{delAguila:2011wk}:

\begin{equation}\label{96}
\begin{aligned}
F^{\gamma}_{L}|_W=\frac{\alpha_W}{4 \pi} \delta^2_\nu \sum_{i} V^{\ell_{a}i*} V^{i \ell} G_{W}(x_i/\omega),
\end{aligned}
\end{equation}

where

\begin{equation}\label{97}
\begin{aligned}
G_{W}(x)&= \frac{1}{2} - \overline{B}_1-6\overline{C}_{00}+\frac{M^2_N}{M^2_W}\left( \overline{C}_{00}+\frac{1}{2} \overline{B}+M^2_W \overline{C}_{0} \right)\\
&=-\Delta^{W}+ \frac{Q^2}{M^2_{W}}G^{(1)}_{W}(x) + \mathcal{O}\left( \frac{Q^2}{M^4_W} \right)\,,
\end{aligned}
\end{equation}
 
 \begin{equation}\label{98}
 \begin{aligned}
 G^{(1)}_{W}(x)= -\frac{1}{6}-\frac{x(-2+7x-11x^2)}{72(1-x)^3}+\frac{x^4}{12(1-x)^4} \log[x],
 \end{aligned}
 \end{equation}

that is:

\begin{equation}\label{99}
\begin{aligned}
F^{\gamma}_{L}\big|_{W} = \frac{\alpha_{W}}{4\pi}\frac{Q^2}{M^2_W}\delta^{2}_{\nu}\sum_{i} V^{\ell_{a} i*}_{\ell}V^{i\ell}_{\ell} G^{(1)}_{W}(x_{i}/\omega).
\end{aligned}
\end{equation}

\begin{description}
   \item[$Z$ penguins] 
\end{description}

In this case there are three pieces: two of them involve only heavy neutrinos in the loop ($F^{Z}_{L}\big|_{X} $, $F^{Z}_{L}\big|_{W} $), and the third contains one heavy and one light neutrino ($F^{Z}_{L}\big|_{hl} $), with either gauge boson,

\begin{equation}\label{100}
\begin{aligned}
F^{Z}_{L} = F^{Z}_{L}\big|_{X}+F^{Z}_{L}\big|_{W}+F^{Z}_{L}\big|_{hl}.
\end{aligned}
\end{equation}

Again we neglect $m^{2}_{\ell}/M^2$ terms. The $Z$ dipole form factors $F^{Z}_{M,E}$ (which are chirality flipping) can be neglected as compared to $F^Z_{L}$. The $X$-based diagrams result in~\cite{delAguila:2011wk}:

\begin{equation}\label{101}
\begin{aligned}
F^{Z}_{L}|_X&= \frac{\alpha_W}{4\pi} \frac{1}{c_W s_W} \sum_{i} V^{\ell_{a}i*} V^{i \ell} \bigg [ \frac{c_W \delta_Z}{\sqrt{3-t^2_W}} I_{X}(x_i) + \delta^2_\nu H_{X}(x_i)  \bigg ],
\end{aligned}
\end{equation}

where

\begin{equation}\label{102}
\begin{aligned}
& I_{X}(x) =  \frac{6x-x^2}{2(1-x)}+ \frac{8x}{2(1-x)^2} \log[x], \\
& H_X(x)=  \frac{x}{4}+\frac{x}{4(1-x)}\log[x].
\end{aligned}
\end{equation}

The $W$ boson diagrams give the following contribution~\cite{delAguila:2011wk}:

\begin{equation}\label{103}
\begin{aligned}
F^{Z}_{L}|_W &= \frac{\alpha_W}{4\pi} \frac{\delta^2_{\nu}}{c_W s_W} \sum_{i} V^{\ell_{a}i*} V^{i \ell} \bigg [ -c^2_W H_{W}(x_i/\omega)+\delta^2_\nu\frac{(2+(1-t^2_W) t^2_\beta)}{8} I_W(x_i/ \omega)\\
& + \left(-\frac{s^2_W}{2}+ \frac{(1-2s^2_W)^2 t^2_{\beta} \delta^2_{\nu}}{8 c^4_W} \right) R_{W}(x_i/\omega)   \bigg ], 
\end{aligned}
\end{equation}

where (the $R_W$ contribution turns out to be negligible)

\begin{equation}\label{104}
\begin{aligned}
&H_{W}(x) =  \frac{1}{8}+ \frac{5x}{4(1-x)} +\frac{5x^2}{4(1-x)^2}\log[x], \\
&I_{W}(x)= -\frac{2x^2+3x}{2(1-x)}-\frac{x^2}{(1-x)^2}\log[x], \\
& R_W(x) = \frac{x}{4(1-x)}+ \frac{x^2}{2 (1-x)^2}\log [x].
\end{aligned}
\end{equation}

Diagrams where the $Z$ couples one heavy to one light neutrino contribute with~\cite{delAguila:2011wk}:

\begin{equation}\label{105}
\begin{aligned}
F^{Z}_{L}\big|_{hl} &=  \frac{ \alpha_{W}}{4 \pi } \frac{ \delta^{2}_{\nu}}{s_{W}c_{W}} \Sigma_{i} V^{\ell_{a}i *}V^{i \ell  } \left[  \hat{C}_{00}(M^2_{W},0;x_{i}/\omega)-\hat{C}_{00}(M^2_{X},0;x_{i}) \right],\\
& = \frac{ \alpha_{W}}{4 \pi } \frac{ \delta^{2}_{\nu}}{s_{W}c_{W}}\Sigma_{i} V^{\ell_{a}i *}V^{i \ell  } \left[ H_{Z}(x_{i}/\omega)- H_{Z}(x_{i})   \right],
\end{aligned}
\end{equation}
where

\begin{equation}\label{106}
H_{z}(x)=\frac{x \log[x]}{4(1-x)}.
\end{equation}

\begin{description}
   \item[$Z'$ penguins]
\end{description}

Here we have two contributions:

\begin{equation}\label{107}
F^{Z'}_{L}= F^{Z'}_{L}|_X+F^{Z'}_{L}|_{W},
\end{equation}
there is no piece analogous to $F^{Z}_{L}\big|_{hl}$ since the $Z'$ has an additional $v^{2}/f^{2}$ suppression from its propagator that makes those terms subleading. The form factors read~\cite{delAguila:2011wk}:

\begin{equation}\label{108}
\begin{aligned}
F^{Z'}_{L}|_X&= \frac{\alpha_W}{4\pi} \frac{1}{s_W \sqrt{3-t^2_W}} \sum_{i} V^{\ell_{a}i*} V^{i \ell} I_{X}(x_i) \,,
\end{aligned}
\end{equation}

\begin{equation}\label{109}
\begin{aligned}
F^{Z'}_{L}|_W&= \frac{\alpha_W}{8\pi} \frac{\delta^2_\nu}{s_W \sqrt{3-t^2_W}} \sum_{i} V^{\ell_{a}i*} V^{i \ell} \left[  I_{W}(x_i/ \omega) + \frac{(1-2s^2_W)}{c^2_W} R_{W}(x_i/ \omega)    \right], 
\end{aligned}
\end{equation}
where $I_X$, $I_W$ and $R_W$ are defined in equation \eqref{102} and \eqref{104}. We note that the pieces with a $(1-2s_W^2)$ prefactor are numerically suppressed and can be neglected.

\begin{description}
   \item[Box diagrams]
\end{description}

\begin{figure}[H]
    \centering
    \includegraphics[width=\textwidth]{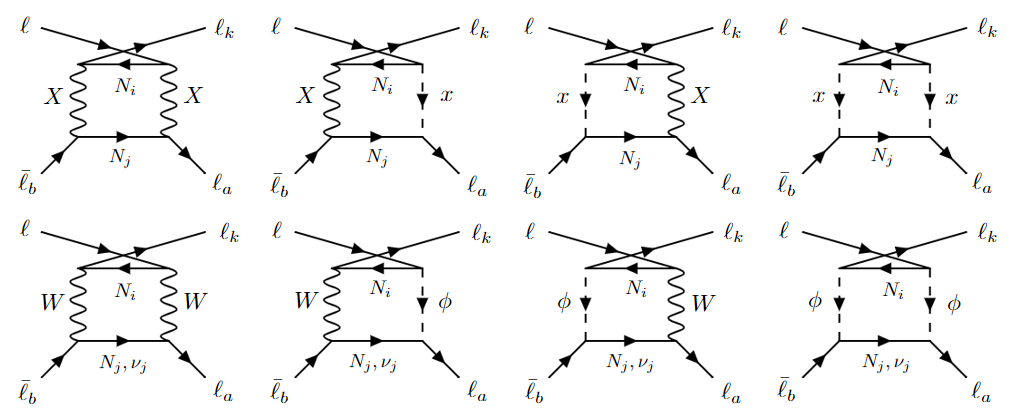}
    \caption{Relevant box diagrams for $\ell \rightarrow \ell_{k} \ell_{a} \bar{\ell}_{b} $ decays.}
    \label{fig:caja}
\end{figure}

Only $W$ and $X$ particles can be involved in the loop (see figure \ref{fig:caja}). Crossed diagrams, not shown in the figure, contribute a factor 2 due to Fierz identities \cite{Nishi:2004st}. In the limit of zero external momenta (all internal masses are much larger than the muon or tau mass) all of them have the same form (being proportional to a scalar integral over the internal momentum).

Neglecting $m_{\ell}/M$, we have contributions only to the $B^{L}_{1}$ form factor, divided in three terms:

\begin{equation}\label{111}
B^L_1=B^L|_X+ B^L|_W+B^L|_{WX},
\end{equation}

where (only the numerically relevant terms below were quoted in ref.~\cite{delAguila:2011wk}, as it happens in other box contributions):

\begin{equation}\label{112}
\begin{aligned}
 B^{L}_{1}|_W= \frac{\alpha_W}{8 \pi} \frac{\delta^4_\nu}{s^2_W} \frac{1}{M^2_W} \sum_{ij}\chi_{ij} \left[ \left(1+\frac{x_i x_j}{4\omega^2}\right) \tilde{d}_{0}(x_i/\omega,x_j/ \omega)  
 + \frac{2 x_i x_j}{\omega^2} d_0 (x_i/\omega,x_j/ \omega) \right],
\end{aligned}
\end{equation}

\begin{equation}\label{113}
\begin{aligned}
 B^{L}_{1}|_X= \frac{\alpha_W}{8 \pi} \frac{(1-2\delta^2_\nu)}{s^2_W} \frac{1}{M^2_X} \sum_{ij}\chi_{ij} \left[ \left(1+\frac{x_i x_j}{4}\right)\tilde{d}_{0}(x_i,x_j)-2x_ix_jd_0(x_i,x_j) \right], 
\end{aligned}
\end{equation}

\begin{equation}\label{114}
\begin{aligned}
 B^{L}_{1}|_{WX}= \frac{\alpha_W}{8 \pi} \frac{\delta^2_\nu}{s^2_W} \frac{1}{M^2_W} \sum_{ij}\chi_{ij} x_i x_j \left[ \frac{1}{2}\tilde{d}_{0}(\omega,x_i,x_j)-2d_0(\omega,x_i,x_j) \right],
\end{aligned}
\end{equation}

and

\begin{equation}\label{115}
\chi_{ij}= V^{\ell_{a}i*} V^{i \ell} |V^{j \ell_a}|^2
\end{equation}
encodes all flavor mixing.

\subsubsection{\texorpdfstring{$\ell \rightarrow \ell_{k} \ell_{a} \bar{\ell}_{a} $}{TEXT} decays}

Now, we analyse the \textit{same-sign} category,   i.e. the decays of the form $\ell (p)\rightarrow \ell_{k} (p_1) \ell_{a}(p_2) \bar{\ell}_{a}(p_3)$. We note that  in this case the amplitude has no crossed penguin diagrams contributions due to swapping $\ell_k$ and $\ell_a$ because it will be a two-loop process. Therefore, the amplitude for the \textit{same-sign} decays has no $p_1 \leftrightarrow p_2$ term in eqs. \eqref{75}. However, for the box amplitudes there are additional diagrams for swapping $\ell_k$ and $\ell_a$. Furthermore, now there is no symmetry factor of $1/2$ in the phase space integration because all three final leptons are distinguishable. The final decay width can be written as \cite{delAguila:2019htj}:

\begin{equation}\label{77a}
\begin{aligned}
& \Gamma\left( \ell \rightarrow \ell_{k} \ell_{a}\bar{\ell}_{a}  \right) = \frac{\alpha^2 m^{5}_{\ell}}{96 \pi} \Big[ 2 \left( \big| A^{L}_{1} \big|^2 + \big| A^{R}_{1} \big|^2 \right)  -4 \left(  A^{L}_{1} A^{* R}_{2} + A^{L}_{2} A^{* R}_{1}+ \textrm{h.c.} \right) \\
& + 4 \left( \big| A^{L}_{2} \big|^2 + \big| A^{R}_{2} \big|^2 \right) \left( 4 \log \left[ \frac{m_{\ell}}{m_{\ell_{a}}} \right] - 7 \right) + \big| F_{LL} \big|^2 + \big| F_{RR} \big|^2 + \big| F_{LR} \big|^2 + \big| F_{RL} \big|^2  \\
&  + \big| \hat{B}_{L1} \big|^2 + \big| \hat{B}_{R1} \big|^2  + \big| \hat{B}_{L2} \big|^2 + \big| \hat{B}_{R2} \big|^2                  + \frac{1}{4} \left( \big| B_{L3} \big|^2 + \big| B_{R3} \big|^2                  \right) +12\left( \big| B_{L4} \big|^2 + \big| B_{R4} \big|^2                  \right) \\
&  + \left( A^{L}_{1}F^{*}_{LL}+  A^{R}_{1}F^{*}_{RR} + A^{L}_{1}F^{*}_{LR} +  A^{R}_{1}F^{*}_{RL} + \textrm{h.c.} \right)  -2 \left( A^{R}_{2}F^{*}_{LL}+  A^{L}_{2}F^{*}_{RR} + A^{R}_{2}F^{*}_{LR} +  A^{L}_{2}F^{*}_{RL} + \textrm{h.c.}   \right) \\
& + \left(  A^{L}_{1} \hat{B}^{L*}_{1} +  A^{R}_{1} \hat{B}^{R*}_{1} +  A^{L}_{1} \hat{B}^{L*}_{2} +  A^{R}_{1} \hat{B}^{R*}_{2} + \textrm{h.c.} \right)-2 \left( A^{R}_{2} \hat{B}^{L*}_{1}  + A^{L}_{2} \hat{B}^{R*}_{1}+ A^{L}_{2} \hat{B}^{R*}_{2} + A^{R}_{2} \hat{B}^{L*}_{2} + \textrm{h.c.} \right) \\
& + \left(  F_{LL} \hat{B}^{L*}_{1}  +  F_{RR} \hat{B}^{R*}_{1} +  F_{LR} \hat{B}^{L*}_{2} +  F_{RL} \hat{B}^{R*}_{2} + \textrm{h.c.}          \right)  \Big], 
\end{aligned}
\end{equation}

with the same simplifying definitions as in eqs. \eqref{78} and \eqref{80}, however the redefinitions of the box form factors are almost the same that in the eqs. \eqref{79} but, in this case we use: 

\begin{equation}
 \begin{aligned}
 & B^{L}_{1} \rightarrow \hat{B}^{L}_{1} = B^{L}_{1}+F'_{LL}, \\
 & B^{R}_{1} \rightarrow \hat{B}^{R}_{1} = B^{R}_{1}+F'_{RR}, \\
 & B^{L}_{2} \rightarrow \hat{B}^{L}_{2} = B^{L}_{2}+ F'_{LR}, \\
 & B^{R}_{2} \rightarrow \hat{B}^{R}_{2} = B^{R}_{2}+ F'_{RL}.
 \end{aligned}
 \end{equation} 
 
As in the example above many of the form factors are zero so the decay width will be simplified. These decays receive contributions from the penguin diagrams in figure \ref{fig:mu3e}. We take the following approximations: 

\begin{equation}\label{116}
\begin{aligned}
\frac{m^{2}_{\ell}}{M^2_X} = \frac{m^{2}_{\ell}}{M^2_W}=\frac{m^{2}_{\ell_{k}}}{M^2_X} = \frac{m^{2}_{\ell_{k}}}{M^2_W}=0 ,
\end{aligned}
\end{equation}

which means that the form factor in equations \eqref{91}-\eqref{109} are the same in this category, also the dipole form factors are the same. On the other hand the contributions from four-point form factors can be written in the generic form \cite{delAguila:2019htj}: 

\begin{equation}\label{117}
\begin{aligned}
\mathcal{F}_{4} = \sum_{ij} \chi^{\ell \ell_{k} \ell_a \ell_a}_{ij} F_{4} \left( M_{N_{i}}, M_{N_{j}}, \ldots \right),
\end{aligned}
\end{equation}
where we have defined the flavor mixing coefficients:

\begin{equation}\label{118}
\begin{aligned}
\chi^{\ell \ell_{k} \ell_a \ell_a}_{ij} = V^{\ell_{k}i*} V^{i \ell} |V^{j \ell_a}|^2 + V^{\ell_{a}i*} V^{i \ell} V^{\ell_{k}j*} V^{j \ell_{a}},
\end{aligned}
\end{equation}
and the second term exchanges $\ell_k$ and $\ell_a$. Thus, the box form factors contributing in this category are represented in  figure \ref{fig:caja}:

\begin{equation}\label{119}
\begin{aligned}
 B^{L}_{1}|_W= \frac{\alpha_W}{ 16 \pi} \frac{\delta^4_\nu}{s^2_W} \frac{1}{M^2_W} \sum_{ij}\chi^{\ell \ell_{k} \ell_a \ell_a}_{ij} \left[ \left(1+\frac{x_i x_j}{4\omega^2}\right) \tilde{d}_{0}(x_i/\omega,x_j/ \omega) 
 + \frac{2 x_i x_j}{\omega^2} d_0 (x_i/\omega,x_j/ \omega) \right], 
\end{aligned}
\end{equation}

\begin{equation}\label{120}
\begin{aligned}
 B^{L}_{1}|_X= \frac{\alpha_W}{ 16 \pi} \frac{(1-2\delta^2_\nu)}{s^2_W} \frac{1}{M^2_X} \sum_{ij}\chi^{\ell \ell_{k} \ell_a \ell_a}_{ij} \left[ \left(1+\frac{x_i x_j}{4}\right)\tilde{d}_{0}(x_i,x_j)-2x_ix_jd_0(x_i,x_j) \right], 
\end{aligned}
\end{equation}

\begin{equation}\label{121}
\begin{aligned}
 B^{L}_{1}|_{WX}= \frac{\alpha_W}{ 16 \pi} \frac{\delta^2_\nu}{s^2_W} \frac{1}{M^2_W} \sum_{ij}\chi^{\ell \ell_{k} \ell_a \ell_a}_{ij} x_i x_j \left[ \frac{1}{2}\tilde{d}_{0}(\omega,x_i,x_j)-2d_0(\omega,x_i,x_j) \right]. 
\end{aligned}
\end{equation}

Another difference between these box form factors and those given in the equations \eqref{112}-\eqref{114} is that here we do not have crossed diagrams, so no factor of $2$ comes from the Fierz identities.

\subsubsection{\texorpdfstring{$\ell \rightarrow \ell_{a} \ell_{a} \bar{\ell}_{b} $}{TEXT} decays}

In the category of \textit{wrong-sign} decays we have the \textit{double flavor} violating three-body process: $\ell(p) \rightarrow \ell_{a}(p_1) \ell_{a}(p_2) \bar{\ell}_{b}(p_3)$, in this case the amplitude does not receive contributions from the three-point form factors\footnote{Contributions of $Z$, $Z'$ and $\gamma$ penguin diagrams start at $2$ loops.}, the box contributions on the other hand are the same that in \textit{same flavors} category. The total decay width is \cite{delAguila:2019htj}:

\begin{equation}\label{77b}
\begin{aligned}
 \Gamma\left( \ell \rightarrow \ell_{a} \ell_{a}\bar{\ell}_{b}  \right) & = \frac{\alpha^2 m^{5}_{\ell}}{192 \pi} \Big[ \big| B^{L}_{1} \big|^2 + \big| B^{R}_{1} \big|^2 +2 \left( \big| B^{L}_{2} \big|^2  + \big| B^{R}_{2} \big|^2 \right) + \frac{1}{4} \left( \big| B^{L}_{3} \big|^2  + \big| B^{R}_{3} \big|^2 \right)\\\
& + \frac{1}{36} \left( \big| B^{L}_{4} \big|^2  + \big| B^{R}_{4} \big|^2 \right) - 3 \left( B^{L}_{3} B^{*L}_{4} + B^{R}_{3} B^{*R}_{4} + \textrm{h.c.}  \right)     \Big],
\end{aligned}
\end{equation}

as in the previous cases, many of the form factors are zero. These kind of processes can receive contributions from box diagrams that conserve lepton number (LNC) like in figure \ref{fig:caja} and diagrams with explicit lepton number violation (LNV), but in our setting we cannot construct box diagrams with LNV vertices because we lack Majorana particles for these  contributions. The relevant box form factors are almost the same that in the equations \eqref{112}-\eqref{114} but the flavor mixing coefficient now differs: 

\begin{equation}\label{122}
\chi^{\ell \ell_{a} \ell_{a} \ell_{b}} = V^{\ell_{a}i *} V^{i \ell}  V^{\ell_{a}j *}  V^{i \ell_{b}}.
\end{equation}

Then, the box form factors are: 

\begin{equation}\label{123}
\begin{aligned}
 B^{L}_{1}|_W= \frac{\alpha_W}{8 \pi} \frac{\delta^4_\nu}{s^2_W} \frac{1}{M^2_W} \sum_{ij}\chi^{\ell \ell_{a} \ell_{a} \ell_{b}} \left[ \left(1+\frac{x_i x_j}{4\omega^2}\right) \tilde{d}_{0}(x_i/\omega,x_j/ \omega)
 + \frac{2 x_i x_j}{\omega^2} d_0 (x_i/\omega,x_j/ \omega) \right], 
\end{aligned}
\end{equation}

\begin{equation}\label{124}
\begin{aligned}
 B^{L}_{1}|_X= \frac{\alpha_W}{8 \pi} \frac{(1-2\delta^2_\nu)}{s^2_W} \frac{1}{M^2_X} \sum_{ij}\chi^{\ell \ell_{a} \ell_{a} \ell_{b}} \left[ \left(1+\frac{x_i x_j}{4}\right)\tilde{d}_{0}(x_i,x_j)-2x_ix_jd_0(x_i,x_j) \right], 
\end{aligned}
\end{equation}

\begin{equation}\label{125}
\begin{aligned}
 B^{L}_{1}|_{WX}= \frac{\alpha_W}{8 \pi} \frac{\delta^2_\nu}{s^2_W} \frac{1}{M^2_W} \sum_{ij}\chi^{\ell \ell_{a} \ell_{a} \ell_{b}} x_i x_j \left[ \frac{1}{2}\tilde{d}_{0}(\omega,x_i,x_j)-2d_0(\omega,x_i,x_j) \right].
\end{aligned}
\end{equation}

In the \textit{wrong sign} decays there is a mixing of the three families in the flavor coefficients, unlike in the \textit{same flavors} and \textit{same sign} decays where only two families of leptons are mixing. 

\subsubsection{\texorpdfstring{$ \mu - e $}{TEXT}  conversion in nuclei}

As we have already said $\mu - e$ nuclei conversion is similar to $\mu \rightarrow e e \bar{e}$ and differs only in that the lower part of the diagrams is coupled to quarks. It does not have crossed penguin and box diagrams because we have a coherent sum of quarks composing the probed nucleus. Also, we do not have identical particles in the final state. We will write the amplitudes as follows \cite{delAguila:2011wk}:

\begin{equation}\label{mue}
\begin{aligned}
\mathcal{M}_{\gamma peng} =& - \frac{e^2}{Q^2} \bar{u}_{\ell_{a}}(p_1) \big[ Q^2 \gamma^{\mu} \left( A^{L}_{1} P_L + A^{R}_{1} P_R \right)+ i m_{\ell} \sigma^{\mu \nu}Q_{\nu} \left( A^{L}_{2} P_L + A^{R}_{2} P_R \right) \big] u_{\ell} (p) \\
& \times \bar{u}_{q}(p_2) Q_q \gamma_{\mu}v_{q}(p_3), \\
\mathcal{M}_{Zpeng} = & \frac{e^2}{M^2_Z} \bar{u}_{\ell_{a}}(p_1) \left[ \gamma^{\mu} \left(F_L P_L + F_R P_R \right) \right]u_{\ell}(p) \bar{u}_{q}(p_2) \gamma_{\mu} \left( Z^{q}_L P_L+ Z^{q}_{R} P_R \right)v_{q}(p_3), \\
\mathcal{M}_{Z' peng} = & \frac{e^2}{M^2_{Z'}} \bar{u}_{\ell_{a}}(p_1) \left[ \gamma^{\mu} \left(F'_L P_L + F'_R P_R \right) \right]u_{\ell}(p) \bar{u}_{q}(p_2) \gamma_{\mu} \left( Z'^{q}_L P_L+ Z'^{q}_{R} P_R \right)v_{q}(p_3), \\
\mathcal{M}^{q}_{box}= & e^2 B^{L}_{1q} \bar{u}_{\ell_{a}}(p_1) \gamma^{\mu}P_L u_{\ell}(p) \bar{u}_{q}(p_2) \gamma^{\mu}P_L v_{q}(p_3).
\end{aligned}
\end{equation}

We have already taken into account that the only non-zero form factor is $B^{L}_{1}$ due to the fact that the SLH couplings are primarily left-handed. This gives for the corresponding conversion width in a nucleus with $Z$ protons and $N$
neutrons~\cite{delAguila:2011wk}:

\begin{equation}
\Gamma \left( \mu N \rightarrow e N  \right) = \alpha^{5} \frac{Z^{4}_{eff}}{Z} |F(q)|^2 m^{5}_{\mu} \Big|2Z \left(A^{L}_1-A^{R}_2 \right) - \left( 2Z+N\right)\bar{B}^{L}_{1u}- \left(Z+2N \right) \bar{B}^{L}_{1d} \Big|^2,
\end{equation}
where $Z_{eff}$ is the nucleus effective charge for the lepton $\ell$ and $F(q)$ the associated form factor. The vertex form factors are as for $\ell \rightarrow \ell_a \ell_a \bar{\ell}_{a}$ and were given in \eqref{76}. We have also defined: 

\begin{equation}
\bar{B}^{L}_{1q} = B^{L}_{1q} + \frac{\left(Z^{q}_{L}+ Z^{q}_{R} \right)F_L}{M^{2}_Z}+  \frac{\left(Z'^{q}_{L}+ Z'^{q}_{R} \right)F'_L}{M^{2}_{Z'}},
\end{equation}

to include the contributions from the $Z' $ penguins. In the case of muons the conversion rate is obtained by dividing by the muon capture rate:

\begin{equation}
\mathcal{R}= \frac{\Gamma \left( \mu \rightarrow e\right)}{\Gamma_{capt}}.
\end{equation}

The nuclei we will consider are $ {}_{48}^{22} \textrm{Ti}$ and $ {}_{197}^{79} \textrm{Au}$, whose relevant parameters are listed in table \ref{tabla12}.

\begin{table}[H]
\begin{center}
\begingroup
\renewcommand{\arraystretch}{1.4}
\begin{tabular}{c c c c c c c}
\toprule
\toprule
Nucleus & $Z$ & $N$ &  $Z_{eff}$ & $F(q)$ & $\Gamma_{capt} \left[ \textrm{GeV} \right]$  \\
\midrule
\midrule
$ {}_{48}^{22} \textrm{Ti}$ & $22$ & $26$ & $17.6$ & $0.54$ & $1.7 \times 10^{-18}$ \\

$ {}_{197}^{79} \textrm{Au}$ & $79$ & $118$ & $33.5$ & $0.16$ & $8.6 \times 10^{-18}$ \\
\bottomrule
\bottomrule
\end{tabular}
\endgroup
\caption{Relevant input parameters for the nuclei under study (see refs. \cite{Kitano:2002mt, Suzuki:1987jf}).}
\label{tabla12}
\end{center}
\end{table}

Only the box form factors need to be recalculated and these are of course embedding-dependent. We stress that we neglect any quark mixing effect for simplicity.

\begin{figure}[H]
    \centering
    \includegraphics[width=\textwidth]{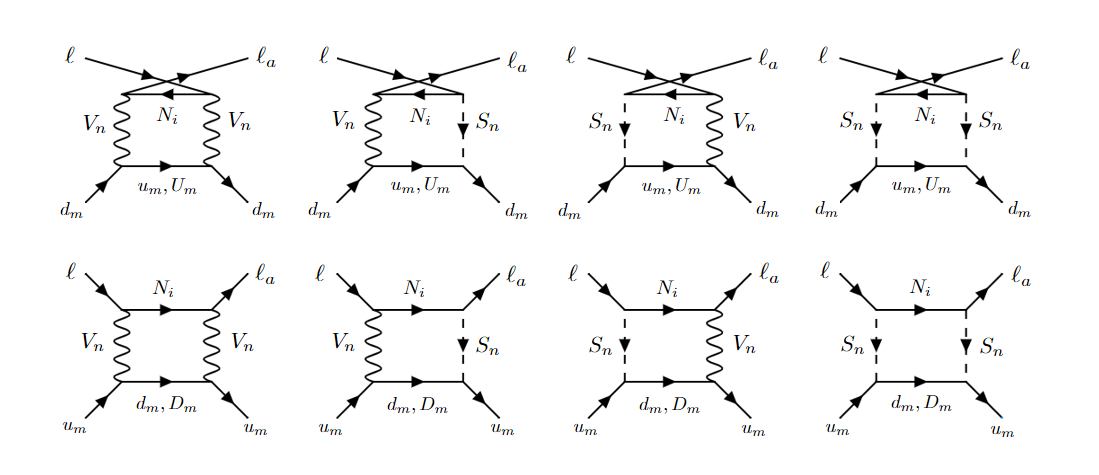}
    \caption{Relevant box diagrams for $\ell N \rightarrow \ell_{a} N $ conversion,  where $V_{n} (S_{n}) = X, W (x, \phi)$, $u_{m} (d_{m}) = u, c (d, s)$ and $U_{m} (D_{m}) = U, C (D, S)$.}
    \label{fig:nuclei}
\end{figure}

In this approximation, only diagrams with a $D$ quark appear in the anomaly-free embedding while only diagrams with a $U$ quark are included for the universal embedding. Diagrams proportional to $\omega$ and with light quarks appear in both embeddings but will be found to be a subleading contribution.

In the anomaly-free embedding we obtain~\cite{delAguila:2011wk}:

\begin{equation}\label{box1}
\begin{aligned}
B^{L}_{1u_{m}}|_{X}& = \frac{\alpha_{W}}{16 \pi} \frac{(1-\delta^2_{\nu})}{s^2_W M^2_X} \sum_{i} V^{\ell_{a}i*} V^{i \ell} \Big[- \left(4+ \frac{x_{D_{m}} x_i}{4} \right) \tilde{d}_{0} \left( x_i, x_{D_{m}} \right) + 2 x_i x_{D_{m}} d_{0} \left( x_i, x_{D_{m}} \right) \\
& -4 |\delta_{d_{m}}|^2 \tilde{d}_{0} \left(x_i,x_{d_{m}} \right) \Big], \\
B^{L}_{1u_{m}}|_{W} &= \frac{\alpha_{W}}{16 \pi} \frac{\delta^{2}_{\nu}}{s^2_W M^2_W} \sum_{i} V^{\ell_{a}i*} V^{i \ell} \Big[ -|\delta_{d_{m}}|^2 \left( 4+\frac{x_{D_{m}} x_i}{4 \omega^2} \right) \tilde{d}_{0}\left(x_i/\omega, x_{D_{m}}/ \omega \right) -4 \tilde{d}_{0} \left(x_i/\omega, x_{d_{m}}/ \omega \right) \\
&+ \frac{x_i x_{D_{m}}}{\omega^2}(\delta^2_{d_{m}}+\delta^{*2}_{d_{m}})d_0 \left(x_i/\omega, x_{D_{m}}/ \omega \right) \Big], \\
B^{L}_{1u_{m}}|_{XW} &= \frac{\alpha_W}{16\pi} \frac{\left(1-\delta^2_\nu \right)}{s^2_W} \frac{1}{M^2_W} \sum_{i} V^{\ell_{a}i*} V^{i \ell} \Big[ -\delta_{\nu} \left(\delta_{d_{m}} +\delta^{*}_{d_{m}}\right)x_{D_{m}} x_i \Big(d_0 \left(\omega, x_i, x_{D_{m}} \right) -\frac{\tilde{d}_0\left(\omega, x_i, x_{D_{m}} \right)}{4} \Big) \\
&+ \omega \delta_{\nu} \left( \delta_{d_{m}}+\delta^{*}_{d_{m}}  \right)\left[ 4 \left( \tilde{d}_{0} \left(\omega, x_i, x_{D_{m}} \right)-  \tilde{d}_{0} \left(\omega, x_i, x_{d_{m}} \right) \right)- x_i x_{D_{m}} d_{0} \left(\omega, x_i, x_{D_{m}} \right)  \right]  \Big],
\end{aligned}
\end{equation}

\begin{equation}\label{box2}
B^{L}_{1d_{m}}=0.
\end{equation}
In the universal embedding we find~\cite{delAguila:2011wk}:

\begin{equation}\label{box3}
B^{L}_{1u_{m}}=0,
\end{equation}

\begin{equation}\label{box4}
\begin{aligned}
B^{L}_{1d_{m}}|_{X}& = \frac{\alpha_{W}}{16 \pi} \frac{(1-\delta^2_{\nu})}{s^2_W M^2_X} \sum_{i} V^{\ell_{a}i*} V^{i \ell} \Big[ \left(1+ \frac{x_{U_{m}} x_i}{4} \right) \tilde{d}_{0} \left( x_i, x_{U_{m}} \right) - 2 x_i x_{U_{m}} d_{0} \left( x_i, x_{U_{m}} \right) \\
& + |\delta_{u_{m}}|^2 \tilde{d}_{0} \left(x_i,x_{u_{m}} \right)  \Big], \\
B^{L}_{1d_{m}}|_{W} &= \frac{\alpha_{W}}{16 \pi} \frac{\delta^{2}_{\nu}}{s^2_W M^2_W} \sum_{i} V^{\ell_{a}i*} V^{i \ell} \Big[|\delta_{u_{m}}|^2 \left(1+ \frac{x_{U_{m}} x_i}{4 \omega^2} \right) \tilde{d}_{0}\left(x_i/\omega, x_{U_{m}}/ \omega \right) +  \tilde{d}_{0} \left(x_i/\omega, x_{u_{m}}/ \omega \right)\\
& + \frac{x_i x_{U_{m}}}{\omega^2}(\delta^2_{u_{m}}+\delta^{*2}_{u_{m}})d_0 \left(x_i/\omega, x_{U_{m}}/ \omega \right)      \Big], \\
B^{L}_{1d_{m}}|_{XW} &= \frac{\alpha_W}{16\pi} \frac{\left(1-\delta^2_\nu \right)}{s^2_W} \frac{1}{M^2_W} \sum_{i} V^{\ell_{a}i*} V^{i \ell} \Big[ -\delta_{\nu} \left(\delta_{u_{m}} +\delta^{*}_{u_{m}}\right)x_{U_{m}} x_i \Big(d_0 \left(\omega, x_i, x_{U_{m}} \right) -\frac{\tilde{d}_0\left(\omega, x_i, x_{U_{m}} \right)}{4} \Big) \\
&+ \omega \delta_{\nu} \left( \delta_{u_{m}}+\delta^{*}_{u_{m}}  \right)\left(  \tilde{d}_{0} \left(\omega, x_i, x_{u_{m}} \right)-  \tilde{d}_{0} \left(\omega, x_i, x_{U_{m}} \right) + x_i x_{U_{m}} d_{0} \left(\omega, x_i, x_{U_{m}} \right)  \right)  \Big], 
\end{aligned}
\end{equation}
with

\begin{equation}\label{box5}
x_{D_{m}} = M^{2}_{D_{m}} / M^2_{X}, \hspace{3mm} x_{U_{m}} = M^{2}_{U_{m}} / M^2_{X}, \hspace{3mm} x_{d_{m}} = m^{2}_{d_{m}} / M^2_{X}, \hspace{3mm} x_{u_{m}} = m^{2}_{u_{m}} / M^2_{X}.
\end{equation}

\subsubsection{\texorpdfstring{$ \ell - \tau $}{TEXT}  conversion in nuclei}

The study of $\ell - \tau$ conversion differs from the well-known $\mu - e$ case. The latter is a low-energy  process, while the former could be probed via a deep inelastic scattering (DIS) of the initial lepton beam. In a DIS the leptons break the nucleons inside the atomic nuclei and interact with the partons (quarks and gluons) leading to any hadronic final state; thus we are  only interested in the $\ell + \mathcal{N}(A,Z) \rightarrow \tau + X$ case, where $X$ could be any hadrons in which we do not focus. One important piece in this analysis corresponds to the parton distribution functions (PDFs) encoding the low-energy non-perturbative QCD effects. Thus, perturbative effects ($\hat{\sigma}$) and the long-distance contributions ($H$) are splitted, via QCD factorization theorems, in the following way: 

\begin{equation}
    \sigma_{\ell - \tau} = \hat{\sigma} \otimes H\,.
\end{equation}

Such oversimplified form of the convolution cannot hold because $ \sigma_{\ell - \tau}$ depends on all partons inside the nucleons, so this calculation is correct up to some scale, which is usually taken as $Q^2 = -q^2$ where $q^2$ is the  momentum transfer of the process. In addition, PDFs are characterized by the Lorentz invariant quantity $\xi$, which represents the fraction of the momentum carried by the interacting parton. Considering both quantities, one should  write:

\begin{equation}
    \sigma_{\ell - \tau} = \hat{\sigma} \left( \xi, Q^{2} \right) \otimes H\left( \xi, Q^{2} \right)\,.
\end{equation}

In this work we are dealing with heavy nuclei (Fe and Pb) and as pointed out in ref.~\cite{EuropeanMuon:1983wih}, binding effects alter the long-distance behavior at different $\xi$ scales. To account for this, we use the fit of the nuclear parton distribution functions (nPDFs) provided by the NCTQ15 project \cite{Kusina:2015vfa} which is incorporated in the ManeParse Mathematica package \cite{Clark:2016jgm}. Also, to consider the running of the quark masses with the scale $Q^2$ we incorporate the RunDec Mathematica package \cite{Chetyrkin:2000yt}. The perturbative cross sections are (no contribution from gluons arises at lowest order in the SLH and there are no quark FCNCs in our setting)~\cite{Husek:2020fru}:

\begin{equation}
    \begin{aligned}
    & \frac{d^{2} \hat{\sigma} \left(\ell q_{i} (\xi P) \rightarrow \tau q_{i} \right)}{d \xi dQ^{2}} = \frac{1}{16 \pi \lambda \left( s(\xi), m^{2}_{\ell}, m^{2}_{i}\right)} \overline{ |\mathcal{M}_{qq} (\xi, Q^{2})|^2}, \\ 
    & \frac{d^{2} \hat{\sigma} \left(\ell \bar{q}_{i}  \rightarrow \tau \bar{q}_{i}(\xi P) \right)}{d \xi dQ^{2}} = \frac{1}{16 \pi \lambda \left( s(\xi), m^{2}_{\ell}, m^{2}_{i}\right)} \overline{ |\mathcal{M}_{\bar{q}\bar{q}} (\xi, Q^{2})|^2},
    \end{aligned}
\end{equation}

where $p_{i} = \xi P$ is the fraction of the nucleus total momentum $P$ carried by the parton, thus we consider $m^{2}_{i} = \xi^{2} M^{2}$. It is necessary to add the same processes but with anti-quarks because the cross section and the non-perturbative behavior is not the same in both cases. The total cross section can be expressed as the sum of the cross section over the nucleons of the nuclei \cite{Gninenko:2018num}:

\begin{equation}
    \sigma \left( \ell + (A,Z) \rightarrow \tau + X  \right) = Z\sigma \left(  \ell + p \rightarrow \tau + X  \right) + (A-Z) \sigma \left(  \ell + n \rightarrow \tau + X  \right), 
\end{equation}

here the nucleon cross section is \cite{Husek:2020fru}, \cite{Gninenko:2018num}:

\begin{equation}
    \begin{aligned}
       \sigma \left(  \ell + N \rightarrow \tau + X  \right) & = \sum_{i} \int^{1}_{\xi_{min}} \int^{Q^{2}_{+}(\xi)}_{Q^{2}_{-}(\xi)} d\xi dQ^{2} \Big[ \frac{d^{2} \hat{\sigma} \left(\ell q_{i} (\xi P) \rightarrow \tau q_{i} \right)}{d \xi dQ^{2}} H_{q_{i}}\left( \xi, Q^{2} \right) \\
       & + \frac{d^{2} \hat{\sigma} \left(\ell \bar{q}_{i}  \rightarrow \tau \bar{q}_{i}(\xi P) \right)}{d \xi dQ^{2}} H_{\bar{q}_{i}}\left( \xi, Q^{2} \right) \Big],
    \end{aligned}
\end{equation}

where $H_{q_{i}}\left( \xi, Q^{2} \right)$ and $H_{\bar{q}_{i}}\left( \xi, Q^{2} \right)$ are the quark and anti-quark PDFs, respectively, and nuclear targets under consideration are Fe with $A=56$, $Z=26$ and Pb with $A=207$, $Z=82$. The integration limits can be found in ref.\cite{Husek:2020fru}. Penguin form factors (quarks and anti-quarks) are the same than for the $\mu-e$ conversion and quark box form factors are computed from the Feynman diagrams in figure \ref{fig:nuclei}  (for anti-quarks we need to invert the lower fermion line). Quark box form factors are the same than in eqs.~\eqref{box1}-\eqref{box5}, and the anti-quarks box form factors can be obtained from those equations as well. When we invert the lower fermion line we exchange the diagrams for the different embeddings: quark diagrams in the Anomaly free (Universal) embedding are then related to antiquark diagrams in the Universal (Anomaly free) embedding, therefore we need to  change $  \{u_{m}, U_{m} \} \leftrightarrow \{d_{m}, D_{m} \}$ in those equations (and the overall mixing coefficient, so that it corresponds to $\ell\to\tau$ transitions), to get the anti-quark box form factors. Again, diagrams with light quarks and those which are proportional to $\omega$ give subleading corrections. The squared amplitude can be computed from eqs. \eqref{mue}, leading to the result (we use the Mandelstam variables $s = \left( p_{\ell} + p_{i} \right)^2$, $t = \left( p_{\ell} - p_{\tau} \right)^2 = -Q^2$, $u = \left( p_{i} - p_{\tau} \right)^2$ ):

\begin{equation}\label{1qq}
    \begin{aligned}
   & \overline{ |\mathcal{M}_{qq} (\xi, Q^{2})|^2}  = -\frac{4e^{4} Q^{2}_{q}}{(Q^{2})^2} \Bigg[ (Q^{2})^2 \big|A_{L1}\big|^{2} \Big[ 2m_{q}\xi M \left( m^{2}_{\ell} + m^{2}_{\tau} + Q^{2} \right) \\
    & + \left( m^{2}_{\tau} + \xi^{2}M^{2} +  Q^{2}-s \right) \left( m^{2}_{\tau} + \xi^{2}M^{2} -u\right)  + \left( m^{2}_{\ell} + \xi^{2}M^{2}-s \right) \left( m^{2}_{\ell} + \xi^{2}M^{2} + Q^2 -u \right) \Big] \\
    & + \frac{\big|A_{R2}\big|^{2} m^2_{\ell}}{2} \Big[ \left( m^2_{\ell} + m^2_{\tau} + Q^2 \right)^{2} \left( \xi^{2}M^{2} + m^2_{q} +Q^2 - 2 \xi Mm_{q} \right) + 4m^2_{\ell}\left( \xi^{2}M^{2} + m^{2}_{\tau} -u \right)\left(\xi^{2}M^{2} + m^2_{\ell} + Q^2 - u  \right)   \\
    & + \left( m^2_{\ell} + m^2_{\tau} + Q^2 \right) \left[ \left( m^2_{\ell} + m^2_{\tau}\right) \left( 6\xi Mm_{q}- \xi^{2}M^{2}-m^2_{q}-Q^2\right) - \left( s-m^2_{\ell}+m^2_{\tau}-u \right)\left( s-m^2_{\tau}+m^2_{\ell}-u \right) \right] \\
    & + 4m^2_{\tau} \left[ \left( s-m^2_{\ell}-\xi^{2}M^{2}  \right) \left( s-m^2_{\tau}-\xi^{2}M^{2} -Q^2 \right) -4\xi Mm_{q}m^2_{\ell}  \right]  \Big] \\
    & + \frac{m^2_{\ell} Q^2}{2} \left( A_{L1}A^{*}_{R2} + A_{R2}A^{*}_{L1}\right) \Big[ \left( 6 \xi M m_{q}- \xi^{2}M^{2}-m^2_{q}-Q^2\right) \left(m^2_{\ell}+Q^2-m^2_{\tau} \right) \\
    & + \left( u-\xi^{2}M^{2}-m^2_{\ell}-Q^2 \right) \left( s-m^2_{\ell}-3\xi^{2}M^{2}-2m^2_{\tau} +2u\right) + \left( s-\xi^{2}M^{2}-m^2_{\tau}-Q^2 \right) \left( u-\xi^{2}M^{2}-m^2_{\tau} \right)     \Big]  \Bigg] \\
    & -4e^4 \Bigg[ \big|\hat{B}^{q}_{L1}\big|^{2} \left( \xi^{2}M^{2}+m^2_{\ell}-s \right) \left( \xi^{2}M^{2}+m^2_{\ell}+Q^2-u \right) + \big|\hat{B}^{q}_{L2}\big|^{2} \left( \xi^{2}M^{2}+m^2_{\tau}-u \right) \left( \xi^{2}M^{2}+m^2_{\tau}+Q^2-s \right) \\
    & +\xi M m_{q} \left( m^2_{\ell} + m^{2}_{\tau} + Q^2\right) \left( \hat{B}^{q}_{L1} \hat{B}^{q*}_{L2} + \hat{B}^{q}_{L2} \hat{B}^{q*}_{L1} \right) \Bigg] \\
    & +4e^4 Q_{q}  \left( A_{L1}\hat{B}^{q*}_{L2} + \hat{B}^{q}_{L2}A^{*}_{L1} \right)  \left[  \left( \xi^{2}M^{2}+m^{2}_{\tau}+Q^2-s \right) \left( \xi^{2}M^{2} + m^2_{\tau}-u \right) + \xi M m_{q} \left( m^{2}_{\ell} + m^{2}_{q}+Q^{2} \right) \right] \\
    & +4e^4 Q_{q}  \left( A_{L1}\hat{B}^{q*}_{L1} + \hat{B}^{q}_{L1}A^{*}_{L1} \right)  \left[  \left( \xi^{2}M^{2}+m^{2}_{\ell}+Q^2-u \right) \left( \xi^{2}M^{2} + m^2_{\ell}-s \right) + \xi M m_{q} \left( m^{2}_{\ell} + m^{2}_{q}+Q^{2} \right) \right] \\
    & -\frac{e^4 Q_{q} m^{2}_{\ell}}{Q^2} \left( A_{R2}\hat{B}^{q*}_{L2} + \hat{B}^{q}_{L2}A^{*}_{R2}  \right) \Bigg[ \left( m^{2}_{\ell}+Q^2-m^{2}_{\tau} \right) \left( \xi^{2}M^{2} +m^2_{q} +Q^2-6 \xi M m_{q}\right) - 4Q^2 \left( s+m^{2}_{\tau} \right) \\ 
    & + \left( u-m^2_{\tau}-\xi^{2}M^{2} \right) \left(\xi^{2}M^{2}+2m^2_{\tau}+Q^2+s+2u \right) + 3 \left( s+u-\xi^{2}M^{2} \right) \left( \xi^{2}M^{2} + m^{2}_{\tau}+Q^2-u \right) \\
    &+ u \left( \xi^{2}M^{2} +m^2_{\tau}+Q^2-u \right) + m^2_{\tau} \left( s+u-m^2_{\tau}-m^{2}_{\ell}-\xi^{2}M^{2} \right) + m^{2}_{\ell} \left( m^2_{\ell}+Q^2+u-m^2_{\tau}-s \right)    \Bigg] \\
    & +\frac{e^4 Q_{q} m^{2}_{\ell}}{Q^2} \left( A_{R2}\hat{B}^{q*}_{L1} + \hat{B}^{q}_{L1}A^{*}_{R2}  \right) \Bigg[ -\left( m^{2}_{\ell}+Q^2+m^{2}_{\tau} \right) \left( \xi^{2}M^{2} +m^2_{q} +Q^2-6 \xi M m_{q}\right)  \\
    & + \left( \xi^{2}M^{2} + m^2_{\tau}-u \right) \left( \xi^{2}M^{2} + 2m^2_{\ell}-m^2_{\tau}+Q^2+s-2u  \right) + 3\left( \xi^{2}M^{2} + m^2_{\ell} -s \right) \left( \xi^{2}M^{2} + m^2_{\ell}+Q^2-u \right)  \\
    & + 2m^2_{\tau} \left( \xi^{2}M^{2} + m^2_{q} +Q^2-6\xi M m_{q}  \right)     \Bigg], \end{aligned}
\end{equation}

where

\begin{equation}
\hat{B}^{q}_{L1} = B^{q}_{L1} + F^{q}_{LL} + F'^{q}_{LL}, \hspace{4mm} \hat{B}^{q}_{L2} = F^{q}_{LR} + F'^{q}_{LR}.
\end{equation}

To find the squared amplitude for anti-quarks we only replace $s \leftrightarrow u$ in eq.~\eqref{1qq}. We note that the dominant contributions come from the $|\hat{B}^q_{L1,2}|^2$ terms.

\section{Numerical Results}
\label{sec:Num}
In this section we show and discuss the numerical results of our 16 LFV processes exposed above. The first step is setting the range for the free parameters of SLH model: $f$, $t_{\beta}$, $M_{N_{i}}$, $\delta_{\nu}$, $V^{\ell i}$ and $\delta_q$, as we motivate in the following. Dependence of the $\mu\to e$ observables on these parameters is studied in detail in ref. \cite{delAguila:2008zu} for the case of two heavy neutrinos. Approximate cancellations between the $\gamma+Z$ penguins and box contributions were already noted in this reference (for $\mu\to e$ transitions). These effects  strongly depend on the specific region in the parameter space of the SLH model and, because of that, we will not dwell into them here.

The scale of compositeness, $f$, can be estimated through the direct search of $Z'$ bosons at LHC \cite{ATLAS:2017fih}, where the lower limit is set as \cite{Mao:2017hpp} $f \gtrsim 7.5$ TeV at $95 \%$ C.L. Following the analysis given in ref.~\cite{Cheung:2018iur} we fix the upper limit $f \lesssim 85$ TeV. Above these energies, SLH loses consistency.

The ratio of the two vevs $t_{\beta} = f_1/f_2$ is another important free parameter of this model. A perturbative unitarity analysis \cite{Cheung:2018iur} binds $1 \leq t_{\beta} \leq 9$. For small $f$ ($10\leq f(\mathrm{TeV})\leq20$), $ t_{\beta}$ can vary freely in this interval, while for $20\leq f(\mathrm{TeV})\leq80$, the approximate relation $t_\beta=\frac{2}{15}f(\mathrm{TeV})-\frac{25}{15}$ holds.

Heavy neutrinos are responsible for the LFV lepton decays, however this ``little" neutrino masses are unknown. We will, nevertheless,  follow ref. \cite{Lami:2016mjf} and take the ratios involving them as: $0.1 \leq x_1 \leq 0.25$, $1.1x_1 \leq x_2 \leq 10x_1$, $1.1x_2 \leq x_3 \leq 10x_2$ (we remind that $x_i$ depends quadratically on the $N_i$ mass), where we include  the cases of a small splitting $x_2 = 1.1x_1$ and a large one $x_2 = 10x_1$ (analogously for $x_3$).

The mixing of the ``little" and light neutrino of the same family is encapsulated in  $\delta_{\nu}$ and, according to data, $\delta_{\nu} \lesssim 0.05$ \cite{delAguila:2011wk, Deppisch:2015qwa, deBlas:2013gla}, that we will take.

We do not have any information of the mixing matrix $V^{\ell i}$ between charged leptons and ``little" neutrinos, which can be parameterized in the standard form \cite{ParticleDataGroup:2020ssz}. According to ref. \cite{Lami:2016mjf}, we have scanned over $−1\leq s_{ij} \leq 1$ ensuring the low-energy restrictions and, in addition, we assumed for simplicity CP conservation (phase in $V^{\ell i}$ set to zero).

Finally, the  mixing between the heavy quarks and the corresponding SM quarks is parameterized by the $\delta_q$ parameters. We follow the arguments of ref.\cite{Han:2005ru} and assume that the mixing effects are suppressed in the down-quark (up-quark) sector for the Anomaly free (Universal) embedding in the $t_{\beta} >1$ regime, so it implies: $\delta_{q} = \mp \delta_{\nu}$, where the upper sign is for the Anomaly free ($q = D,S$) and the lower sign is for the Universal embedding ($q = U,C$). We also follow the proposal of ref.\cite{delAguila:2011wk} and take the reference values for the ratios of heavy quark masses as $x_U = x_D = x_C = x_S \equiv 1$.

As we have already mentioned, there are still no experimental limits for $\ell - \tau$ conversion, but if we consider the expected sensitivity of NA64 experiment we can express the conversion probability as the ratio \cite{Gninenko:2018num}:

\begin{equation}
    \mathcal{R} = \frac{\sigma \left( \ell + N \rightarrow \tau + X \right)}{\sigma \left( \ell + N \rightarrow \ell + X \right)} \sim 10^{-13}-10^{-12},
\end{equation}

where the denominator is the dominant contribution to the inclusive $\ell + N$ processes due to the bremsstrahlung of leptons off nuclei \cite{Gninenko:2018num}:

\begin{equation}
    \begin{aligned}
    & \sigma \left(e+Fe \rightarrow e+X \right) = 0.129\times10^{5} \hspace{2mm} \textrm{GeV}^{-2}, \hspace{5mm} \sigma \left(\mu+Fe \rightarrow \mu+X \right) = 0.692 \hspace{2mm} \textrm{GeV}^{-2},\\
   &\sigma \left(e+Pb \rightarrow e+X \right) = 1.165\times10^{5} \hspace{2mm} \textrm{GeV}^{-2}, \hspace{5mm} \sigma \left(\mu+Pb \rightarrow \mu+X \right) = 6.607 \hspace{2mm} \textrm{GeV}^{-2}.
    \end{aligned}
\end{equation}

As representative energy for the initial electron or muon beam we take $E_{e} = 100$ GeV and $E_{\mu} = 150$ GeV. We do the analysis within a Monte Carlo simulation with all channels sharing the free parameters enumerated above. In the following table we summarize our mean values, the present experimental bounds  \cite{ParticleDataGroup:2020ssz} and the future expected sensitivities  (whose values were taken from ref. \cite{Hernandez-Tome:2019lkb} and references therein). 

\begin{table}[H]
\begin{center}
\begingroup
\renewcommand{\arraystretch}{1.4}
\begin{tabular}{c c c c c}
\toprule
\toprule
LFV decays & Experimental Limits & Our mean values &  Future sensitivity \\
\midrule
\midrule
$\mu \rightarrow e \gamma$ & $4.2 \times 10^{-13}$ & $2.1
\times 10^{-14}$ & $6 \times 10^{-14}$  \\
$\mu \rightarrow e e \bar{e}$ & $1.0 \times 10^{-12}$ & $5.7
\times 10^{-15}$ & $10^{-16}$ \\
$\tau \rightarrow e \gamma$ & $3.3 \times 10^{-8}$ & $5.6
\times 10^{-12}$ & $3 \times 10^{-9}$\\
$\tau \rightarrow \mu \gamma$ & $4.4 \times 10^{-8}$ & $2.3
\times 10^{-12}$ & $10^{-9}$\\
$\tau \rightarrow e e \bar{e}$ & $2.7 \times 10^{-8}$ & $3.2
\times 10^{-12}$ & $(2-5) \times 10^{-10}$\\
$\tau \rightarrow \mu \mu \bar{\mu}$ & $2.1 \times 10^{-8}$ & $1.6
\times 10^{-12}$ & $(2-5) \times 10^{-10}$\\
$\tau \rightarrow e \mu \bar{\mu}$ & $2.7 \times 10^{-8}$ & $2.1
\times 10^{-12}$ & $(2-5) \times 10^{-10}$ \\
$\tau \rightarrow  \mu e \bar{e}$ & $1.8 \times 10^{-8}$ & $1.0
\times 10^{-12}$ & $(2-5) \times 10^{-10}$\\
$\tau \rightarrow  \mu \mu \bar{e}$ & $1.7 \times 10^{-8}$ & $3.8
\times 10^{-18}$ & $(2-5) \times 10^{-10}$\\
$\tau \rightarrow  e e \bar{\mu}$ & $1.5 \times 10^{-8}$ & $5.6
\times 10^{-18}$ & $(2-5) \times 10^{-10}$ \\
$ \mu Ti \rightarrow e Ti$ & $4.3 \times 10^{-12}$ & $6.8
\times 10^{-14}$ (AF), $8.6
\times 10^{-14}$ (U) & $10^{-18}$
\\
$ \mu Au \rightarrow e Au$ & $7.0 \times 10^{-13}$ & $8.2
\times 10^{-14}$ (AF), $1.1
\times 10^{-13}$ (U) &-\\
$ e Fe \rightarrow \tau Fe$ & - & $9.2
\times 10^{-20}$ (AF), $9.3
\times 10^{-20}$ (U) &-\\
$ e Pb \rightarrow \tau Pb$ & - & $1.6
\times 10^{-19}$ (AF), $1.6
\times 10^{-19}$ (U) &-\\
$ \mu Fe \rightarrow \tau Fe$ & - & $6.2
\times 10^{-16}$ (AF), $6.2
\times 10^{-16}$ (U) &-\\
$ \mu Pb \rightarrow \tau Pb$ & - & $9.6
\times 10^{-16}$ (AF), $9.8
\times 10^{-16}$ (U)&-\\
\bottomrule
\bottomrule
\end{tabular}
\endgroup
\caption{Mean values of branching ratios and conversion rates (where AF stands for Anomaly Free embedding and U stands for Universal embedding) against current upper limits at 90 \% confidence level and future sensitivities.}
\label{tabla13}
\end{center}
\end{table}

In the case of muon decays, our results are below the experimental limit by one ($\mu\to e\gamma$) and three ($\mu\to3e$) orders of magnitude, the mean values of nuclei conversion in both embeddings are below the upper bounds by one or two orders of magnitude. For the case of Au nuclei conversion in the Universal embedding our mean value is only a factor $\sim7$ below the experimental limit.

We turn now to LFV transitions involving the tau flavour. For the cases of $\tau \rightarrow \ell \gamma$ ($\ell = e, \mu$), same flavor and same sign decays, our results are below the experimental limits by four orders of magnitude (not for wrong sign decays, which are six orders of magnitude further  suppressed). For the case of $\ell - \tau$ conversion, we find that the mean values with electrons are too small for the expected sensitivity of the  NA64 experiment. However, the analogous processes with muons are only slightly below their forecasted  sensitivity, and in principle could be tested with future experiments.

From these results we verify that muon physics is the best candidate to test LFV; our mean values are of the same order ($\mu \rightarrow e \gamma$) or higher ($\mu \rightarrow 3e$, $\mu Ti \rightarrow e Ti$) than the future sensitivity and will set the strongest limits. For tau decays, our mean values are still below the future sensitivity and only next generations of $B$ factories could be able to search for them, according to the SLH model. Still, $\mu\to\tau$ conversion in nuclei appears promising as a discovery tool to first measure LFV involving the tau flavor. Consequently, it can play a significant role in characterizing the underlying new physics causing charged LFV, as can be checked from the correlations amid processes, that we discuss next.

\begin{figure}[H]
\begin{subfigure}{.5\textwidth}
  \centering
  \includegraphics[width=\textwidth]{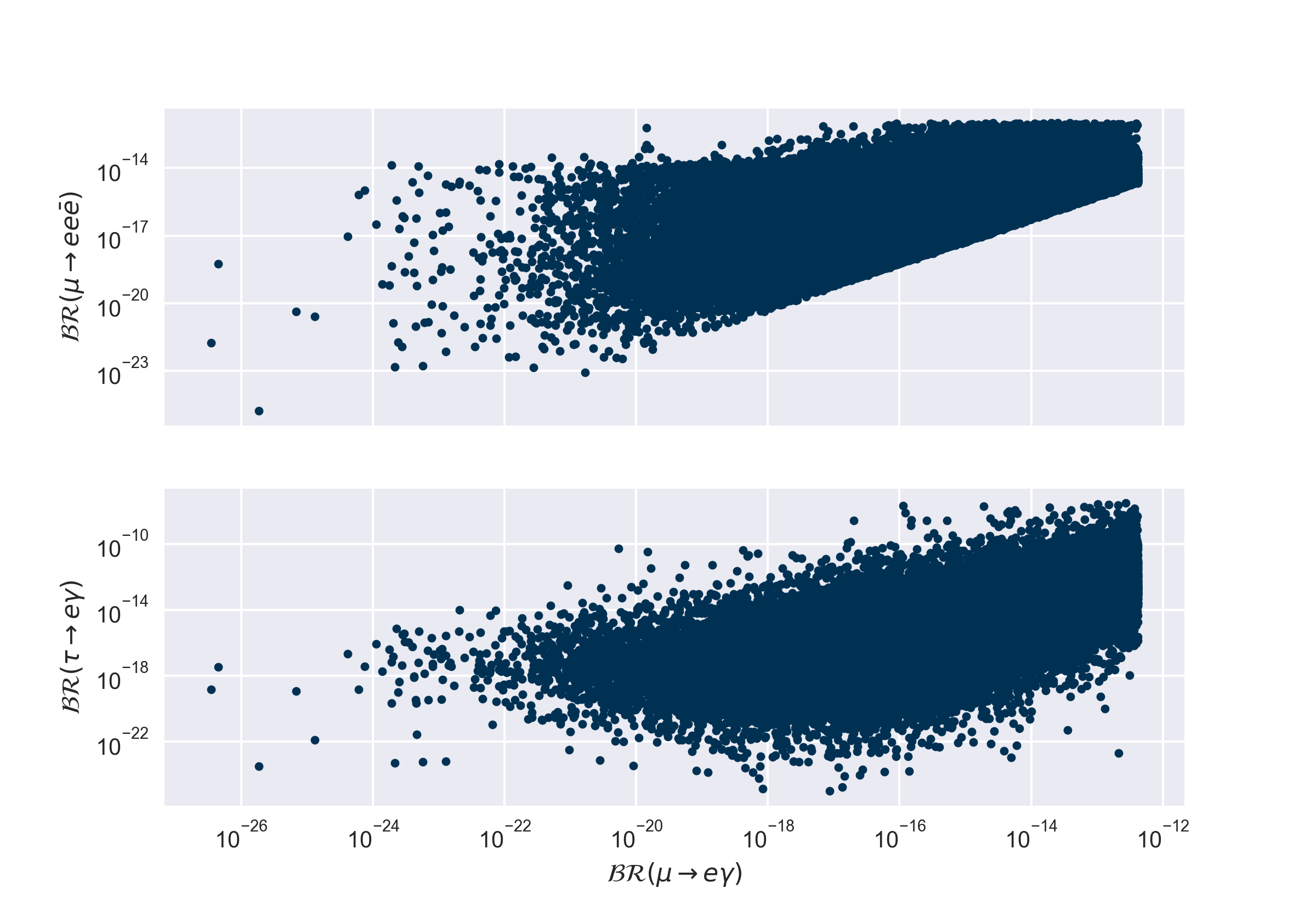}  
  \caption{ $\mathcal{BR}(\mu \rightarrow e \gamma)$ vs. $\mathcal{BR}(\mu \rightarrow e e \bar{e})$, $\mathcal{BR}(\tau \rightarrow e \gamma)$}
  \label{fig:sub-uno}
\end{subfigure}
\begin{subfigure}{.5\textwidth}
  \centering
  \includegraphics[width=\textwidth]{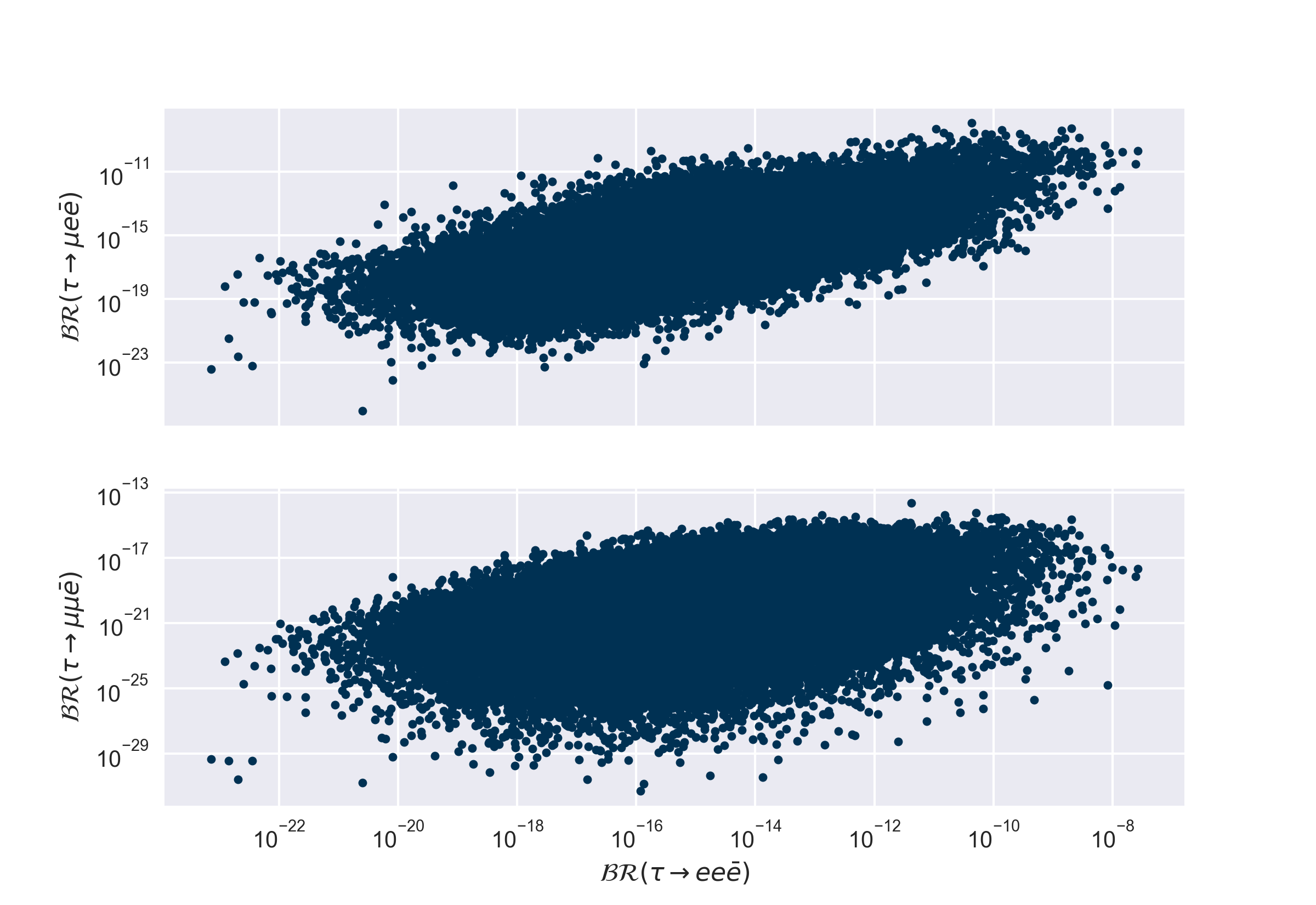}  
  \caption{ $\mathcal{BR}(\tau \rightarrow ee \bar{e})$ vs. $\mathcal{BR}(\tau \rightarrow \mu \mu \bar{e}) $, $\mathcal{BR}(\tau \rightarrow \mu e \bar{e})$}
  \label{fig:sub-dos}
\end{subfigure}
\caption{Scatter plots for $\ell\to\ell'\gamma$ and some $\ell\to 3\ell'$ decays.}
\label{fig:fig1}
\end{figure}

\begin{figure}[H]
\begin{subfigure}{.5\textwidth}
  \centering
  \includegraphics[width=\textwidth]{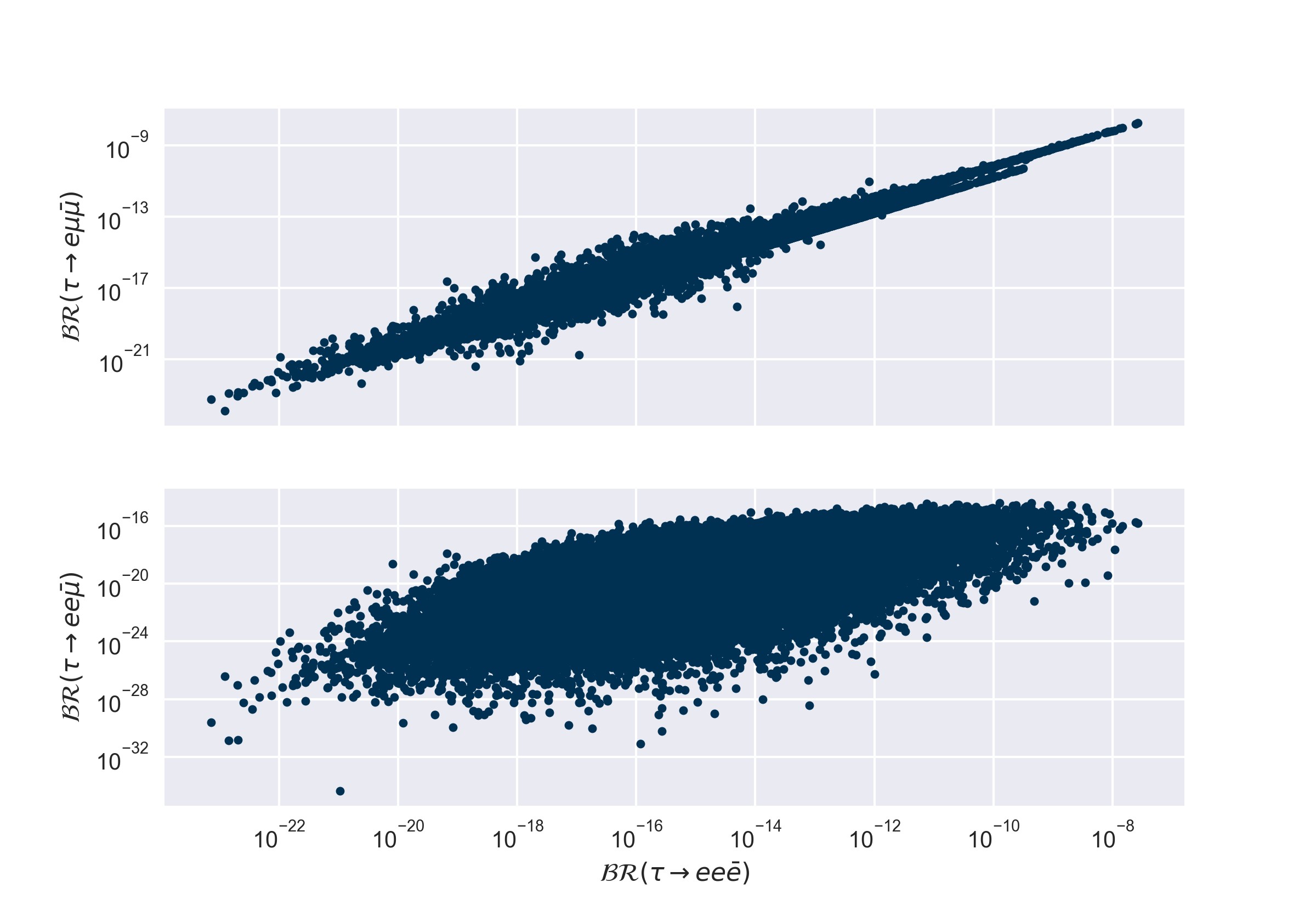}  
  \caption{ $\mathcal{BR}(\tau \rightarrow ee \bar{e})$ vs. $\mathcal{BR}(\tau \rightarrow e e \bar{\mu})$, $\mathcal{BR}(\tau \rightarrow e \mu \bar{\mu})$}
  \label{fig:sub-tres}
\end{subfigure}
\begin{subfigure}{.5\textwidth}
  \centering
  \includegraphics[width=\textwidth]{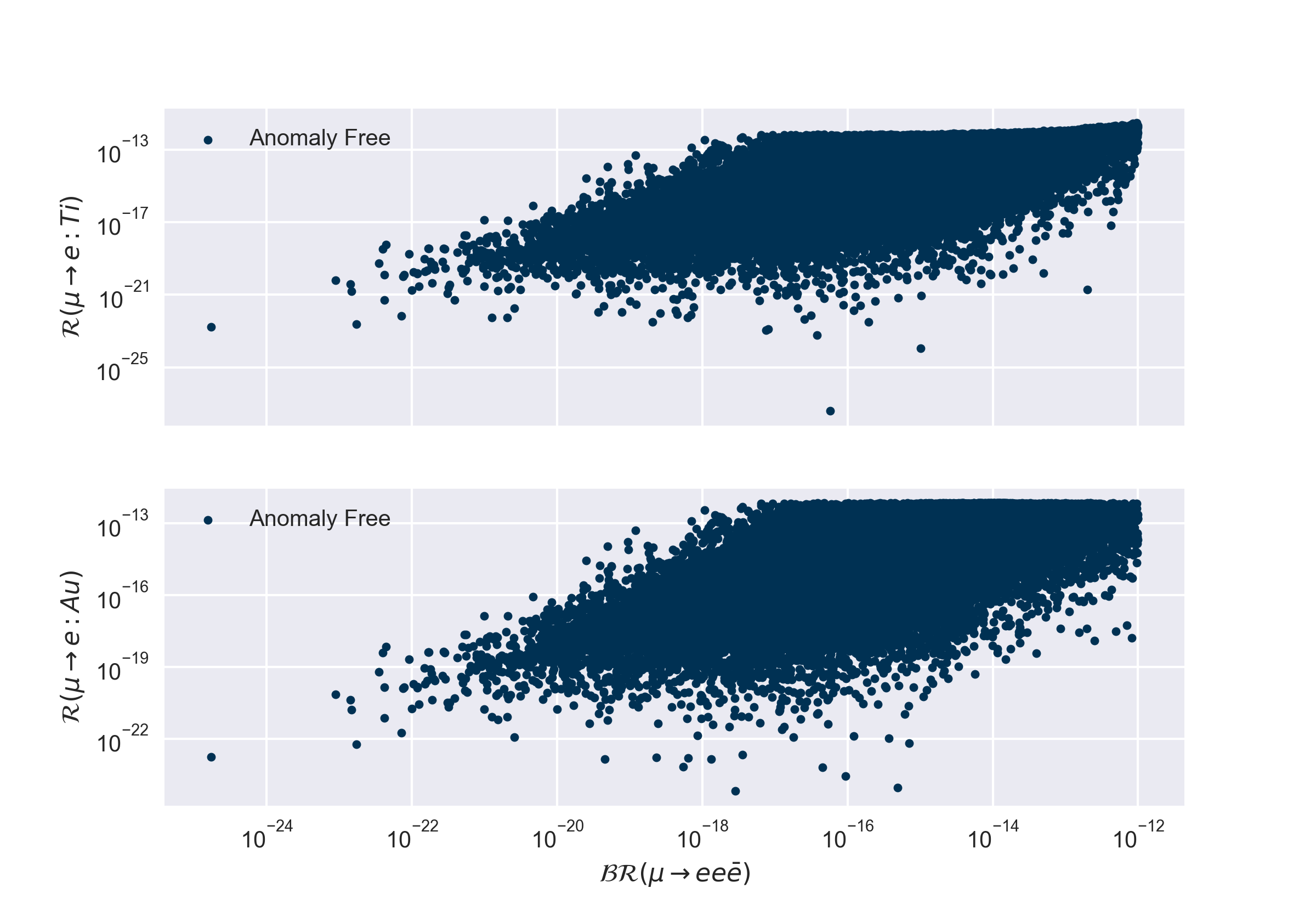}  
  \caption{ $\mathcal{BR}(\mu \rightarrow ee \bar{e})$ vs. $\mathcal{R}(\mu \rightarrow e: Ti) $, $\mathcal{R}(\mu \rightarrow  e: Au)$}
  \label{fig:sub-cuatro}
\end{subfigure}
\caption{Scatter plots for $\ell\to3\ell'$ decays and $\ell \to \ell'$ nuclei conversion.}
\label{fig:fig2}
\end{figure}

\begin{figure}[H]
\begin{subfigure}{.5\textwidth}
  \centering
  \includegraphics[width=\textwidth]{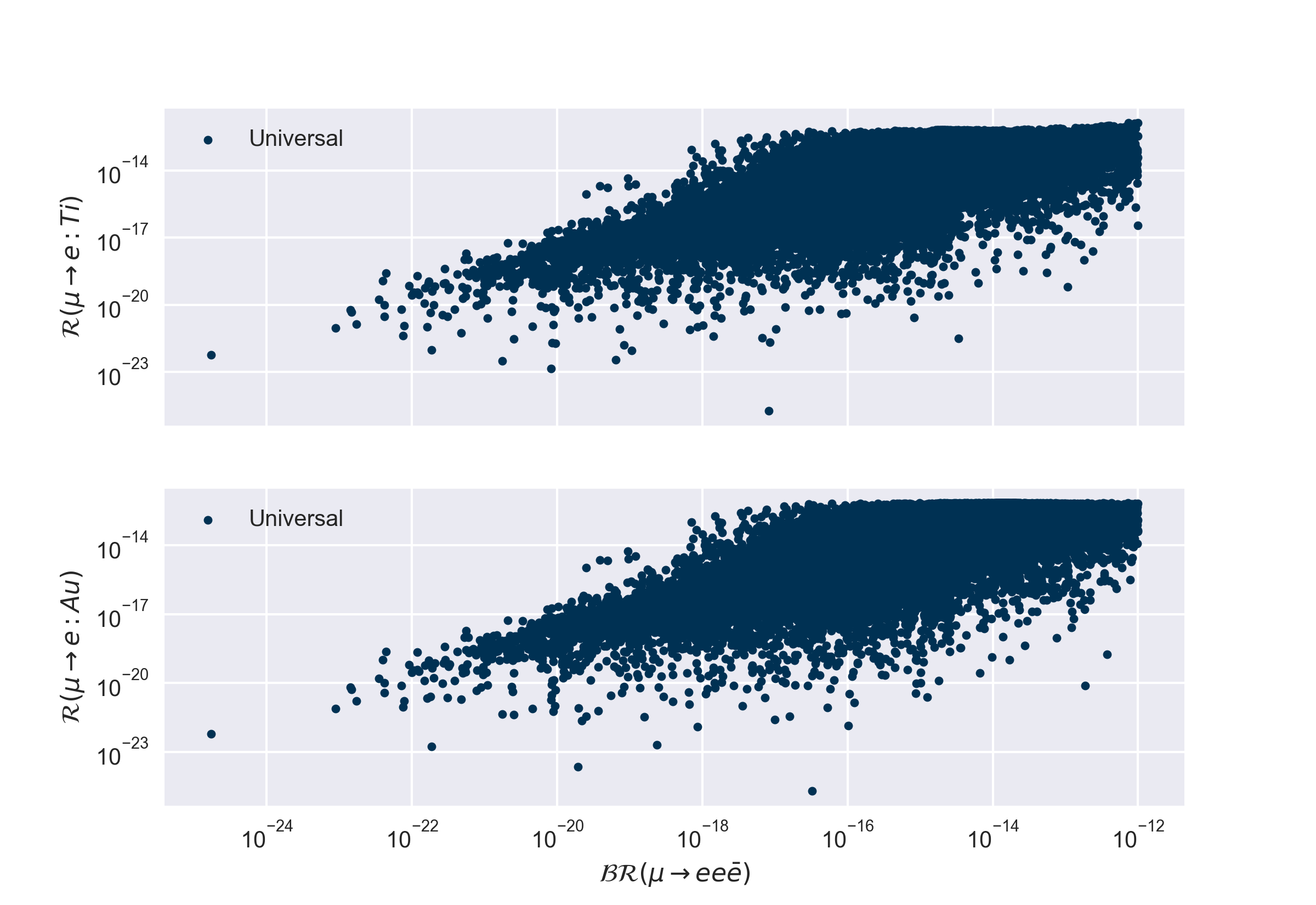}  
  \caption{ $\mathcal{BR}(\mu \rightarrow ee \bar{e})$ vs. $\mathcal{R}(\mu \rightarrow e: Ti) $, $\mathcal{R}(\mu \rightarrow  e: Au)$}
  \label{fig:sub-cinco}
\end{subfigure}
\begin{subfigure}{.5\textwidth}
  \centering
  \includegraphics[width=\textwidth]{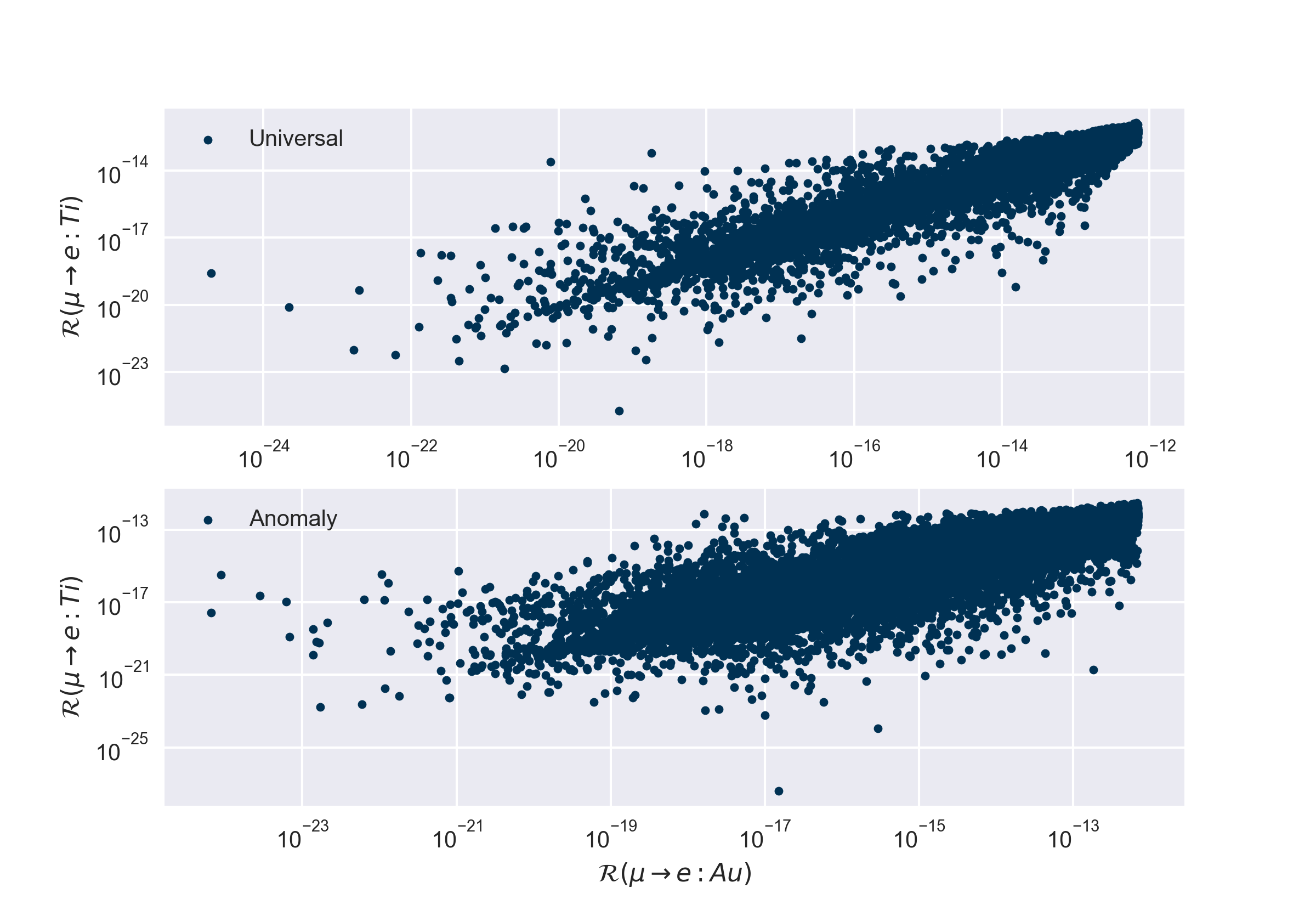}  
  \caption{ $\mathcal{R}(\mu \rightarrow e: Ti) $ vs. $\mathcal{R}(\mu \rightarrow  e: Au)$}
  \label{fig:sub-seis}
\end{subfigure}
\caption{Scatter plots for $\mu \to 3e$ and $\mu \to e$ nuclei conversion.}
\label{fig:fig3}
\end{figure}

We show in figs.~\ref{fig:fig1}-\ref{fig:fig3} a selected set of  scatter plots comparing the different processes. Fig.\ref{fig:sub-uno} plots $\mathcal{BR}(\mu \rightarrow e \gamma)$ vs. $\mathcal{BR}(\mu \rightarrow ee\bar{e})$, which are moderately  correlated. A similar, though softened, trend is  observed in the accompanying figure $BR ( \mu \to e \gamma)$ vs. $BR ( \tau \to e \gamma)$ (despite they differ in the flavor  coefficients). Figure \ref{fig:sub-dos} and \ref{fig:sub-tres} should be understood together. In these plots, we show the correlation between same flavor decay against the same sign and wrong sign decays, respectively. We see that $\mathcal{BR}(\tau \rightarrow ee \bar{e})$ keeps a big correlation with $\mathcal{BR}(\tau \rightarrow e \mu \bar{\mu})$ and the opposite happens with $\mathcal{BR}(\tau \rightarrow \mu e \bar{e})$. This is also caused by the corresponding flavor coefficients, as expected. The comparison between same flavor and wrong sign decays is quite different, because wrong sign are decays with only box contributions, so no correlation is expected in those plots (however the lower panel of fig. \ref{fig:sub-tres} exhibits a small one, albeit this will be very challenging to probe, given the big suppression of the wrong-sign decays in our setting). We do not show other analogous plots including, for instance, the $\tau\to3\mu$ decay.

In plots \ref{fig:sub-cuatro} and \ref{fig:sub-cinco} we show the scatter plot for the $\mathcal{BR}(\mu \rightarrow 3e)$ against $\mathcal{R}(\mu \rightarrow e)$ nuclei conversion. The result for $Ti$ and $Au$ nuclei are almost the same, and in general the outcome for both embeddings is very similar. However, results for nuclei conversion in $Ti$ show a bigger correlation than the results for $Au$, being the dependence  stronger in the Anomaly free embedding. Plot \ref{fig:sub-seis} shows the comparison of nuclei conversion in both embeddings (which is the reason why we draw the x-axis in both plots). Our results are alike in both, with stronger correlation in the Universal embedding.

\begin{figure}[H]
\begin{subfigure}{.5\textwidth}
  \centering
  \includegraphics[width=\textwidth]{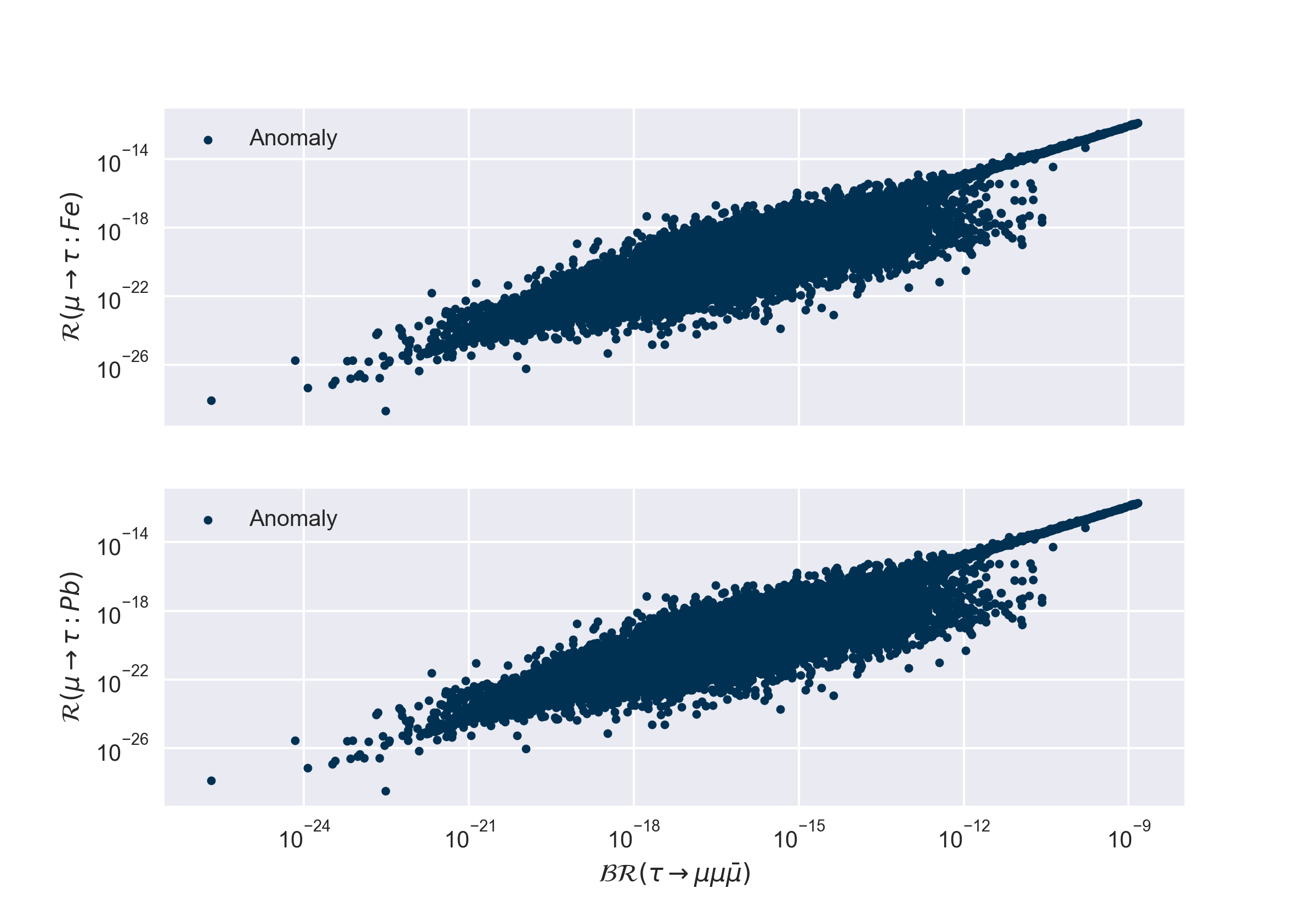}  
  \caption{ $\mathcal{BR}(\tau \rightarrow \mu \mu \bar{\mu})$ vs. $\mathcal{R}(\mu \rightarrow \tau: Fe) $, $\mathcal{R}(\mu \rightarrow  \tau: Pb)$}
  \label{fig:sub-siete}
\end{subfigure}
\begin{subfigure}{.5\textwidth}
  \centering
  \includegraphics[width=\textwidth]{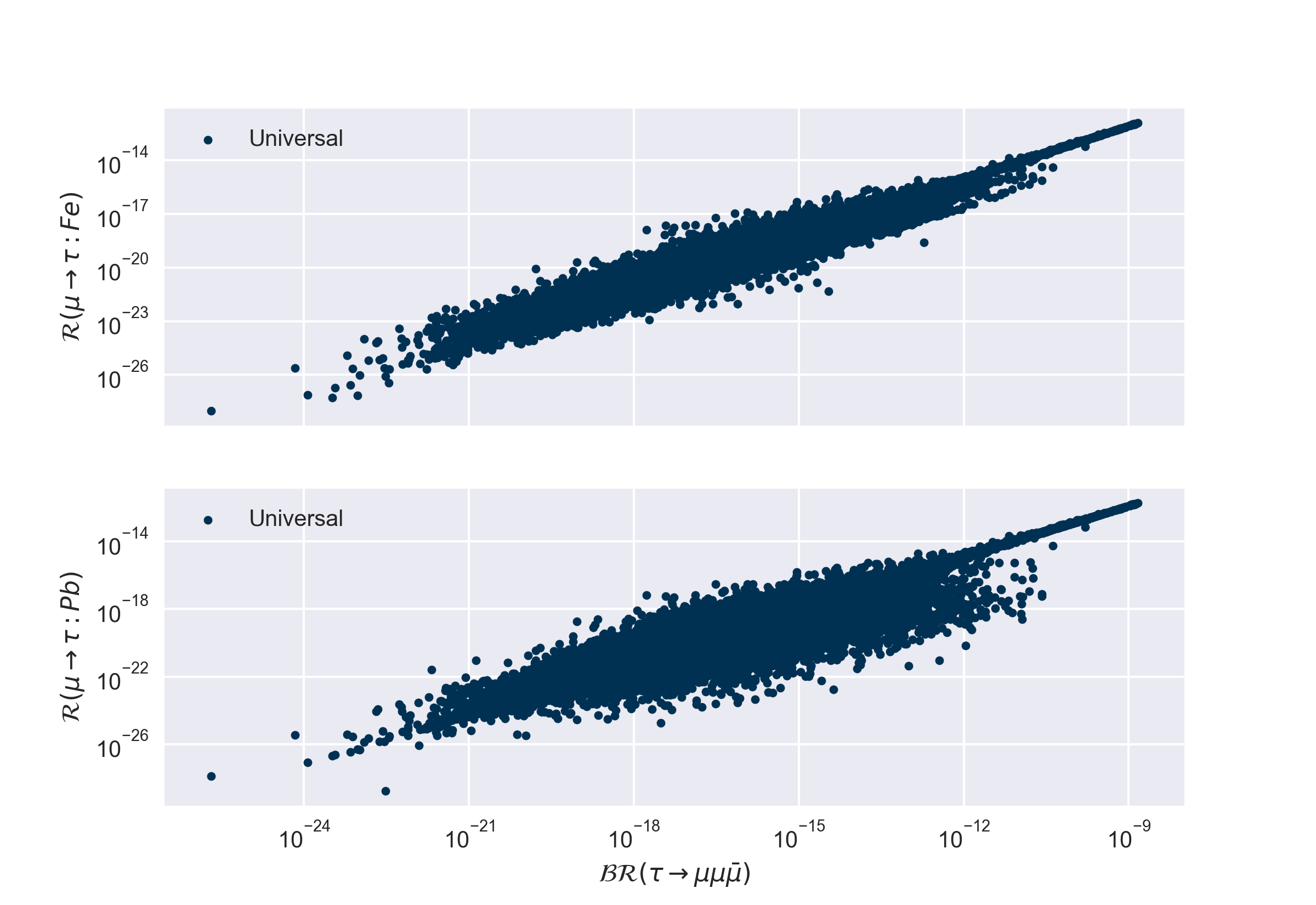}  
  \caption{ $\mathcal{BR}(\tau \rightarrow \mu \mu \bar{\mu})$ vs. $\mathcal{R}(\mu \rightarrow \tau: Fe) $, $\mathcal{R}(\mu \rightarrow  \tau: Pb)$}
  \label{fig:sub-ocho}
\end{subfigure}
\caption{Scatter plots for $\tau \to 3\mu$ and $\mu \to \tau$ nuclei conversion.}
\label{fig:fig4}
\end{figure}

\begin{figure}[H]
\centering
\includegraphics[width=7.5cm]{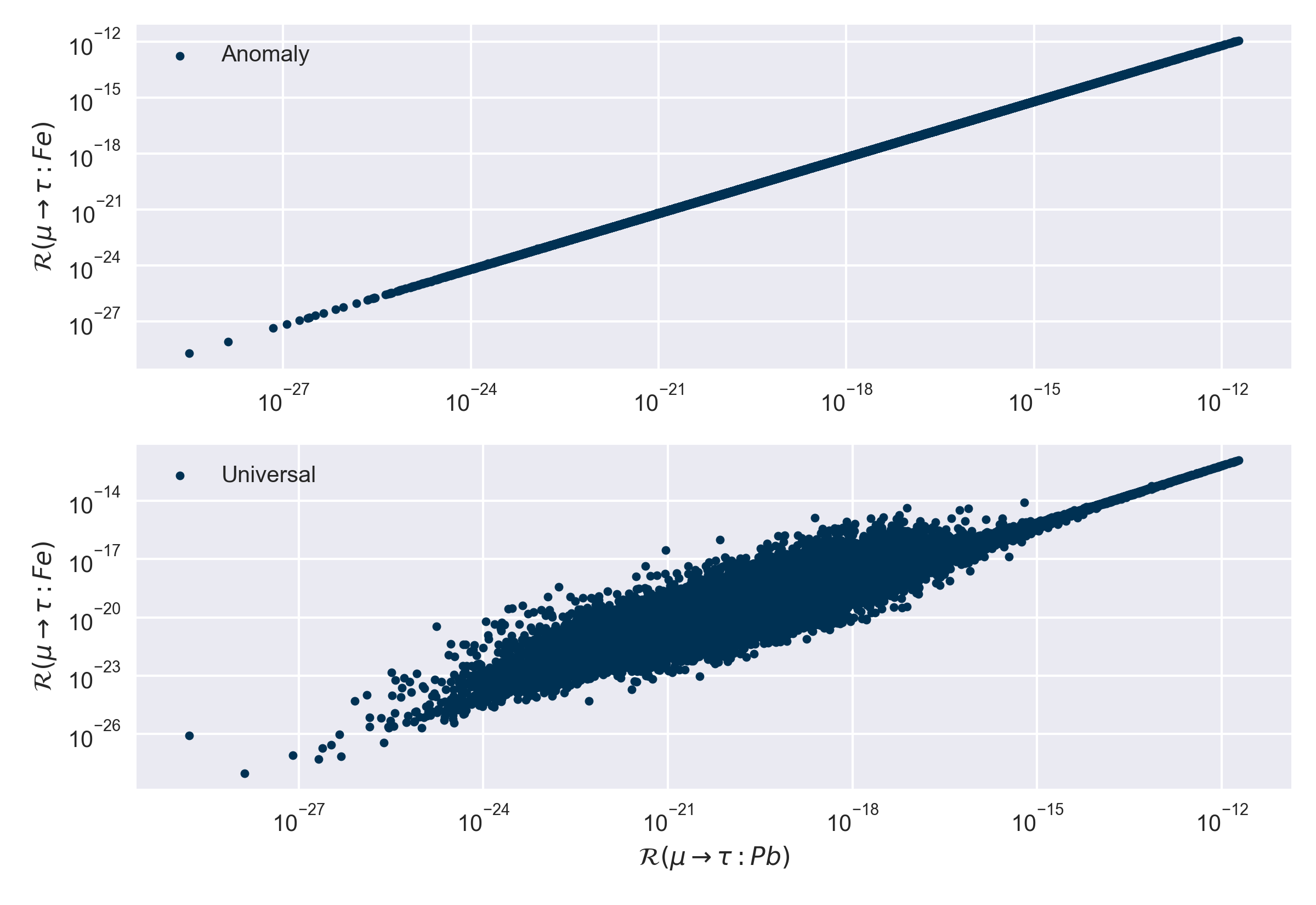}
\caption{ $\mathcal{R}(\mu \rightarrow \tau: Pb) $ vs. $\mathcal{R}(\mu \rightarrow  \tau: Fe)$}
\label{fig:fig5}
\end{figure}

The scatter plots in fig.~\ref{fig:fig4} show the comparison of $\mathcal{BR}(\tau \rightarrow 3\mu)$ with $\mu \to \tau$ nuclei conversion, the general behavior in both embeddings is the same, but small differences lie in the non-perturbative behavior of quarks and anti-quarks inside the heavy nuclei under consideration. However, as we can see in table \ref{tabla13} these differences are negligible in the expected probabilities.  Again, the strongest correlations are found in the Universal embedding. Finally, the scatter plot in figure \ref{fig:fig5} compares the $\mu \to \tau$ conversion in both embeddings. We see a perfect correlation in the Anomaly free embedding that is a bit degraded for the universal case. We do not show analogous correlations (even if with three orders of magnitude smaller probabilities) for $e \to \tau$ nuclei conversion.

\section{The CDF \texorpdfstring{$M_{W}$}{TEXT} measurement within the SLH model}\label{sec:MW}

As a timely topic, we will finally  discuss the implications, within the SLH, of the new measurement of the $W$ boson mass given by the CDF collaboration with a result $M_{W} = (80.4335 \pm 0.0094)$ GeV \cite{CDF:2022hxs}, that shows a discrepancy of $7 \sigma$ with the SM prediction $M_{W} = (80.357 \pm 0.006)$ GeV and is also in tension with respect to the world average $M_{W} = (80.379 \pm 0.012)$ GeV \cite{ParticleDataGroup:2020ssz}. Including the CDF measurement, the new world average would be $M_{W} = (80.4242 \pm 0.0087)$ GeV. As already mentioned before, SLH does not have $SU(2)$ custodial symmetry, and the tree-level SM relation $\rho =1$ is no longer valid:

\begin{equation}
\rho = 1 + \frac{v^2}{8f^2}\left(1-t^2_W \right)^2\,.
\end{equation}

In the SM the $\rho$ parameter is $\rho \equiv 1$ at tree level, and the EWPD, upon the inclusion of radiative corrections, yields $\rho \equiv 1.00038 \pm 0.00020$ \cite{ParticleDataGroup:2020ssz}, such deviation from the SM value can be encoded as: 
 
\begin{equation}
 \delta \rho = \frac{v^2}{8f^2}\left(1-t^2_W \right)^2 \equiv \alpha T,
\end{equation}
 
where $T$ together with $S$ and $U$ are the oblique parameters \cite{Peskin:1990zt, Peskin:1991sw} which can parametrize potential new physics affecting electroweak two-point Green  functions. We are going to use the formalism given in references \cite{Marandella:2005wd, Strumia:2022qkt} to show how heavy $Z'$ bosons can modify the oblique parameters and, since SLH is not an Universal theory\footnote{An universal theory is such that Z' couples to the fermions universally (which does not occur in SLH) or proportionally to the SM vectors. }, these corrections cannot be fully represented  with four universal parameters: $\hat{S}$, $\hat{T}$, $Y$, $W$. Nevertheless, the corresponding  effects can be well approximated with this formalism. SM corresponds to $\hat{T} = \hat{S} = W = Y = 0$ and these new parameters are related to the usual oblique ones as $S = 4s^2_{W}\hat{S}/\alpha$ and $T = \hat{T}/\alpha$, the $U$ parameter corresponds to dimension-eight operators, and because $f \gg v$, we can neglect it.

A generic model with a $Z'$ boson is determined by few quantities: gauge coupling $g_{Z'}$, Mass $M_{Z'}$, the couplings to the Higgs boson $Z'_{H}$, and to the left and right fermion multiplets $Z'_{L}$, $Z'_{e}$, $Z'_{Q}$, $Z'_{u}$, $Z'_{d}$ (but we can omit  quark data because they are less precise that the leptons ones).  From ref.~\cite{Marandella:2005wd} we get that a generic model with $Z'$ boson contributes to the universal parameters as \cite{Marandella:2005wd}: 

\begin{equation}\label{Z}
 \begin{aligned}
 &  \hat{S} \approx \frac{M^2_{W}}{M^2_{Z'}} \left(b_W - b_{y}/t_{W} \right) \left( b_W - b_{y}t_{W}-\frac{2g_{Z'}Z'_{H}}{g}  \right), & W &\approx \frac{M^2_{W}}{M^2_{Z'}}b^2_W, & b_{W} &= \frac{2g_{Z'}}{Y_{e}g} \left(Z'_{e}Y_{L} - Z'_{L}Y_{e}  \right)\,,\\
& \hat{T} \approx \frac{M^2_{W}}{M^2_{Z'}}\left[ \left(c_{y}t_{W}+ \frac{2g_{Z'}Z'_{H}}{g} \right)^2-b^2_W  \right], & Y &\approx \frac{M^2_{W}}{M^2_{Z'}}b^2_y, & b_{y} &= \frac{g_{Z'}Z'_{e}}{g' Y_{e}}.
\end{aligned}
\end{equation}

In section \ref{sec:feynman} we can find all the necessary coefficients: 
\begin{equation}\label{z1}
\begin{aligned}
& g_{Z'} = \frac{g}{\sqrt{1-\frac{t^2_W}{3}}}, & Z'_{L} &= Z'_{H}= \frac{1}{2\sqrt{3}}-\frac{\sqrt{3}}{2}s^2_{Z'}, & \frac{g'}{g} = t_{W} \\
& s^2_{Z'} = \frac{t^2_W}{3}, & Z'_{e} &= \sqrt{3}s^2_{Z'}. 
\end{aligned}
\end{equation}

With eqs.~\eqref{z1}, eqs.~\eqref{Z} reduce to:

\begin{equation}
\hat{T} \approx 0, \hspace{4mm} \hat{S} \approx \frac{4 M^2_{W}}{M^2_{Z'}\left(3-t^2_W \right)} = 4W = \frac{4Y}{t^2_W}\,.
\end{equation}

From the above equations, we see that corrections to the $T$ parameter due to the $Z'$ boson are negligible, however we could estimate them using the fact that $\hat{T} = \alpha T = \delta \rho$. We present our results taking into account the PDG average and the update including the new measurement of the $W$ boson mass.

\begin{table}[H]
\begin{center}
\begingroup
\renewcommand{\arraystretch}{1.4}
\begin{tabular}{c c c c c}
\toprule
\toprule
   & SM & EWPD &  $M_{W} = 80.357$ GeV & $M_{W} = 80.4242$ GeV \\
\midrule
\midrule
$\rho$  & 1  & $1.00038 \pm 0.00020$                                                &  $1.0004758$   &  $1.0016013$  \\
$\hat{T}$ & 0  &  -  & $5
\times 10^{-4}$ &  $1.6
\times 10^{-3}$   \\ 
$\hat{S}$ & 0  &  -  &  $7
\times 10^{-5}$   &  $7
\times 10^{-5}$     \\
$T$    & 0  & \begin{tabular}[c]{@{}c@{}}$0.03 \pm 0.12$\\ $ 0.05 \pm 0.06$\end{tabular} & $0.07$ 
& $0.22$
\\
$S$    & 0  & \begin{tabular}[c]{@{}c@{}}$-0.01 \pm 0.10$\\ $ 0.00 \pm 0.07$\end{tabular} & $0.008$
& $0.008$
\\ 
\bottomrule
\bottomrule
\end{tabular}
\endgroup
\end{center}
\caption{Values of oblique parameters according to EWPD and using instead $M_W$ as in the PDG \cite{ParticleDataGroup:2020ssz}, or from the CDF measurement \cite{CDF:2022hxs}. Two values are given for $T$ and $S$. The upper one is obtained fitting also $U$ (for which $0.02\pm0.11$ is obtained) and the second one setting $U=0$ \cite{ParticleDataGroup:2020ssz}.}
\label{tabla14}
\end{table}

\begin{figure}[H]
\begin{subfigure}{.5\textwidth}
  \centering
  \includegraphics[width=\textwidth]{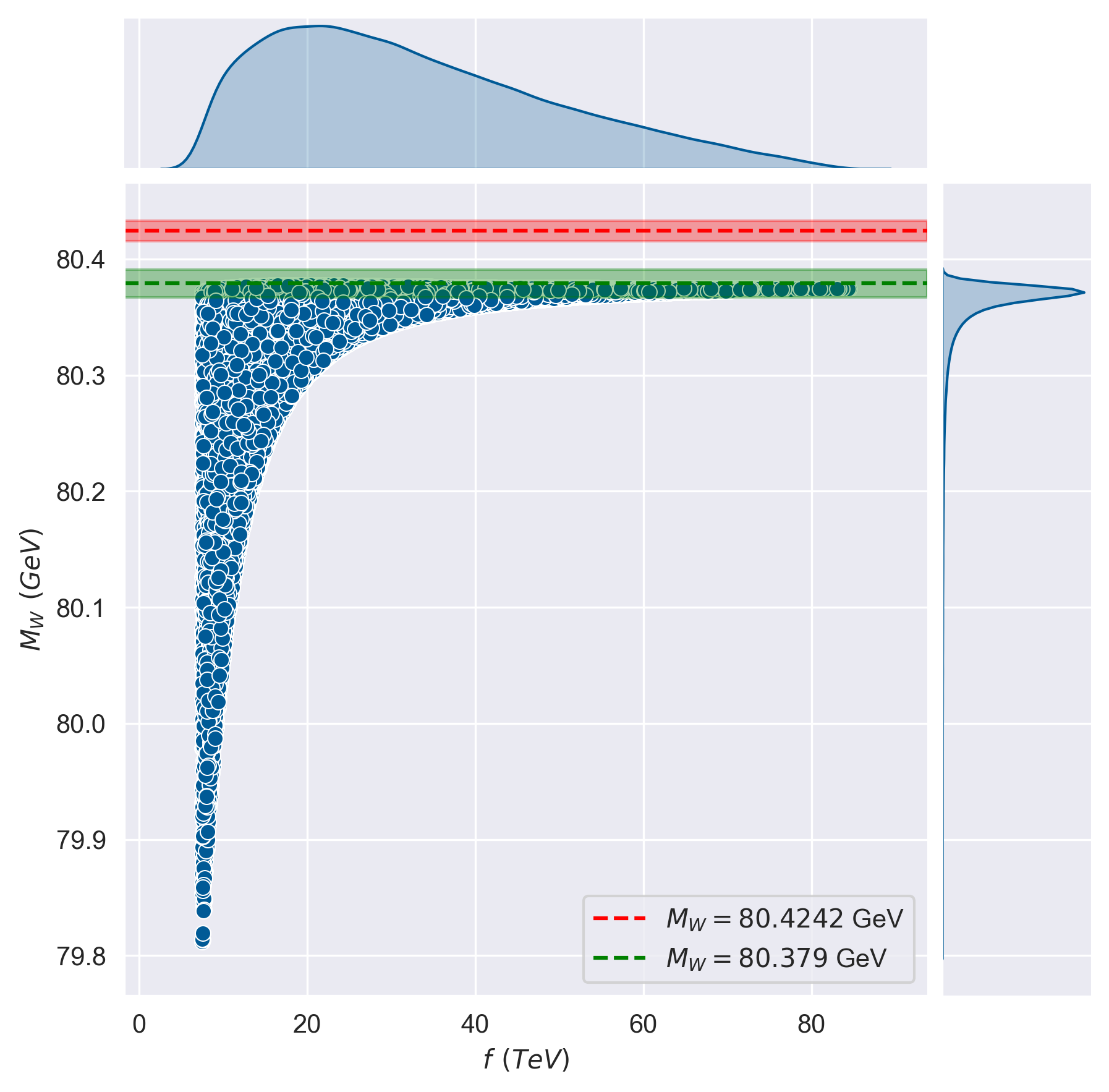}  
  \caption{Scatter plot using $M_W = 80.379$ GeV. }
  \label{fig:sub-treintaydos}
\end{subfigure}
\begin{subfigure}{.5\textwidth}
  \centering
  \includegraphics[width=\textwidth]{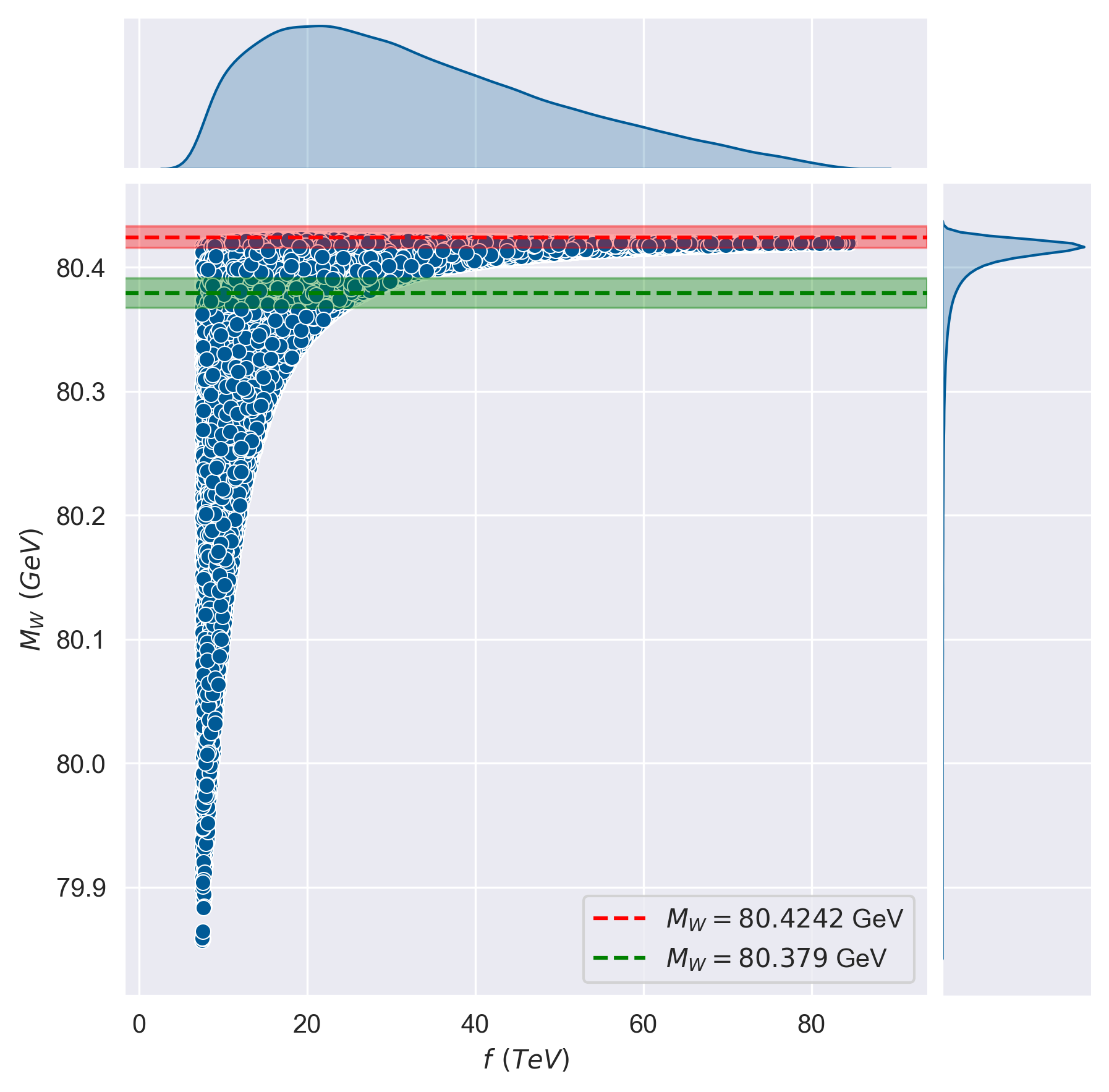}  
  \caption{Scatter plot using $M_W = 80.4242$ GeV. }
  \label{fig:sub-treintaytres}
\end{subfigure}
\caption{Corrections to the $W$ boson mass provided by the SLH compared to its measurement.}
\label{fig:fig7}
\end{figure}

\begin{figure}[H]
\begin{subfigure}{.5\textwidth}
  \centering
  \includegraphics[width=\textwidth]{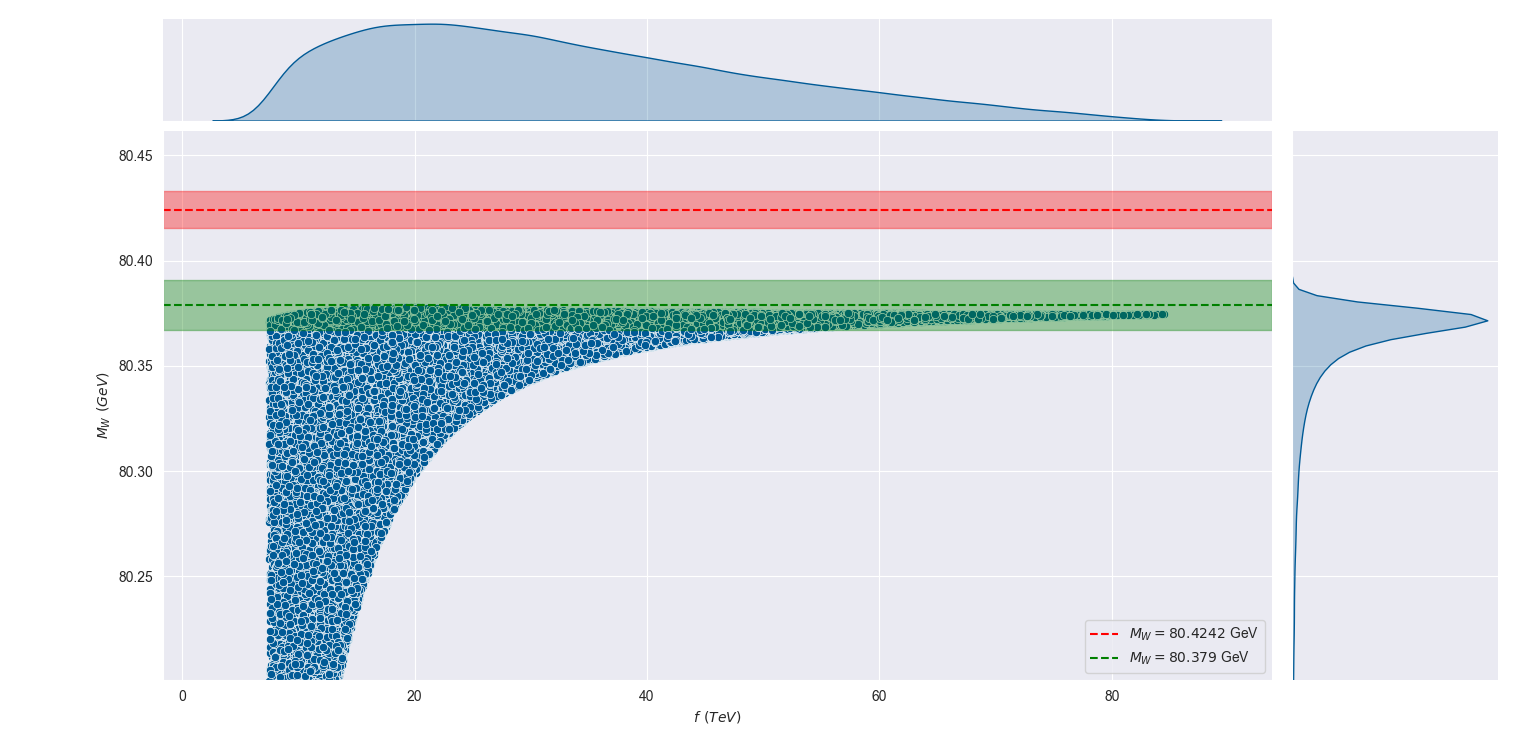}  
  \caption{Scatter plot using $M_W = 80.379$ GeV. }
  \label{fig:sub-treintaycuatro}
\end{subfigure}
\begin{subfigure}{.5\textwidth}
  \centering
  \includegraphics[width=\textwidth]{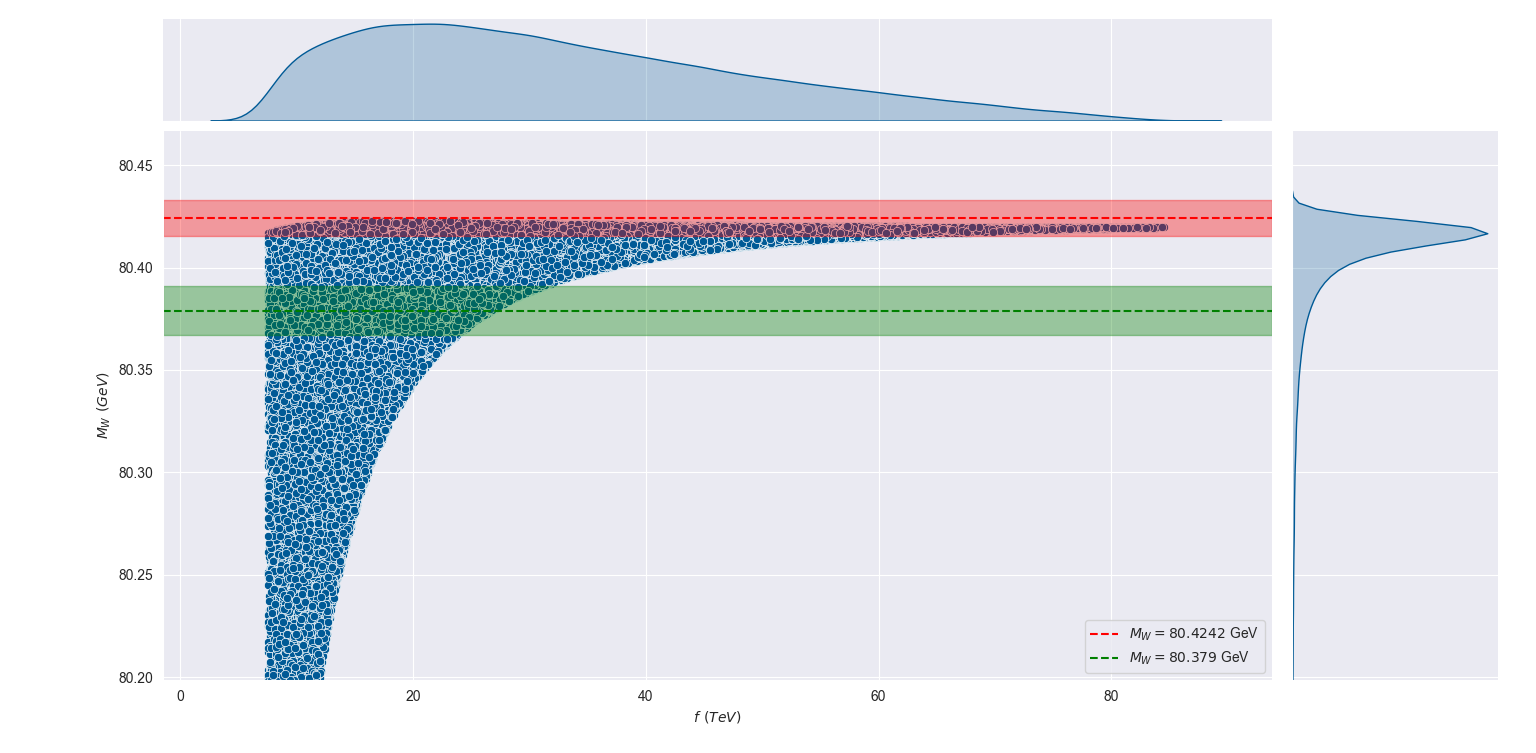}  
  \caption{Scatter plot using $M_W = 80.4242$ GeV. }
  \label{fig:sub-treintaycinco}
\end{subfigure}
\caption{Zoom in on figure \ref{fig:fig7}.}
\label{fig:fig8}
\end{figure}

Plots \ref{fig:fig7} and \ref{fig:fig8} show the corrections that SLH provided to the $W$ boson mass for different values of $f$ \footnote{$\tan\beta$ is also varied, although not shown. All pairs $(f,\tan\beta)$ considered satisfy experimental limits on the LFV processes studied before. Collider limits and unitarity bounds are also respected.}. When we use as input the PDG average, the corrections to $M_W$ agree within the  uncertainties. In the supplementary axes of plot \ref{fig:sub-treintaydos}, we show the distribution for the values of $M_W$ and $f$. For $M_W$ most of the values are around the PDG average, and for  $f$ most of them are within  $\left[8.5 , 40\right]$ TeV. Then, in plot \ref{fig:sub-treintaycuatro} we  zoom in to show that SLH reproduces the world average for the $W$ mass in the range $f \in \left[16,22 \right]$ TeV.  Also for larger $f$, values of $M_W$ are inside the uncertainties. For plots \ref{fig:sub-treintaytres} and  \ref{fig:sub-treintaycinco} we use as input the new world average  including the CDF II result. For the range $f \in \left[8,27 \right]$ TeV, the PDG world average and the new one  can be reproduced in the SLH model but the marginal distribution shows that getting the PDG average is unlikely. In the range $f \in \left[23,84 \right]$ TeV, the $W$ mass is always below the central value, but still within its uncertainties.

\begin{figure}[H]
\begin{subfigure}{.5\textwidth}
  \centering
  \includegraphics[width=\textwidth]{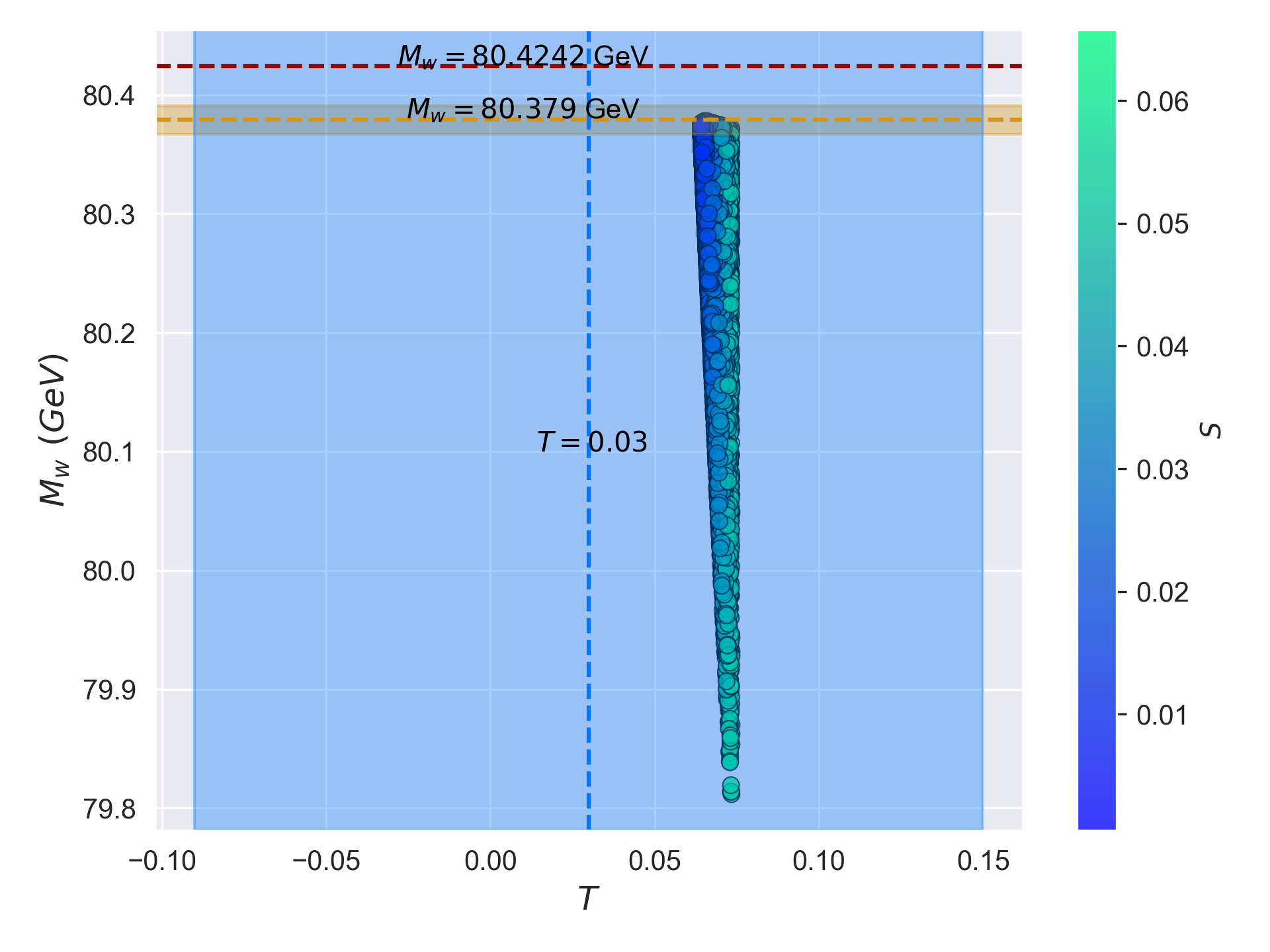}  
  \caption{Scatter plot using $M_W = 80.379$ GeV. }
  \label{fig:sub-treintayseis}
\end{subfigure}
\begin{subfigure}{.5\textwidth}
  \centering
  \includegraphics[width=\textwidth]{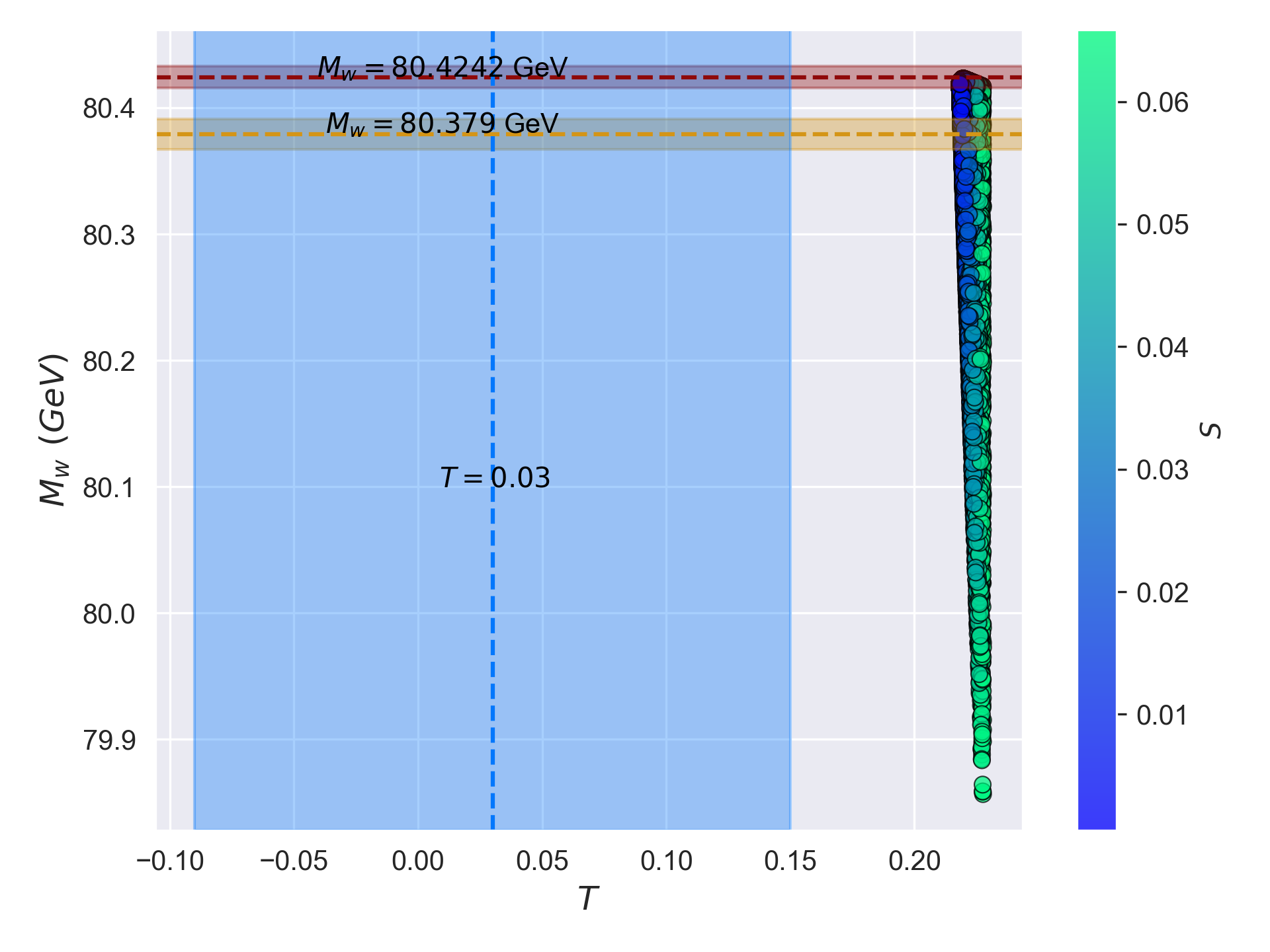}  
  \caption{Scatter plot using $M_W = 80.4242$ GeV. }
  \label{fig:sub-treintaysiete}
\end{subfigure}
\caption{Correction to the oblique parameters $S$ and $T$ in the SLH. }
\label{fig:fig9}
\end{figure}

Finally, figure \ref{fig:fig9}  shows the corrections to the oblique parameters $S$ and $T$ in the SLH, and for the different values of the $M_W$. In figure \ref{fig:sub-treintayseis} we use the PDG value, and show that the oblique parameters agree with the EWPD within  uncertainties. This means that although SLH modifies the $\rho$ and $T$ parameters, it is only slightly, without conflicting with the SM. For the $S$ parameter all  values agree with the SM as well. Now in plot \ref{fig:sub-treintaysiete} we show the corrections to the oblique parameters with the average including the CDF II measurement, and show that  $T$ is outside the EWPD (corrections to the $S$ parameter are negligible, as shown in  table \ref{tabla14}) confidence interval.  

\section{Conclusions}
\label{sec:Con}

One of the virtues of the SLH model is its minimality,  extending the SM gauge group to $SU(3)_{c} \times SU(3)_{L} \times U(1)_{x}$ attempting to understand the hierarchy problem related to the Higgs boson mass value in presence of generic new physics coupling to it. As a simple group model, it has a small number of unknown parameters and new heavy particles. These  allow the appearance of lepton flavor violation processes driven by three heavy neutrinos, with signals that could in principle be probed in current and forthcoming experiments. In this work we have examined the most relevant LFV processes: $\ell \rightarrow \ell_{a}$, $\ell \rightarrow \ell_{k} \ell_{a}\bar{\ell}_{b}$ and $\ell N \rightarrow \ell_{a} N$, which arise at one loop within the SLH. We have computed the relevant observables as an expansion in $v/f$, keeping the results at $\mathcal{O}(v^2/f^2)$. To carry out our numerical calculations, we have floated the free parameters within the allowed region, ensuring that all  experimental upper limits were satisfied.

As is well-known, processes with muons would most likely be the discovery channels for LFV. Within the SLH,  this can be expected either from conversions in nuclei, $\mu\to e\gamma$ or $\mu\to3e$. However, those with taus (not considered exhaustively in previous SLH analyses) will then be needed for characterizing the underlying new physics. We have found that -in analogy with the role of $\mu\to e$ conversion in nuclei amid $\mu\to e$ transitions-, $\ell\to\tau$ conversion in nuclei are synergic with the $\tau\to\ell\gamma$ and $\tau\to3\ell$ decays in probing LFV transitions involving taus. We hope that our work and other recent related studies motivate the experimental collaborations (Belle-II, NA64, EIC, muon collider, etc.) to  pursue the corresponding dedicated searches. Finally, we verified that -although the SLH modifies the $\rho$ parameter and can in principle accommodate the recent CDF $M_W$ measurement-, this is in tension with electroweak precision data.

\section{Acknowledgments}
We thank J. I. Illana, G. Hernández Tomé, E. Arganda, F.~Fortuna, I.~Pacheco and E.~Gutiérrez for useful discussions on this topic. E.~R. acknowledges financial support from CONACyT graduate grants program No. 728500. P.~R is indebted to Cátedras Marcos Moshinsky (Fundación Marcos Moshinsky) and CONACyT (`Paradigmas y Controversias de la Ciencia 2022', Project No. 319395).

\appendix
\section{Loop functions}
\label{sec:jiji}

A general one-loop tensor integral in $D$ dimensions with $N$ legs and $P$ 
integration momenta in the numerator is represented as \cite{Denner:1991kt}:

\begin{equation}\label{in}
    \frac{i}{16 \pi^2} T^{N}_{\mu_{1},\ldots,\mu_{P}} \equiv \mu^{4-D} \int \frac{d^{D} q}{(2 \pi)^{D}} \frac{q_{\mu_{1} \ldots q_{\mu_{P}}}}{D_{0}D_{1}\ldots D_{N-1}},
\end{equation}
where

\begin{equation}
    D_{0} = q^2-m^2_{0} + i \epsilon, \hspace{4mm} D_{i} = (q+k_i)^2-m^2_{i} + i \epsilon, \hspace{4mm} i = 1,\ldots,N-1.
\end{equation}
The vectors $K_{i}$ are the sum of external momenta $p_{i}$:

\begin{equation}
    k_1 = p_1, \hspace{4mm} k_2 = p_1+p_2, \hspace{2mm} \ldots, \hspace{2mm} k_{N-1} = \sum^{N-1}_{i=1}p_{i}.
\end{equation}

These tensor integrals are invariant under permutations of propagators $D_i$ and symmetric in the Lorentz indices. Generally we define $T^{1} = A$, $T^2=B$, etc. The scalar integrals are $A_0$, $B_0$, etc. Lorentz covariance allows decomposing  equation \eqref{in} into a linear combination of tensors constructed with the metric and the external momenta, however this basis is not unique, we could use any set of linearly independent momenta and $g_{\mu \nu}$ \cite{Passarino:1978jh}. For this work we use the decompositions:

\begin{equation}
    \begin{aligned}
    & B_{\mu} = k_{1 \mu}B_{1}, \\
    & C_{\mu} = k_{1 \mu}C_{1} + k_{2 \mu}C_{2}, \\
    & C_{\mu \nu} = g_{\mu \nu} C_{00} + \sum^2_{i,j=1} k_{i \mu}k_{j \nu}C_{ij}, \\
    & D_{\mu} = \sum^{3}_{i =1}k_{i \mu}D_{i}, \\
    & D_{\mu \nu} = g_{\mu \nu} D_{00} + \sum^3_{i,j=1} k_{i \mu}k_{j \nu}D_{ij}.
    \end{aligned}
\end{equation}

These functions have been calculated for the argument configuration required by the processes under study, obtaining the  results presented in the following. All of them agree with those collected in the appendix B of ref. \cite{delAguila:2008zu}.

\subsection*{Two-point functions}
There appear:

\begin{equation}
\begin{aligned}
\frac{i}{16 \pi^2} \{B_{0}, B^{\mu} \}(args) = \mu^{4-D}\int \frac{d^{D}q}{\left(2\pi \right)^{D}} \frac{\{1,q^{\mu} \}}{\left(q^2-m^{2}_{0} \right)\left[\left(q+p\right)^2-m^{2}_{1} \right]}.
\end{aligned}
\end{equation}
Their tensor coefficients depend on the invariant quantities (args)$= \left(p^2,m^{2}_{0},m^{2}_{1} \right)$. Functions $B\equiv B \left(0,M^{2}_{1},M^{2}_{2}\right)$ and $\overline{B}\equiv B \left(0,M^{2}_{2},M^{2}_{1}\right)$ read 

\begin{equation}
\begin{aligned}
B_{0}= \overline{B}_{0} = \Delta_{\epsilon}+1-\frac{M^{2}_{1} \log \left( \frac{M^{2}_{1}}{\mu^{2}} \right)-M^{2}_{2} \log \left( \frac{M^{2}_{2}}{\mu^{2}} \right)}{M^{2}_{1}-M^{2}_{2}}, 
\end{aligned}
\end{equation}

\begin{equation}
\begin{aligned}
B_{1}=& -\frac{\Delta_{\epsilon}}{2}+ \frac{4M^{2}_{1}M^{2}_{2}-3M^{4}_{1}-M^{4}_{2}+2 M^{4}_{1} \log\left(\frac{M^{2}_{1}}{\mu^{2}} \right)+ 2 M^{2}_{2} \left( M^{2}_{2}-2M^{2}_{1} \right) \log\left(\frac{M^{2}_{2}}{\mu^{2}} \right)}{4 \left(M^{2}_{1}-M^{2}_{2} \right)^{2}} \\
&= - \overline{B}_{0}-\overline{B}_{1},
\end{aligned}
\end{equation}
with $\Delta_{\epsilon}= \frac{1}{\epsilon}-\gamma+ \log(4\pi)$ encoding  the ultraviolet divergences in $D = 4$ dimensions.

\subsection*{Three-point functions}
In this case there arise: 

\begin{equation}
    \begin{aligned}
        \frac{i}{16 \pi^2}  \{ C_{0},C^{\mu}, C^{\mu \nu}   \}(args) = \mu^{4-D} \int \frac{d^{D}q}{\left(2\pi \right)^{D}} \frac{ \{ 1, q^{\mu}, q^{\mu} q^{\nu} \} }{\left(q^2-m^{2}_{0} \right) \left[ \left(q+p_1 \right)^2 -m^{2}_{1}\right]\left[ \left(q+p_2 \right)^2 -m^{2}_{2}\right]},
    \end{aligned}
\end{equation}
with $p^{2}=p^{2}_{1}=0$ and $Q^{2}= \left( p-p_1\right)^{2}$, so that  only the following general types are necessary for us: $C= C \left(0,Q^{2},0,M^{2}_{1},M^{2}_{2},M^{2}_{2} \right)$ (we define the mass ratio $x= M^{2}_{2}/M^{2}_{1}$):

\begin{equation}
    \begin{aligned}
        C_{0}= \frac{1}{M^{2}_{1}} \left[ \frac{1-x+ \log (x)}{(1-x)^2} + \frac{Q^{2}}{M^{2}_{1}} \left( \frac{-2-3x+6x^2-x^3-6x \log(x)}{12(1-x)^4}   \right) \right] + \mathcal{O}(Q^4),
    \end{aligned}
\end{equation}

\begin{equation}
\begin{aligned}
C_{1}=C_{2}= \frac{1}{M^{2}_{1}} \frac{4x-3-x^2-2 \log(x)}{4(1-x)^3}+ \mathcal{O}(Q^2),
\end{aligned}
\end{equation}

\begin{equation}
\begin{aligned}
C_{11}=C_{22}= 2C_{12}=\frac{1}{M^{2}_{1}} \frac{11-18x+9x^2-2x^3 +6 \log(x)}{18(1-x)^4}+ \mathcal{O}(Q^2),
\end{aligned}
\end{equation}

\begin{equation}
\begin{aligned}
C_{00}=- \frac{1}{2}B_{1}-\frac{Q^2}{M^{2}_{1}} \left( \frac{11-18x+9x^2-2x^3+6 \log(x)}{72(1-x)^4}\right) + \mathcal{O}(Q^4).
\end{aligned}
\end{equation}

Now, defining $\overline{C}=C \left(0,Q^{2},0,M^{2}_{2},M^{2}_{1},M^{2}_{1} \right)$: 

\begin{equation}
    \begin{aligned}
       \overline{C}_{0}= \frac{1}{M^{2}_{1}} \left[ \frac{-1+x-x \log(x)}{(1-x)^2} + \frac{Q^{2}}{M^{2}_{1}} \left( \frac{-1+6x-3x^2-2x^3+ 6x^2 \log(x)}{12(1-x)^4}   \right) \right] + \mathcal{O}(Q^4),
    \end{aligned}
\end{equation}

\begin{equation}
\begin{aligned}
\overline{C}_{1}= \overline{C}_{2}= \frac{1}{M^{2}_{1}} \frac{1-4x+3x^2-2x^2 \log(x)}{4(1-3)^3}+ \mathcal{O}(Q^2),
\end{aligned}
\end{equation}

\begin{equation}
\begin{aligned}
\overline{C}_{11}=\overline{C}_{22}= 2\overline{C}_{12}=\frac{1}{M^{2}_{1}} \frac{-2+9x-18x^2+11x^3-6x^3\log(x)}{18(1-x)^4}+ \mathcal{O}(Q^2),
\end{aligned}
\end{equation}

\begin{equation}
\begin{aligned}
\overline{C}_{00}=- \frac{1}{2}\overline{B}_{1}-\frac{Q^2}{M^{2}_{1}} \left( \frac{-2+9x-18x^2+11x^3-6x^3 \log(x)}{72(1-x)^4}\right) + \mathcal{O}(Q^4).
\end{aligned}
\end{equation}
Alternatively, defining $\hat{C}= \left(0,Q^{2},0,M^{2}_{1},M^{2}_{2},0 \right)$, which is symmetric under the exchange $M_{1} \leftrightarrow M_{2}$,

\begin{equation}
\begin{aligned}
\hat{C}_{00}= \frac{1}{8} \left( 3+2\Delta_{\epsilon} -2 \log \left( \frac{M^{2}_{1}}{\mu^{2}} \right) \right)+ \frac{x\log(x)}{4(1-x)} + \mathcal{O}(Q^2).
\end{aligned}
\end{equation}

\subsection*{Four-point functions}

Those needed are all ultraviolet convergent:

\begin{equation}
    \begin{aligned}
        \frac{i}{16 \pi^2}  \{ D_{0},D^{\mu}, D^{\mu \nu}   \}(args) = \mu^{4-D} \int \frac{d^{D}q}{\left(2\pi \right)^{D}} \frac{ \{ 1, q^{\mu}, q^{\mu} q^{\nu} \} }{\left(q^2-m^{2}_{0} \right) \left[ \left(q+p_1 \right)^2 -m^{2}_{1}\right]\left[ \left(q+p_2 \right)^2 -m^{2}_{2}\right]\left[ \left(q+p_3 \right)^2 -m^{2}_{3}\right]},
    \end{aligned}
\end{equation}
with $(\textrm{args})= \left(p^{2}_{1},p^{2}_{2},p^{2}_{3},p^{2}_{4}, \left(p_1+p_2 \right)^2,\left(p_2+p_3 \right)^2; m^{2}_{0},m^{2}_{1},m^{2}_{2},m^{2}_{3} \right)$. Zero external momenta will be set, so we only need:

\begin{equation}
\frac{i}{16 \pi^2}D_0 = \mu^{4-D} \int \frac{d^{D}q}{\left( 2 \pi \right)^{D}} \frac{1}{\left(q^2-m^{2}_{0} \right)\left(q^2-m^{2}_{1} \right)\left(q^2-m^{2}_{2} \right)\left(q^2-m^{2}_{3} \right)},
\end{equation}

\begin{equation}
\frac{i}{16 \pi^2}D_{00} = \frac{ \mu^{4-D}}{4} \int \frac{d^{D}q}{\left( 2 \pi \right)^{D}} \frac{q^2}{\left(q^2-m^{2}_{0} \right)\left(q^2-m^{2}_{1} \right)\left(q^2-m^{2}_{2} \right)\left(q^2-m^{2}_{3} \right)}.
\end{equation}
In terms of the mass ratios $x=m^2_{1}/m^2_{0}$, $y=m^2_{2}/m^2_{0}$, $z=m^2_{3}/m^2_{0}$ the previous  integrals read:

\begin{equation}
\begin{aligned}
d_{0} (x,y,z)\equiv m^{4}_{0} D_0 &=  \frac{x \log[x]}{\left(1-x \right)\left(x-y \right)\left(x-z \right)} -\frac{y \log[y]}{\left(1-y \right)\left(x-y \right)\left(y-z \right)} \\
& + \frac{z \log[z]}{\left(1-z \right)\left(x-z \right)\left(y-z \right)} 
\end{aligned}
\end{equation}

\begin{equation}
\begin{aligned}
\tilde{d}_{0} (x,y,z)\equiv 4m^{2}_{0} D_{00} &=  \frac{x^2 \log[x]}{\left(1-x \right)\left(x-y \right)\left(x-z \right)} -\frac{y^2 \log[y]}{\left(1-y \right)\left(x-y \right)\left(y-z \right)} \\
& + \frac{z^2 \log[z]}{\left(1-z \right)\left(x-z \right)\left(y-z \right)}.
\end{aligned}
\end{equation}
The case in which $z \rightarrow 1$ (argument omitted below) is also necessary, reading:

\begin{equation}
\begin{aligned}
d_{0} (x,y)&= - \Big[  \frac{x \log[x]}{\left(1-x \right)^2\left(x-y \right)} -\frac{y \log[y]}{\left(1-y \right)^2\left(x-y \right)} + \frac{1}{\left(1-x\right)\left(1-y \right)} \Big],
\end{aligned}
\end{equation}

\begin{equation}
\begin{aligned}
\tilde{d}_{0} (x,y) &= - \Big[ \frac{x^2 \log[x]}{\left(1-x \right)^2\left(x-y \right)} -\frac{y^2 \log[y]}{\left(1-y \right)^2\left(x-y \right)} + \frac{1}{\left(1-x \right)\left(1-y \right)} \Big].
\end{aligned}
\end{equation}


\printbibliography

\end{document}